\documentclass[twocolumn,tighten]{aastex62}

\usepackage{graphicx}
\usepackage{ulem}
\usepackage{amsmath}
\usepackage{wrapfig}

\usepackage{tikz}
\usepackage{ulem}

\definecolor{green}{rgb}{0.0, 0.5, 0.0}
\definecolor{brightube}{rgb}{0.82, 0.62, 0.91}

\def\Msun{{M_\odot}}

\newcommand\lsim{\mathrel{\rlap{\lower4pt\hbox{\hskip1pt$\sim$}}
        \raise1pt\hbox{$<$}}}
\newcommand\gsim{\mathrel{\rlap{\lower4pt\hbox{\hskip1pt$\sim$}}
        \raise1pt\hbox{$>$}}}

\begin{document}
\shorttitle{IMBHs escaping from YSCs}
\shortauthors{Gonz\'{a}lez Prieto et al.}

\title{Intermediate-mass Black Holes on the Run from Young Star Clusters}

\correspondingauthor{Elena Gonz\'{a}lez Prieto}
\email{elena.prieto@northwestern.edu}

\author[0000-0002-0933-6438]{Elena Gonz\'{a}lez Prieto}
\affil{Center for Interdisciplinary Exploration \& Research in Astrophysics (CIERA) and Department of Physics \& Astronomy, Northwestern University, Evanston, IL 60208, USA}

\author[0000-0002-4086-3180]{Kyle Kremer}
\affiliation{TAPIR, California Institute of Technology, Pasadena, CA 91125, USA}
\affiliation{The Observatories of the Carnegie Institution for Science, Pasadena, CA 91101, USA}

\author[0000-0002-7330-027X]{Giacomo Fragione}
\affil{Center for Interdisciplinary Exploration \& Research in Astrophysics (CIERA) and Department of Physics \& Astronomy, Northwestern University, Evanston, IL 60208, USA}

\author[0000-0001-5285-4735]{Miguel A.S.~Martinez}
\affil{Center for Interdisciplinary Exploration \& Research in Astrophysics (CIERA) and Department of Physics \& Astronomy, Northwestern University, Evanston, IL 60208, USA}

\author[0000-0002-9660-9085]{Newlin C.~Weatherford}
\affil{Center for Interdisciplinary Exploration \& Research in Astrophysics (CIERA) and Department of Physics \& Astronomy, Northwestern University, Evanston, IL 60208, USA}

\author[0000-0002-0147-0835]{Michael Zevin}
\affiliation{Kavli Institute for Cosmological Physics, The University of Chicago, 5640 South Ellis Avenue, Chicago, IL 60637, USA}
\affiliation{Enrico Fermi Institute, The University of Chicago, 933 East 56th Street, Chicago, IL 60637, USA}

\author[0000-0002-7132-418X]{Frederic A.~Rasio}
\affil{Center for Interdisciplinary Exploration \& Research in Astrophysics (CIERA) and Department of Physics \& Astronomy, Northwestern University, Evanston, IL 60208, USA}

\begin{abstract}

The existence of black holes (BHs) with masses in the range between stellar remnants and supermassive BHs has only recently become unambiguously established. GW190521, a gravitational wave signal detected by the LIGO/Virgo Collaboration, provides the first direct evidence for the existence of such intermediate-mass BHs (IMBHs). This event sparked and continues to fuel discussion on the possible formation channels for such massive BHs. As the detection revealed, IMBHs can form via binary mergers of BHs in the ``upper mass gap'' ($\approx40$--$120\,\Msun$). Alternatively, IMBHs may form via the collapse of a very massive star formed through stellar collisions and mergers in dense star clusters. In this study, we explore the formation of IMBHs with masses between $120$ and $500\,\Msun$ in young, massive star clusters using state-of-the-art Cluster Monte Carlo (\texttt{CMC}) models. We examine the evolution of IMBHs throughout their dynamical lifetimes, ending with their ejection from the parent cluster due to gravitational radiation recoil from BH mergers, or dynamical recoil kicks from few-body scattering encounters. We find that \textit{all} of the IMBHs in our models are ejected from the host cluster within the first $\sim 500$~Myr, indicating a low retention probability of IMBHs in this mass range for globular clusters today. We estimate the peak IMBH merger rate to be $\mathcal{R} \approx 2 \, \rm{Gpc}^{-3}\,\rm{yr}^{-1}$ at redshift $z \approx 2$.

\vspace{1cm}
\end{abstract}

\section{Introduction}
\label{sec:intro}

The first binary black hole (BBH) detection in 2015 \citep{Abbott2016} revolutionized the field of gravitational wave (GW) physics. Since then, the growing catalog of GW events has sparked debates about the environments that produce these sources \citep[e.g.,][]{GWTC2_gr2020,GWTC2_grb2020,LIGO2020_O3}. Several features of these detections challenge our current understanding of stellar evolution and BH formation \citep[e.g.,][]{LIGO2020_O3populations}. An example of this is GW190521 \citep{GW190521}, a BBH merger with a remnant mass of $\sim150\,\Msun$ whose component masses both exceed the range expected from standard isolated stellar evolution. This event represents the first \textit{direct} detection of an intermediate-mass BH (IMBH). As the number of GW detections grows and second-generation detectors such as the Laser Interferometer Space Antenna (LISA) come online, detailed stellar evolution models for massive stars are essential to our understanding of the observed BH population.

The evolutionary stages of massive stars, the progenitors of compact objects such as BHs, are expected to leave strong features on the BH mass spectrum. Massive stars with core masses between roughly $64$ and $135\,\Msun$ (${150 \lesssim M_{\rm ZAMS}/ \Msun \lesssim 260}$ at ${Z \sim 0.1\,Z_\odot}$) undergo pair-instability supernovae (PISNe) that completely destroy the star, leaving no remnant behind \citep[e.g.,][]{Fowler1964, Ober1983, Bond1984, Belczynski2016b, Woosley2017, Woosley2019}. At lower core masses, between $32$ and $64\,\Msun$ (${70 \lesssim M_{\rm ZAMS}/ \Msun \lesssim 150}$), pulsational pair-instability supernovae (PPISNe) induce severe mass loss episodes, which limit the remnant mass \citep[e.g.,][]{HegerWoosley2002}. These processes should produce an ``upper mass gap'' in the BH mass distribution in the range \mbox{$M \approx 40$--$120\,\Msun$} \citep[e.g.,][]{Spera17, limongi2018, Takahashi2018, Stevenson2019, Marchant2019, Farmer2019, Mapelli2020, Renzo2020, Bel2020_pisne}. The exact boundaries of this gap are highly uncertain, with recent studies showing that the uncertainty on the carbon reaction rate can shift the lower boundary to $\sim 90\,\Msun$ \citep{Farmer_2020,Costa2020}; in this case, pair-instability-gap BBH mergers like GW190521 may be feasible through isolated binary evolution \citep{Belczynski_2020}. A second prominent feature expected from these evolutionary processes is a slight excess of BHs with masses in the range $30$--$40\,\Msun$ \citep{Stevenson2019}. In its latest observing run, the LIGO-Virgo-Kagra (LVK) Collaboration reports a significant over-density of BHs at $M \approx 30\,\Msun$ \citep{LVK21}.

Several formation channels for BHs in the upper mass gap have been discussed, including hierarchical mergers of lower-mass BHs \citep[e.g.,][]{MillerHamilton2002,McKernan2012,Rodriguez2018b,Rodriguez2019,AntoniniGieles2019,GerosaBerti2019,FragioneLoeb2020,FragioneSilk2020,FragioneKocsis2022,Zevin22}, stellar mergers in dense star clusters \citep[e.g.,][]{PortegiesZwartMcMillan2002,DiCarlo19,Rizzuto,Kremer20,Banerjee_2020,Banerjee_2020_b, Weatherford21, Gonzalez_2021, Banerjee2022}, BH growth through gas accretion in star-forming environments \citep[e.g.,][]{Roupas2019}, BH formation through gravitational instabilities in the early universe \citep[e.g.,][]{LoebRasio1994,Carr2016}, remnants of population III stars \citep[e.g.,][]{Madau_2001,BrommLarson2004}, BH-star collisions in dense star clusters \citep{Giersz2015,Rizzuto2022}, and IMBH formation from BH-star collisions in galactic nuclei \citep[e.g.,][]{Rose2022}.

In this paper we explore the formation of  ``low-mass'' IMBHs with masses between $120$ and $500\,\Msun$. This covers the lower range of IMBH masses (typically defined as \mbox{$10^{2}$--$10^{5}\,\Msun$)}. We explore the formation of these IMBHs through repeated stellar collisions. Two distinct paths exist within this scenario, the first involving collisions between main sequence stars and evolved stars that have a small enough core to avoid the pair instability. \cite{Spera19} suggests that such a merger produces a star with the same initial core as the giant, but with an oversized hydrogen envelope.This merger remnant could avoid the regime where pair instability occurs in the core and collapse into a BH more massive than those formed through single-star evolution.The second path produces an IMBH via the collapse of a progenitor whose core is above the pair instability threshold due to multiple previous massive giant collisions, as seen in \cite{Kremer20}. We consider the general repeated stellar collisions channel described above, as well as IMBHs resulting from BBH mergers. In contrast with previous work on formation of very massive stars \mbox{($M > 1000 \,\Msun$)} in the classic collisional runaway scenario \citep[e.g.,][]{Zwart,Gurkan,Giersz2015,Mapelli2016}, the progenitors of our IMBHs are products of only a few collisions, and thus more than one IMBH may form in a cluster at a time. Our study is limited to IMBHs with \mbox{$M < 500\,\Msun$} because  treatment of such massive stars/BHs would require physics beyond what is currently implemented in \texttt{CMC}  \citep[e.g., loss cone physics,][]{Umbreit2012U}. 

In studies of Young Massive Clusters (YMCs) with $N \approx 10^{3}$, \cite{DiCarlo19} showed that BHs with masses up to $\sim 120\,\Msun$ may form through BBH mergers but are unlikely to be retained in their host cluster. In a later study with $N \approx 10^4$, \cite{DiCarlo21} showed that repeated stellar collisions can produce stars massive enough to directly collapse into IMBHs. In our simulations of globular cluster (GC) progenitors with $N \approx 10^{5} $, IMBHs also form through repeated stellar collisions, thus hinting that IMBH formation in star clusters requires massive and highly dense stellar systems, where multiple consecutive stellar collisions are common \citep{Gonzalez_2021, Kremer20}.

The exact properties of massive-star merger products remain highly uncertain, but progress is ongoing. Notably, \cite{Costa22} and \cite{Ballone22} recently used hydrodynamic simulations to model the mass loss and chemical composition of stellar collision products. Evolving the collision products using the stellar evolution codes \texttt{Parsec} and \texttt{MESA}, they showed that the collision products can indeed avoid pair instability and collapse into BHs in the upper mass gap. We note, however, that these studies were limited to specific collision scenarios and further study is necessary to determine if this outcome can be generalized to different masses and stellar types.

The paper is organized as follows. In Section~\hyperref[sec:methods]{2}, we detail the cluster modeling methodology. In Sections~\hyperref[sec:formation]{3} and \hyperref[sec:companions]{4} we examine IMBH formation and evolution in the models, including an analysis of IMBH binary properties. We also explore in detail the mechanisms of IMBH ejection from the cluster and BBH mergers in Sections~\hyperref[sec:ejection]{5} and \hyperref[sec:mergers]{6}, respectively. Finally, we conclude in Section~\hyperref[sec:discussion]{7} with a discussion of the implications and uncertainties in our analysis.

\vspace{0.5cm}
\section{Methods}
\label{sec:methods}

We perform numerical simulations using \texttt{CMC} (for \texttt{Cluster Monte Carlo}), a Hénon-type Monte Carlo code that models the evolution of stellar clusters \citep[][for the most recent review]{Pattabiraman2013,Kremer20CMC, Rodriguez_2022}. This code incorporates prescriptions for various physical processes including two-body relaxation \citep{Joshi2000}, stellar/binary evolution using the population synthesis code \texttt{COSMIC} \citep{Breivik19}, direct integration of small-$N$ strong encounters using \texttt{Fewbody} \citep{Fregeau_2007}, and stellar collisions \citep{Fregeau_2007}. 

The present study is based on the set of models listed in Table~\hyperref[table:models]{1}.
All models consist of $8 \times 10^5$ objects, corresponding to an initial total cluster mass of  $\approx 5 \times 10^5\,\Msun$. The metallicity is set to $0.002$ ($0.1\,Z_\odot$) and the initial conditions are King models with concentration parameter $W_0=5$. Stellar masses are sampled from a \citet{Kroupa2001} initial mass function (IMF) in the range $0.08$--$150\,\Msun$. We set the virial radius of the cluster to $1$~pc.

As we are focused specifically on massive BHs, we require that at least one BH with a mass greater than $50\,\Msun$ be present after the first $100$~Myr of each model; if no such objects form in a given model or if they are all quickly ejected, we stop the model prematurely. Additionally, we note that one of our models, \texttt{1e}, produced a massive BH ($M \sim 460\,\Msun$) which caused a small time step and thus became prohibitively computationally expensive. We argue that this IMBH will most likely not remain in the cluster since BHs of similar masses in other models are ejected.

For the purpose of this paper and for consistency with previous studies, we define ``pair-instability gap'' (or ``upper mass gap'') as BHs  with masses in the range $40.5$--$120\,\Msun$, as determined by our assumed prescriptions for pair-instability physics \citep[for details, see][]{Belczynski2016b,Kremer20}. Furthermore, we use the term ``IMBH'' to refer specifically to BHs with $M>120\,\Msun$, beyond our assumed upper boundary for the pair-instability gap and ``massive BH'' as a general term to refer to any BH with mass greater than $40.5\,\Msun$.

In our models we assume that BHs formed from stellar collapse have zero effective spin. This assumption is consistent with predictions of near-zero BH natal spins from theoretical models of angular momentum transport in massive stars \citep{Fuller&Ma, Fuller19}. We discuss the caveats of this assumption in Section~\hyperref[sec:caveats]{7.3}. Additionally, BHs formed from BH mergers are assigned spins from a distribution that is isotropically distributed on a sphere.

\subsection{Binary Fraction}
\label{sec:binary fraction}

Studies of the Galactic field demonstrate that the binary fractions of O- and B- type stars are nearly $100\%$ \citep[e.g.,][]{Sana2012,MoeDiStefano2017}. Furthermore, YMCs, the likely progenitors of GCs, have binary fractions comparable to those seen in the field \citep[e.g.,][]{Sana2009}. Thus, it is possible that even though GCs currently have low binary fractions ($\lesssim10\%$), their primordial binary fractions may have been higher in the past \citep[e.g.,][]{Ivanova_2005,Milone2012}.

The importance of binaries in star cluster evolution is well known \citep{HeggieHut2003,Chatterjee_2010,Chatterjee2013}. In particular, they act as a dynamical energy source that effectively heats the cluster and slows gravothermal contraction \citep[e.g.,][]{HeggieHut2003}. In addition, binaries increase stellar collision rates \citep[e.g.,][]{Fregeau_2007, Bacon1996} and BH merger rates \citep[e.g.,][]{Chatterjee2017}. \cite{Gonzalez_2021} demonstrated that increasing the primordial binary fraction for massive stars from 0 to 1, as suggested by observations \citep{Sana2012}, doubled the number of massive-star collisions and produced BHs with masses within and above the upper mass gap. In this follow-up work, we examine in more detail the influence of binary fractions on BH formation, and study the long-term dynamical evolution and retention of the IMBHs formed in the models.

We define the high-mass binary fraction, $f_{\rm{b,high}}$, as the fraction of objects with masses above $15\,\Msun$ that have a companion at the time of cluster formation.  For all models, the low-mass ($<15\,\Msun$) binary fraction is fixed at $0.05$. This value is motivated by observations of low binary fractions in GCs \citep[e.g.,][]{Milone2012} and detailed studies of the evolution of the binary fraction in dense star clusters \citep[e.g.,][]{Fregeau2009}. The $f_{\rm{b,high}}$ is set to 0.5, 0.75, and 1. We run 15 realizations each for the models with $f_{\rm{b,high}} = (0.5, 1)$ and 25 realizations for those with $f_{\rm{b,high}} = 0.75$.

 For low-mass binaries, primary masses are drawn randomly from the Kroupa IMF, secondary masses are drawn assuming a flat mass ratio distribution in the range $0.1$--$1\,\Msun$ \citep[e.g.][]{Duquennoy&Mayor}, and initial orbital periods are drawn from a log-uniform distribution $dn / d \log P \propto P$. For the secondaries of the massive stars ($> 15\,\Msun$), a flat mass ratio distribution in the range [0.6,1] is assumed and initial orbital periods are drawn from the distribution $dn / d \log P \propto P^{-0.55}$ \citep[e.g.,][]{Sana2012}. For all binaries, the initial orbital periods are drawn from the Roche limit (using the stars' ZAMS masses and radii) to the hard-soft boundary and eccentricities are assumed to be thermal \citep{Heggie1975}.

\startlongtable
\begin{deluxetable*}{l| c c c c c c c| c c c}
\tabletypesize{\scriptsize}
\tablewidth{0pt}
\setlength{\tabcolsep}{1.2\tabcolsep}  
\tablecaption{List of cluster models \label{table:models}}
\tablehead{
    \colhead{$\rm Model$} &
	\colhead{$f_{\rm{b,high}}$} &
	\colhead{$t_{\rm{BH,ejec}}$} &
	\colhead{$N_{\rm{BH}}$} &
	\colhead{$N_{\rm{PI gap}}$} &
	\colhead{$N_{\rm{IMBH}}$} &
	\multicolumn{2}{c}{$M_{\rm{BH,max}}$} &
	\multicolumn{3}{c}{$\rm BBH \, Mergers$} \\
	\colhead{} &
	\colhead{} &
	\colhead{$\rm [Gyr]$} &
	\colhead{($N > 2 \rm gen $)} &
	\colhead{($N > 2\rm gen$)} &
	\colhead{($N > 2\rm gen$)} &
	\colhead{$(\rm stellar \, ev \, $[$M_{\odot}$])} &
	\colhead{$\rm (BH \, merger \, $[$M_{\odot}$])} & 
	\colhead{$N_{\rm total}$} & 
	\colhead{$N_{\rm PI gap}$} & 
	\colhead{$N_{\rm IMBH}$} 
    }
\startdata
1a & 0.50 & 3.2 & 2560 (52)  & 40 (38)  & 2 (1)  & 137 & 132 & 143 & 41 & 1\\ 
1b & 0.50 & 1.3 & 2543 (46)  & 29 (27)  & 0 (0)  & 90 & 82 & 144 & 43 & 0\\ 
1c & 0.50 & 1.0 & 2570 (31)  & 14 (13)  & 1 (0)  & 182 & 211 & 140 & 45 & 1\\ 
1d & 0.50 & 1.5 & 3002 (55)  & 25 (23)  & 1 (0)  & 301 & 106 & 163 & 36 & 1\\ 
1e & 0.50 & 1.2 & 3120 (22)  & 5 (3)  & 1 (0)  & 430 & 459 & 44 & 6 & 1\\ 
1f & 0.50 & 1.3 & 2549 (44)  & 30 (25)  & 0 (0)  & 104 & 77 & 149 & 37 & 0\\ 
1g & 0.50 & 1.5 & 2567 (33)  & 21 (17)  & 0 (0)  & 93 & 117 & 130 & 46 & 0\\ 
1h & 0.50 & 1.2 & 2548 (43)  & 22 (21)  & 1 (0)  & 125 & 93 & 140 & 36 & 1\\ 
1i & 0.50 & 3.3 & 2549 (56)  & 31 (26)  & 1 (0)  & 136 & 155 & 154 & 43 & 1\\ 
1j & 0.50 & 2.2 & 2585 (48)  & 23 (20)  & 0 (0)  & 71 & 105 & 146 & 43 & 0\\ 
1k & 0.50 & 2.1 & 2562 (41)  & 32 (26)  & 1 (0)  & 146 & 163 & 130 & 40 & 1\\ 
1l & 0.50 & 1.5 & 2553 (39)  & 25 (20)  & 0 (0)  & 105 & 114 & 138 & 38 & 0\\ 
1m & 0.50 & 1.2 & 3163 (46)  & 33 (26)  & 2 (0)  & 367 & 139 & 130 & 36 & 2\\ 
1n & 0.50 & 0.5 & 2545 (30)  & 23 (20)  & 0 (0)  & 91 & 79 & 107 & 37 & 0\\ 
1o & 0.50 & 0.3 & 2545 (27)  & 22 (15)  & 2 (2)  & 104 & 153 & 104 & 34 & 0\\   
\hline
2a & 0.75 & 2.2 & 2636 (69)  & 42 (39)  & 2 (2)  & 96 & 124 & 204 & 44 & 0\\ 
2b & 0.75 & 1.1 & 2663 (51)  & 32 (26)  & 1 (0)  & 170 & 252 & 188 & 51 & 1\\ 
2c & 0.75 & 2.2 & 2709 (60)  & 39 (35)  & 1 (0)  & 194 & 257 & 186 & 50 & 1\\ 
2d & 0.75 & 2.0 & 2656 (64)  & 38 (32)  & 0 (0)  & 72 & 111 & 186 & 52 & 0\\ 
2e & 0.75 & 1.1 & 2648 (50)  & 30 (27)  & 1 (0)  & 146 & 175 & 176 & 40 & 1\\ 
2f & 0.75 & 1.2 & 2669 (45)  & 26 (21)  & 1 (1)  & 109 & 120 & 169 & 44 & 0\\ 
2g & 0.75 & 2.2 & 2671 (60)  & 34 (29)  & 2 (0)  & 185 & 91 & 186 & 49 & 1\\ 
2h & 0.75 & 1.3 & 2620 (53)  & 44 (41)  & 0 (0)  & 83 & 99 & 160 & 47 & 0\\ 
2i & 0.75 & 1.1 & 2622 (46)  & 35 (29)  & 0 (0)  & 93 & 117 & 176 & 54 & 0\\ 
2j & 0.75 & 1.7 & 2672 (46)  & 31 (27)  & 0 (0)  & 85 & 100 & 172 & 54 & 0\\ 
2k & 0.75 & 1.5 & 2626 (45)  & 27 (26)  & 0 (0)  & 85 & 77 & 169 & 41 & 0\\ 
2l & 0.75 & 1.8 & 2639 (57)  & 37 (34)  & 0 (0)  & 76 & 86 & 182 & 39 & 0\\ 
2m & 0.75 & 0.7 & 3206 (47)  & 25 (21)  & 2 (1)  & 288 & 325 & 137 & 35 & 3\\ 
2n & 0.75 & 1.2 & 2628 (49)  & 39 (31)  & 0 (0)  & 97 & 116 & 165 & 49 & 0\\ 
2o & 0.75 & 0.6 & 2627 (35)  & 23 (20)  & 0 (0)  & 65 & 118 & 141 & 42 & 0\\ 
2p & 0.75 & 0.7 & 2631 (41)  & 34 (28)  & 0 (0)  & 94 & 110 & 141 & 43 & 0\\ 
2q & 0.75 & 0.2 & 2612 (31)  & 29 (21)  & 0 (0)  & 89 & 85 & 113 & 29 & 0\\ 
2r & 0.75 & 0.7 & 2613 (35)  & 21 (18)  & 0 (0)  & 72 & 107 & 145 & 36 & 0\\ 
2s & 0.75 & 1.7 & 2634 (70)  & 42 (38)  & 0 (0)  & 101 & 99 & 197 & 49 & 0\\ 
2t & 0.75 & 0.7 & 2724 (40)  & 25 (20)  & 1 (0)  & 193 & 222 & 156 & 46 & 1\\ 
2u & 0.75 & 0.5 & 2634 (40)  & 33 (24)  & 1 (1)  & 113 & 142 & 140 & 51 & 0\\ 
2v & 0.75 & 0.6 & 2659 (38)  & 30 (22)  & 3 (2)  & 126 & 151 & 144 & 43 & 1\\ 
2w & 0.75 & 0.6 & 2678 (37)  & 28 (24)  & 2 (0)  & 197 & 230 & 140 & 47 & 2\\ 
2x & 0.75 & 0.6 & 2685 (36)  & 20 (17)  & 1 (0)  & 199 & 222 & 134 & 37 & 1\\ 
2y & 0.75 & 0.6 & 2629 (39)  & 31 (25)  & 0 (0)  & 94 & 77 & 147 & 33 & 0\\ 
\hline
3a & 1.0 & 0.1 & 2711 (31)  & 25 (21)  & 1 (0)  & 168 & 187 & 116 & 26 & 1\\ 
3b & 1.0 & 0.2 & 2703 (26)  & 23 (15)  & 0 (0)  & 98 & 69 & 115 & 28 & 0\\ 
3c & 1.0 & 2.6 & 2709 (51)  & 28 (21)  & 0 (0)  & 75 & 107 & 190 & 45 & 0\\ 
3d & 1.0 & 1.7 & 3249 (70)  & 36 (25)  & 2 (0)  & 280 & 342 & 222 & 48 & 3\\ 
3e & 1.0 & 3.6 & 2737 (62)  & 33 (32)  & 0 (0)  & 64 & 111 & 230 & 47 & 0\\ 
3f & 1.0 & 0.6 & 3322 (57)  & 33 (25)  & 3 (1)  & 306 & 334 & 190 & 43 & 3\\ 
3g & 1.0 & 1.7 & 3109 (76)  & 42 (38)  & 3 (0)  & 261 & 98 & 210 & 52 & 1\\ 
3h & 1.0 & 2.1 & 3654 (79)  & 37 (29)  & 2 (0)  & 315 & 348 & 237 & 47 & 2\\ 
3i & 1.0 & 2.1 & 2739 (60)  & 44 (39)  & 0 (0)  & 94 & 120 & 209 & 54 & 0\\ 
3j & 1.0 & 0.6 & 2725 (56)  & 36 (31)  & 0 (0)  & 94 & 111 & 179 & 39 & 0\\ 
3k & 1.0 & 0.3 & 2734 (38)  & 29 (20)  & 1 (0)  & 181 & 67 & 184 & 45 & 1\\ 
3l & 1.0 & 0.6 & 2685 (43)  & 44 (33)  & 0 (0)  & 106 & 108 & 174 & 54 & 0\\ 
3m & 1.0 & 1.7 & 2749 (78)  & 39 (33)  & 2 (2)  & 103 & 154 & 231 & 48 & 0\\ 
3n & 1.0 & 0.6 & 3257 (49)  & 26 (19)  & 2 (0)  & 285 & 118 & 174 & 40 & 1\\ 
3o & 1.0 & 1.9 & 3986 (78)  & 45 (36)  & 3 (2)  & 474 & 511 & 208 & 43 & 1\\
\enddata 
\tablecomments{List of all cluster models included in this study. In column $2$ we indicate the model's primordial high-mass binary fraction. Column $3$ lists the time at which the last BH with mass greater than $50\,M_{\odot}$ escapes the cluster. Column $4$ indicates the total number of BHs formed through both stellar collapse or BH merger, with only the number that were formed through BH mergers in parentheses. Columns $5$--$6$ indicate the number of BHs formed with masses in the pair-instability gap ($40.5$--$120\,M_{\odot}$) and number of IMBHs, respectively. Similarly, the number of BHs in these mass ranges formed through BH mergers is noted in parentheses. Columns $7$--$8$ list the masses of the most massive BH formed through stellar collapse and BH mergers, respectively. Columns $9$--$11$ list the total number of binary BH mergers between two stellar-mass BH components, mergers with at least one component in the pair-instability mass-gap, and mergers with at least one IMBH, respectively.}
\end{deluxetable*}

\clearpage
\section{IMBH Formation}
\label{sec:formation}

\begin{figure}
\begin{center}
\includegraphics[width=0.9\linewidth]{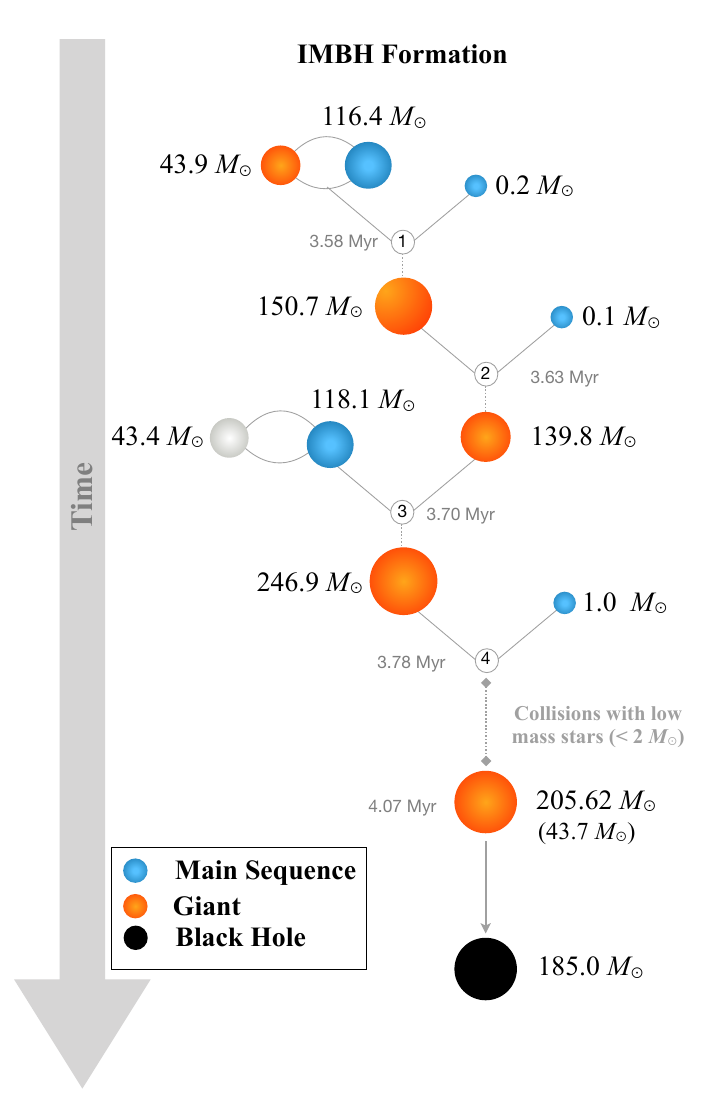}
\caption{\footnotesize \label{fig:cartoon1} Formation history for one of the IMBHs formed in our cluster models. The color indicates the stellar type and binaries are marked with a loop. The times of each collision are indicated, as well as the masses at the time of the next collision. Lastly, the core mass of the progenitor is indicated in parenthesis. Circles in gray indicate stars that are present in the dynamical interaction but do not contribute to the merger product.}
\end{center}
\end{figure}

Of the total 49 IMBHs formed in 55 simulations, 34 form via the collapse of a massive star grown through repeated stellar collisions and 15 result from BBH mergers of lower-mass BHs (typically involving two or fewer consecutive BBH mergers). The most massive IMBHs form through the collapse of massive stars, since the low escape velocity of GCs makes it difficult to retain BH merger products (as discussed in more depth in Section~\hyperref[sec:ejection]{5}). Figure~\hyperref[fig:cartoon1]{1} shows the formation history of an exemplary IMBH from our models. Initially, during a binary--single interaction, all of the masses merge and form a massive remnant \citep{Gurkan}. The collision product rejuvenates accordingly and evolves into a giant star that participates in another single-single (SS) collision with a main-sequence star. This is followed by a merger with the component of a binary and $\sim 10$ collisions with low-mass main-sequence stars. The final product, a giant with a total mass of roughly $ 205\,\Msun$ and core mass of $\approx 43.7\,\Msun$, collapses into a BH of $\approx 185\,\Msun$.

\begin{figure}
\begin{center}
\includegraphics[width=1\linewidth]{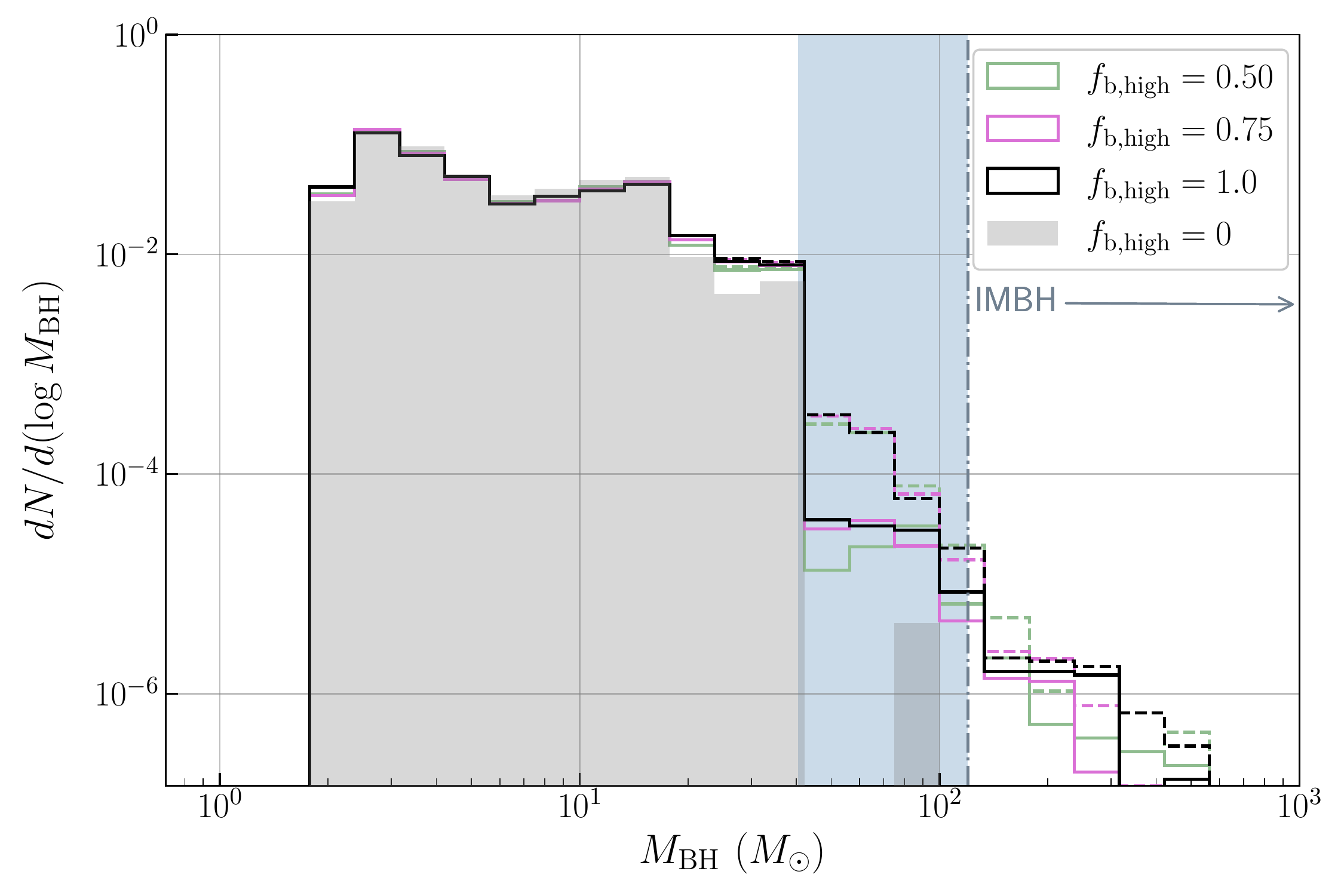}
\caption{\footnotesize \label{fig:bhspectrum} The normalized BH mass spectrum for our models listed in Table~\hyperref[table:models]{1}. The solid lines presented BHs formed through the collapse of a massive star while the dashed lines are products of binary BH mergers. The colors indicate the value for the high-mass binary fraction. The blue shaded region indicates the upper mass gap and the dashed/dotted line the beginning of the IMBH region. The shaded histogram is taken from \cite{Gonzalez_2021}}
\end{center}
\end{figure}

In Figure~\hyperref[fig:bhspectrum]{2} we show the normalized mass spectrum of all BHs formed in our models. We distinguish between BHs formed via stellar collapse (solid curves) and those produced via BH mergers (dashed curves). As $f_{\rm{b,high}}$ increases, the masses of BHs formed through stellar collapse skew slightly higher, though this effect is more notable when comparing to the models with zero primordial binaries studied by \cite{Gonzalez_2021}, reproduced here as the gray-shaded distribution. At the $f_{\rm{b,high}}$ examined here, however, the difference is only slight. This trend reflects how binaries increase collision rates in clusters; the extra collisions induced by a higher binary fraction produce a few more stars massive enough to collapse into BHs. Since BH mergers are even rarer than stellar collisions, the merger product mass is less noticeably affected by $f_{\rm{b,high}}$. Even so, Table~\hyperref[table:models]{1} demonstrates that the $f_{\rm{b,high}} = 1$ model does in fact produce the highest number of IMBHs. On average, the models with  $f_{\rm{b,high}} = 0.5$ produce $\approx 0.8$ IMBHs per simulation, while in the case of $f_{\rm{b,high}} = 1$, this number increases to $\approx 1.3$.

We find no clear relation between the formation of an IMBH and the effects it may have on the mass-gap population of BHs. In forming a very massive IMBH, some models exhibit a smaller number of mass-gap BHs, but this is not true in all cases. This reveals that IMBH formation is a random process that affects cluster evolution in ways dependent on the IMBH's specific formation pathway (for example, the number of massive star collisions that were needed to form the IMBH progenitor). This of course, is not the case for the formation of IMBHs with masses $M > 500\,\Msun$, as this stellar runaway process would leave a stronger impact on the stellar-mass BH population.

\vspace{0.5cm}

\section{Companions}
\label{sec:companions}

\begin{figure}
\begin{center}
\includegraphics[width = 1\linewidth]{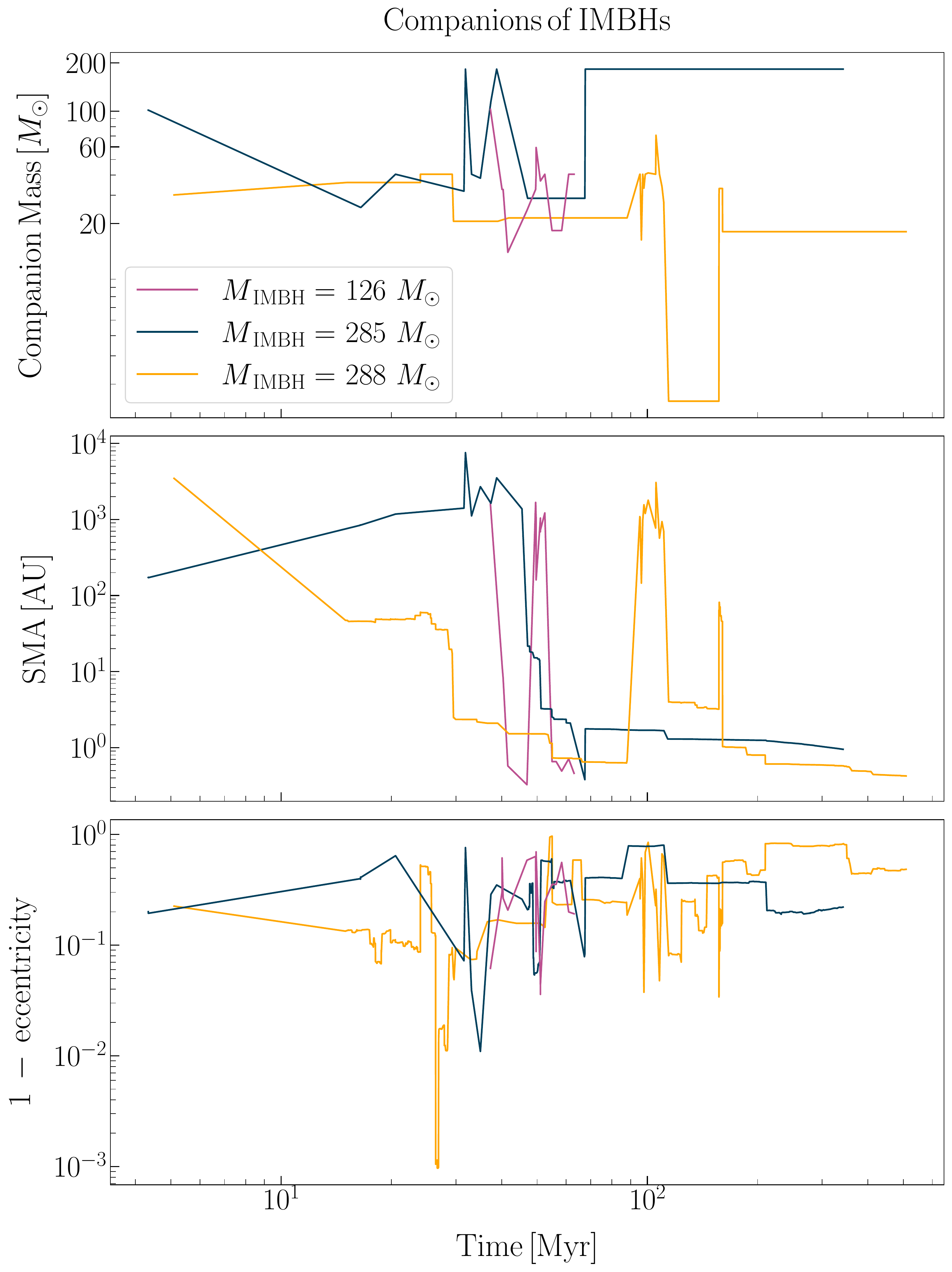}
\caption{\footnotesize \label{fig:BH_Companions} The top panel shows the BH companion masses over time for three separate IMBHs. The middle panel shows the semi-major axes of the different binaries that the IMBH inhabits and the bottom panel plots the eccentricity.}
\end{center}
\end{figure}

In this section we take a closer look at the companions of IMBHs in our models. To do this, we show the time evolution of IMBH binaries in several representative cases in Figure~\hyperref[fig:BH_Companions]{3}. These include a typical IMBH quickly ejected from the cluster (magenta), an IMBH--IMBH binary (blue), and the IMBH that remains in the cluster for the longest period of time ($\sim 500$~Myr) across all models (yellow). The dynamical history is shown in three panels, the first showing the companion masses, the second the semi-major axis of the binary, and the third a measure of the eccentricity.

In the first case, the IMBH never stays for long with a consistent binary companion and is quickly ejected from the cluster (at $t \sim 65$~Myr) by a scattering interaction. Repeated exchange interactions during binary--binary and binary--single encounters cause the IMBH to appear in harder, more eccentric binaries until it finally escapes from the cluster with its companion. This particular escaping binary, due to its high eccentricity, has a short inspiral time and will merge within a Hubble time, making it a potentially detectable GW source. This case is typical of most of the IMBHs generated in our models.

The second case that we consider is an IMBH that forms a binary with another IMBH. The primary IMBH initially inhabits multiple wide binaries, with semi-major axes of thousands of AU. It eventually settles into a wide binary with another IMBH of $183\,\Msun$, which steadily hardens through hundreds of interactions before finally gaining enough energy to escape the cluster. Once again, this ejected binary merges within a Hubble time.

Lastly, we examine a case where an IMBH survives in-cluster for a longer period of time. This IMBH is very dynamically active, constantly involved in gravitational scattering interactions. Although it inhabits binaries with other BHs most of the time, it has a main-sequence companion of roughly $1.6\,\Msun$ during the later stages, before it exchanges into a binary with another BH, with which it merges during a binary--binary encounter. The GW recoil kick imparted on the IMBH causes it to escape the cluster in a binary that will merge within a Hubble time.

Approximately $65\%$ of the IMBHs in our models are ejected within the first $100$~Myr. The reason for these ejections is explored in Section \hyperref[sec:ejection]{5}.  The characteristic examples above all feature numerous close scattering interactions, demonstrating that IMBHs formed in clusters are very dynamically active. The most common companions to the IMBHs are other BHs, with only a few instances of main-sequence companions.

Binaries between a star and an IMBH are of interest due to their possible detection using radial velocity measurements of star clusters \citep{Giesers19, Kamann21}. However, in our models, we only observe one star-IMBH binary with a low-mass main-sequence star ($M < 2\,\Msun$) that is constantly perturbed by other objects in the cluster (every $\sim 0.003$~Myr, on average). Thus, indirect measurements of IMBHs with masses like those explored in this paper using this method seems unlikely.

An additional observational signature is the tidal disruption of a star by an IMBH, which could produce strong electromagnetic signatures \citep{Ramirez_tdes, Chen_tdes, Kremer22}. We count approximately 20 candidates ($95 \%$ of which are on the main sequence) for such events in our simulations \citep[i.e., instances where a star passes within the classical disruption radius of an IMBH; see][]{Kremer22}. K{\i}ro\u{g}lu et al. (2022, in preparation) investigates tidal disruptions involving stars and IMBHs of similar masses using the smoothed-particle hydrodynamical simulation \texttt{StarSmasher}.

\begin{figure}
\begin{center}
\includegraphics[width=1\linewidth]{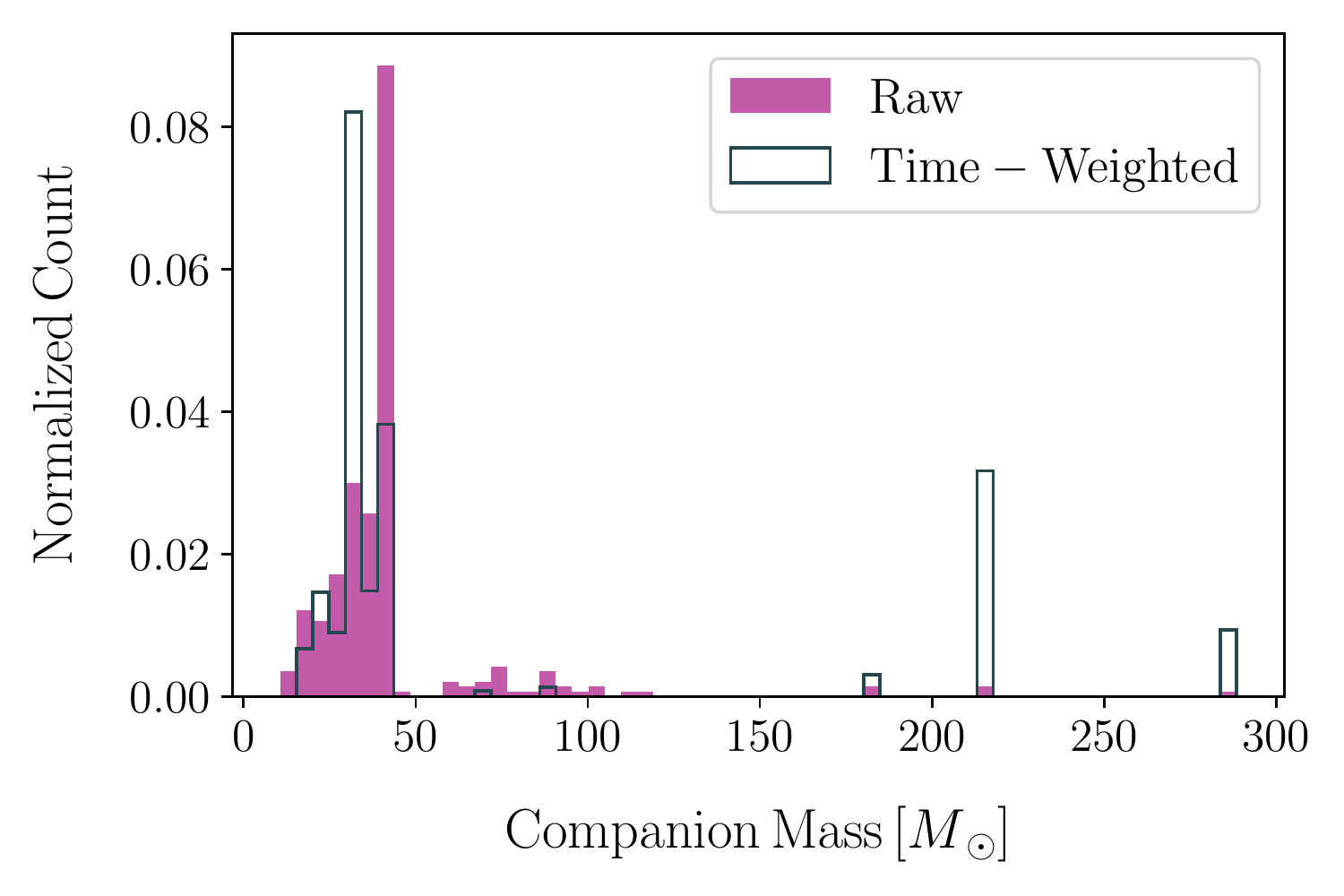}
\caption{\footnotesize \label{fig:companion_dist} In magenta we show the raw distribution of the masses of the BH companions of IMBHs. The blue outline is the companion mass distribution weighted by the lifetime of the binary. }
\end{center}
\end{figure}

To study the BHs that accompany the IMBHs, we plot the distribution of their masses as the filled magenta histogram in Figure~\hyperref[fig:companion_dist]{4}. The blue outline shows the same distribution, weighted by the amount of time that each of these companions spends in the binary with the IMBH. As can be inferred from the plot, the most common companions are BHs with masses roughly in the range $ 30$--$40\,\Msun$. This is because BHs in this mass range are massive enough to survive dynamical interactions with other BHs (i.e., not be replaced in the binary during exchange interactions).

\section{IMBH Ejection}
\label{sec:ejection}

IMBH ejection from star clusters primarily occurs through two distinct mechanisms: recoil kicks from BH mergers and dynamical kicks from strong binary-mediated gravitational scattering encounters.
In the first case, which dominates, anisotropic emission of GWs during a BH merger gives the merger remnant a recoil kick determined by the mass ratio of the system and the initial spins of the BHs \citep{Merritt2004,Campanelli2007, Lousto2008}. In general, near-equal mass binaries and low-spinning (as well as aligned) BHs will receive smaller kicks and thus are most likely to be retained in the cluster. Our assumption that BHs are born with zero spin increases the retention likelihood for first-generation mergers. However, since the merger remnant typically receives a significant spin, any later mergers are more likely to eject the product from the cluster, lowering the rate of 2G+ mergers in GCs \cite[see, e.g.,][for an overview of repeated BH mergers in dense star clusters]{Rodriguez2019}. Overall, GW recoil kicks account for $64\%$ of the IMBH ejections in our models.\footnote{Of course, these fractions are specific to the cluster mass adopted in our simulations and will change for clusters with different masses and hence different escape velocities. For example, see \cite{Mahapatra2021} for a study of BH retention in both nuclear and globular star clusters. Of the IMBH ejections in our models, $84\%$ occur during $\rm1G + 1G$ mergers (first-generation), $10\%$ during $\rm1G + 2G$ mergers, and $6\%$ during $\rm 2G + 2G$ mergers. The low mass ratios---and correspondingly strong GW recoil kicks---typical of mergers between IMBHs and smaller companions are partially responsible for the high fraction of ejections during first-generation mergers. These mergers, which can occur in isolated binaries or during binary--single or binary--binary interactions, are also promising GW sources, which we discuss in Section~\hyperref[sec:mergers]{6}.}

Strong gravitational scattering encounters can also eject IMBHs from clusters. Recall that scattering interactions in star clusters typically harden binaries \citep{Heggie1975}, where passing objects (singles or binaries) on average gain kinetic energy as the binary components sink deeper into their mutual potential well. This can be conceptualized as an exchange between the binary binding energy and the orbital energy of the passing objects in the global cluster potential. Conservation of momentum, however, guarantees that any velocity change experienced by the passing object is countered by a change to the velocity of the binary's center of mass. So potential energy from the binary components not only gives a recoil velocity to the passing object, but also to the binary itself. These recoil kicks can be large enough to eject an IMBH involved in such an encounter.

\begin{figure}
\begin{center}
\includegraphics[width=1\linewidth]{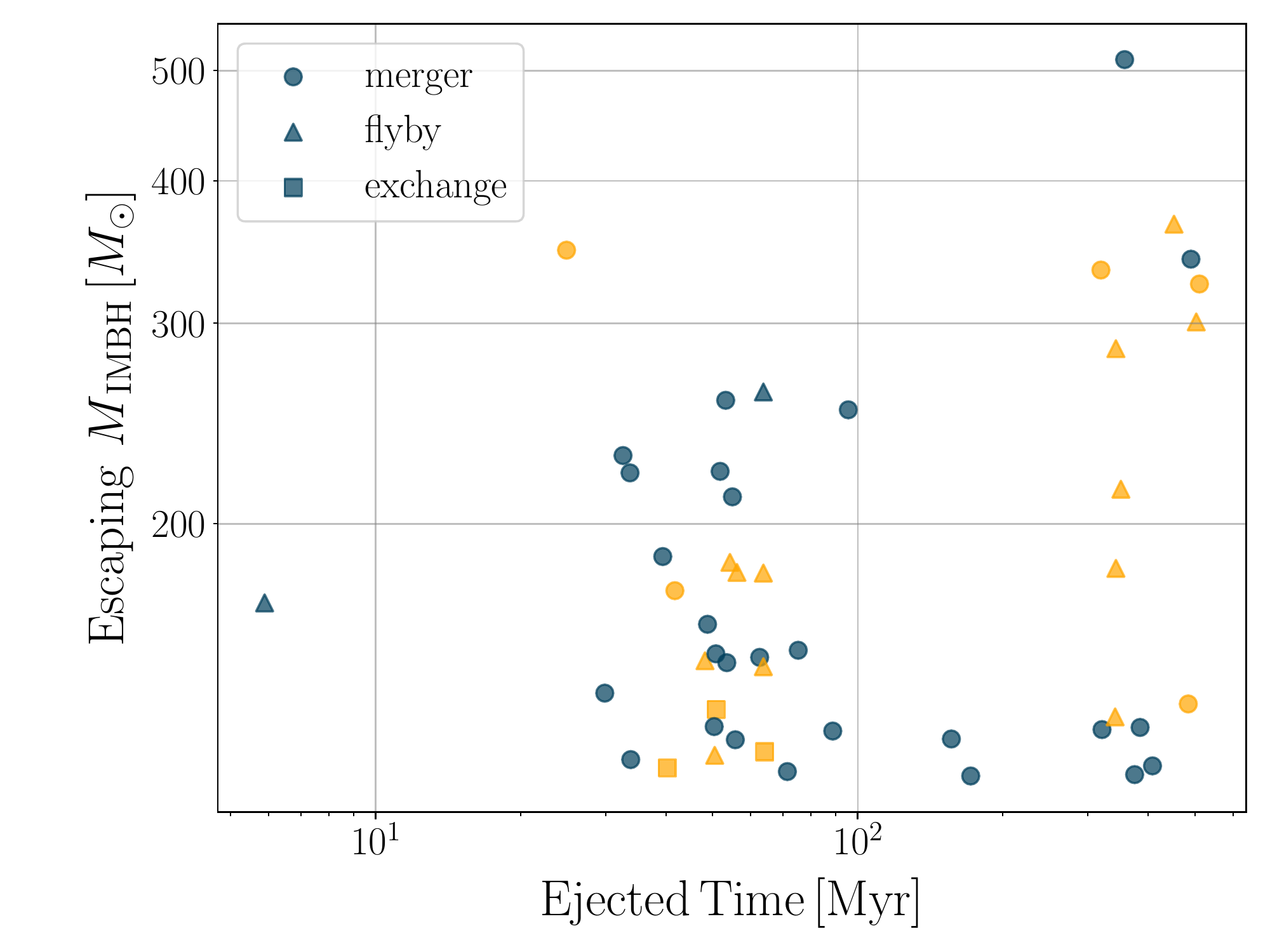}
\caption{\footnotesize \label{fig:ejected}Mass of IMBHs as a function of ejection time from the cluster. The IMBHs escaping in a binary are indicated in yellow, while those escaping as single objects are marked in blue. The different shapes indicate the dynamical interactions responsible for the ejection of the IMBHs. }
\end{center}
\end{figure}

Figure~\hyperref[fig:ejected]{5} plots the masses of ejected IMBHs in our models as a function of the time of escape from the cluster. Circles indicate IMBHs ejected by GW recoil kicks received during BH mergers while triangles indicate flyby interactions, where the initial configuration of objects remains unchanged. In this case, if the IMBH is the passing object, it will extract energy from a binary during the flyby and as a result gain a velocity kick, sometimes larger than the escape speed of the cluster. This, in turn, hardens the binary (usually a BH binary), sometimes to the point of merger. A similar output can be expected if the IMBH is originally a member of the binary instead and experiences a flyby with a single BH. Lastly, the squares represent exchange interactions during which the IMBH is exchanged into a binary (yellow) or out of its original binary (blue). This case is favored when the intruder's mass exceeds that of either binary component. Furthermore, the new binary resulting from the exchange has a higher binding energy as a result. Whether the final single star or binary receive kick velocities large enough to eject them from the cluster depends on the initial velocity of the single star as well as the semi-major axis of the binary. It is clear from Figure~\hyperref[fig:ejected]{5} that the merger channel is the most efficient at ejecting massive BHs from clusters.

\begin{figure}
\begin{center}
\includegraphics[width=1\linewidth]{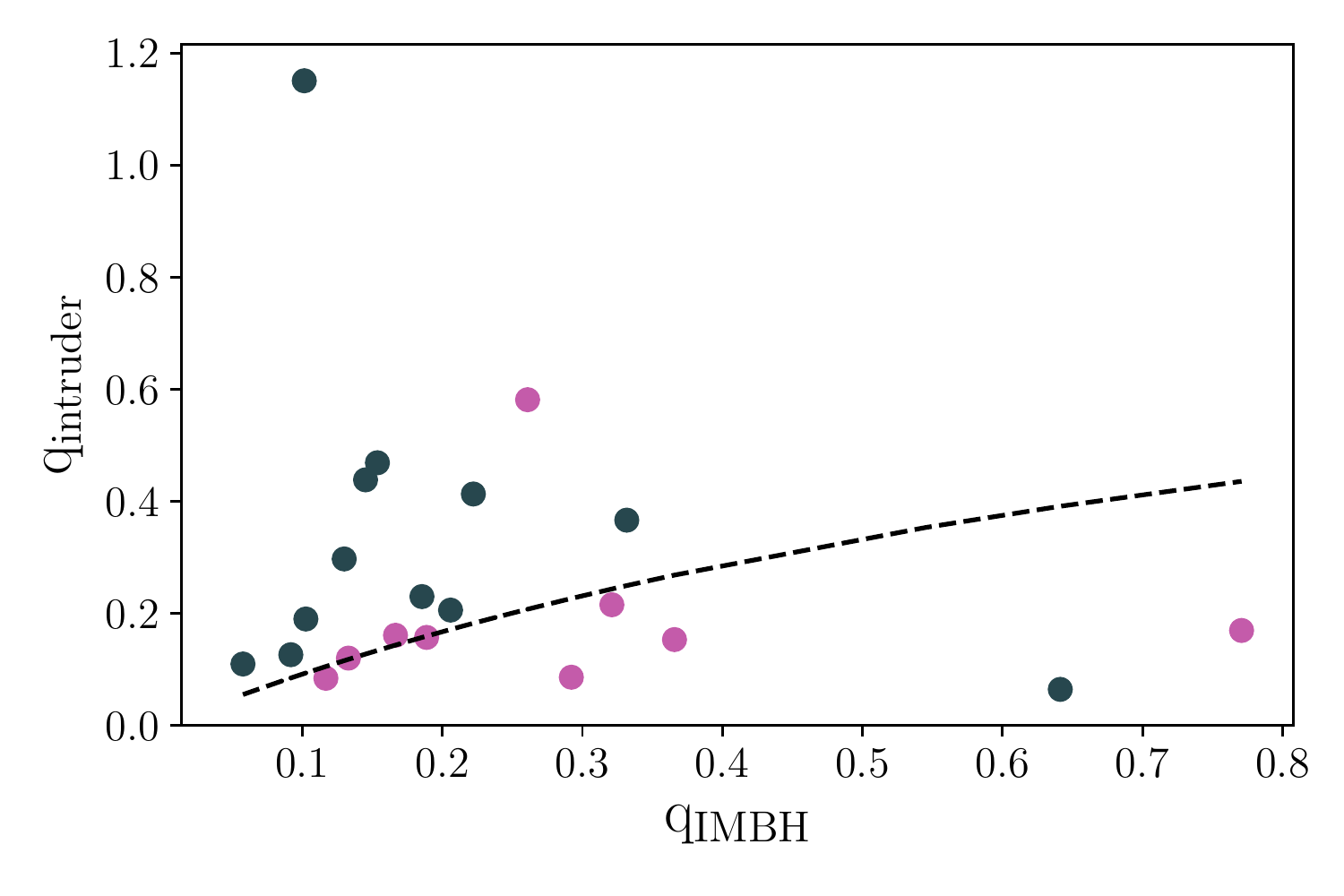}
\caption{\footnotesize \label{fig:last_intruder} The mass ratios of an interaction involving an IMBH-BH binary and an intruder. The horizontal axis shows the mass ratio between the IMBH companion and the IMBH while the vertical axis shows the mass ratio with the intruder. The color indicates if the intruder is a single object (magenta) or a binary (blue). The dashed line indicates an intruder with the same mass as the IMBH companion for cases where the intruder is a single object. }
\end{center}
\end{figure}

In Figure~\hyperref[fig:last_intruder]{6} we show the properties of the systems responsible for IMBH ejection during binary--single and binary--binary interactions if the IMBH was originally in a binary, which is often the case. Along the horizontal axis we plot the mass ratio for the binary containing the IMBH, defined as \mbox{$\rm q_{IMBH} = \frac{m_{comp}}{m_{IMBH}}$}. In the case of binary--single interactions (shown in magenta), the vertical axis, $\rm q_{intruder}$, is the mass ratio of the single intruder to total mass of the binary \mbox{($\rm q_{intruder} = \frac{m_{intruder}}{m_{comp} + m_{IMBH}}$)}. For binary--binary interactions (shown in blue), $\rm q_{intruder}$ is the ratio between the total mass of the intruder binary and the IMBH-BH binary. The dashed line indicates the case where the intruder mass is equal to the mass of the IMBH companion \mbox{($\rm m_{intruder} = m_{comp}$)}.

This plot shows that, in general, $\rm m_{intruder}\gtrsim m_{comp}$ is necessary to eject the
IMBH---though this is not always the case, since some binaries have already been influenced dynamically by previous interactions.

\subsection{Singly Ejected IMBHs}
\label{sec: singly ejected}
Out of the 48 escaping IMBHs, 28 escape the cluster as single BHs. Table~\hyperref[table:2]{2} lists the distribution of ejection methods for singly ejected IMBHs in each model. The second column indicates the number of singly ejected IMBHs per model. The third column records the number of BHs ejected due to GW recoil kicks  received during a merger. These mergers can occur in isolated binaries or during strong dynamical interactions, and this difference is taken into account in the table. It is important to note that dynamical hardening during earlier scattering interactions may have significantly hastened mergers labeled as isolated. Lastly, the rightmost column counts IMBHs ejected due to velocity kicks received during dynamical interactions (shown as flybys or exchanges in Figure~\hyperref[fig:ejected]{5}). 

\begin{deluxetable}{c c c c c c}[h!]
\tabletypesize{\scriptsize}
\setlength{\tabcolsep}{1.7\tabcolsep}  
\centering
\tablecaption{Singly Ejected IMBHs}  
\label{table:2}
\tablehead{
	\colhead{$f_{\rm b,high}$} &
	\colhead{$N_{\rm bh}$} &
	\multicolumn{2}{c}{$N_{\rm GW recoil}$} &
	\colhead{$N_{\rm dyn}$} \\
	\colhead{} &
	\colhead{(single)} &
	\colhead{(iso. bin.)} &
	\colhead{(dyn.)} &
	\colhead{}
}
\startdata
    0.50 & 6 & 5 & 1 & 0 \\ 
    0.75 & 13 & 9 & 3 & 1 \\ 
    1.0 & 9 & 6 & 2 & 1 \\ 
\enddata
\tablecomments{For each set of models with distinct $f_{\rm b,high}$ (column 1), column 2 lists the total number of singly escaping IMBHs from that model set. Columns 3--4 list the number of IMBHs ejected due to the large GW recoil kicks received during an isolated binary merger or dynamically-induced merger, respectively. Column 5 lists the number of IMBHs ejected by large velocity kicks from dynamical interactions.}
\end{deluxetable}

\begin{figure}[h!]
\begin{center}
\includegraphics[width=1\linewidth]{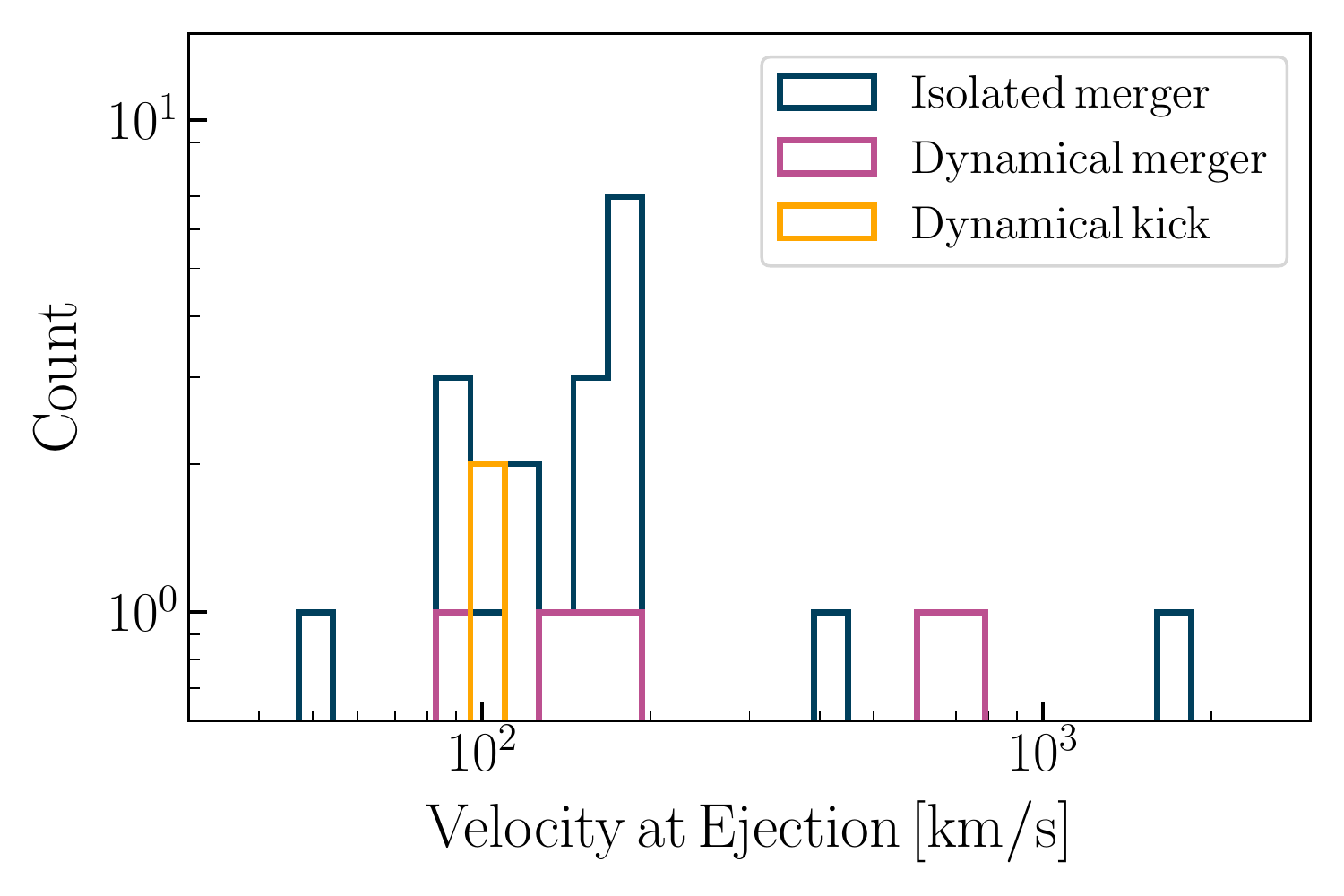}
\caption{\footnotesize \label{fig:kicks} The escape velocities [initial speed + kick velocity] of single IMBHs ejected from the clusters in km/s. The sub-populations indicate IMBHs ejected due to strong GW recoil kicks during isolated BH mergers (blue) and dynamical mergers (magenta), as well as IMBHs receiving velocity kicks during strong dynamical interactions (yellow). }
\end{center}
\end{figure}

To further study the properties of these ejected IMBHs, Figure~\hyperref[fig:kicks]{7} shows the distribution of velocity kicks that singly-escaping IMBHs received for each of the ejection scenarios. The highest IMBH ejection velocities result from BH mergers (blue and magenta). The bimodality of the GW recoil velocity distribution is attributable to $\rm 2G+$ mergers, which have higher recoil kicks. The scarcity of high ejection velocities reflects the early ejection of most IMBHs, which are then unable to participate in $\rm 2G+$ mergers. Since the extreme mass ratio typical of an IMBH encounter with smaller objects damps any dynamical kick, BHs ejected due to dynamical interactions experience
lower ejection velocities (though still well above the typical GC escape speed of $\sim 60\,\mathrm{km/s}$). 

\subsection{IMBHs Escaping in Binaries}
\label{sec:escaping binaries}
\begin{figure}[h!]
\begin{center}
\includegraphics[width=1\linewidth]{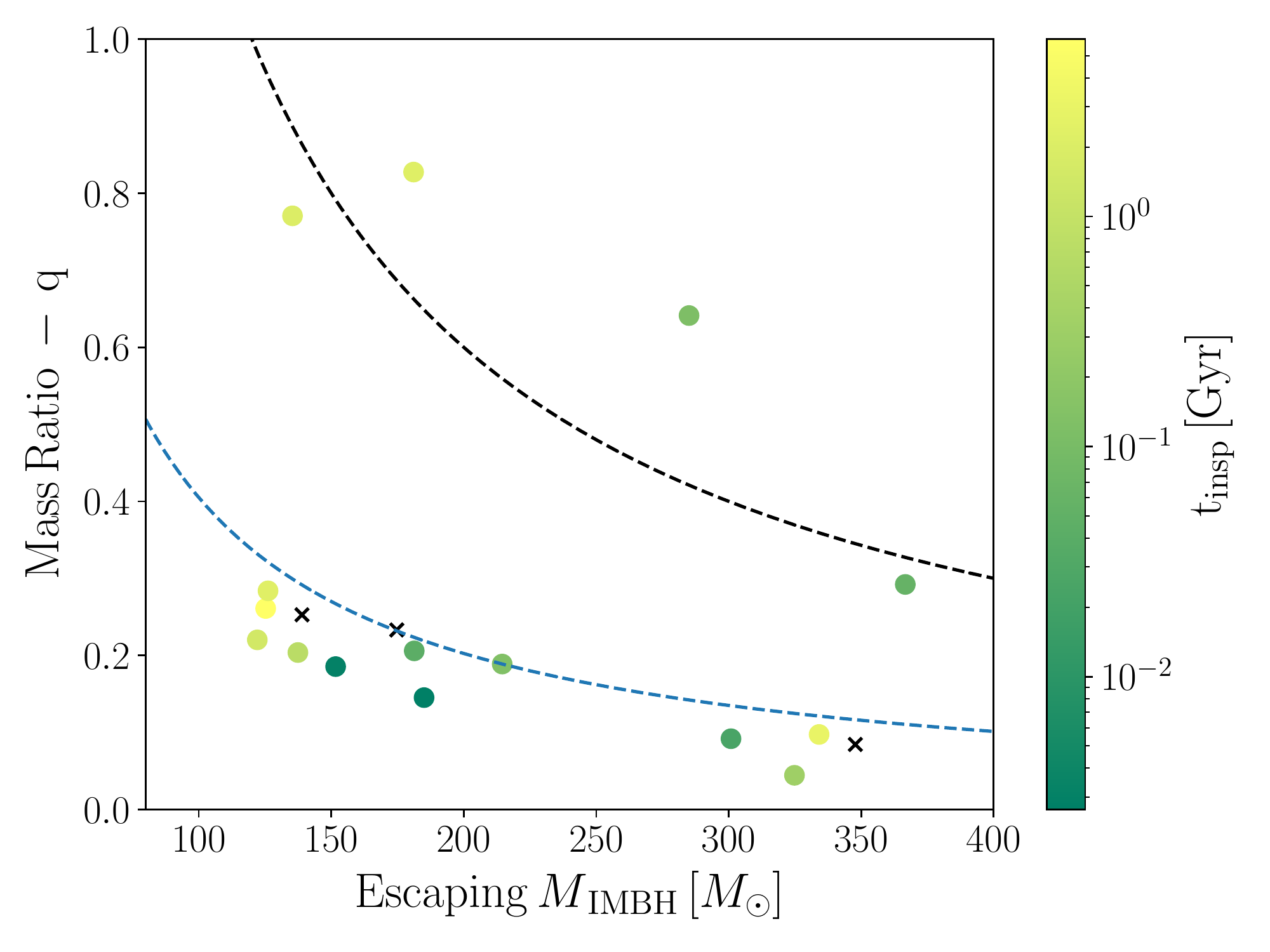}
\caption{\footnotesize \label{fig:imbh-q} The mass ratio as a function of the IMBH mass in escaping binaries. The color gradient indicates the inspiral times. The crosses label binaries that do not merge within a Hubble time ($\rm t_{inspiral} \geq 12$~Gyr). The inspiral times are calculated using the Peters equation \citep{Peters}, taking into account the eccentricity and separation of the binary. The dashed blue and black curves indicate companions with masses larger than $40$ and $120\,\Msun$, respectively. }
\end{center}
\end{figure}

The remaining $20$ IMBHs escape in binaries, and depending on their inspiral times, can be interesting sources of GWs. To study the IMBHs that are ejected in binaries, we plot the binary mass ratio (q) as a function of the escaping IMBH mass in Figure~\hyperref[fig:imbh-q]{8}. We see that the majority of the IMBHs ejected in binaries have inspiral times that would make them potential GW sources detectable by current and future detectors such as LIGO/Virgo/Kagra and LISA. Furthermore, four IMBHs escape the cluster in a binary with another massive BH (as shown by the four points above the blue dashed line), while most are ejected with stellar mass BHs.

\begin{figure}
\begin{center}
\includegraphics[width=0.8\linewidth]{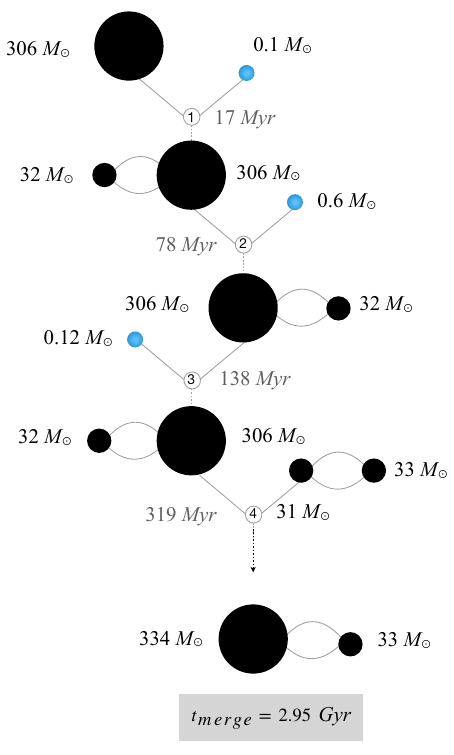}
\caption{\footnotesize \label{fig:map} An example of an escaping BH binary that merges within a Hubble time.}
\end{center}
\end{figure}

The short inspiral times can be attributed to binary hardening through dynamical interactions in the cluster. An example of this is shown in Figure~\hyperref[fig:map]{9}, which illustrates the dynamical history of a case in which the binary merges with a short inspiral time after ejection from the cluster. Initially, the IMBH is a single object that forms a binary with a $31\,\Msun$ BH. After $\approx 280$~Myr, during an interaction with another BH binary, the IMBH exchanges into a new binary with a smaller separation before experiencing a velocity kick large enough to eject it from the cluster. This binary eventually merges in $\approx 3$~Gyr.

\section{BBH Mergers}
\label{sec:mergers}

The total number of BBH mergers is listed column 9 on Table~\hyperref[table:models]{1}. Columns 10 and 11 list the number of mergers where at least one component is in the upper mass gap or an IMBH, respectively. Note that the ratio of mass gap mergers to stellar mergers is higher than estimates in previous \texttt{CMC} studies \citep[e.g.,][]{Rodriguez2019,Kremer20} and GW observations \citep[e.g.,][]{LIGO2020_O3} simply because the models were interrupted at early times. Thus, we do not account for stellar-mass BH mergers that occur at later evolutionary stages of these clusters.

As we have discussed in previous sections, IMBHs in our models tend to
form binaries with other BHs deep in the core of their host cluster.
This leads, in some occasions, to in-cluster IMBH-BH mergers. Additionally, $\approx 42 \%  $ of the IMBHs are ejected in binaries with other BHs, most of which will merge within a Hubble time. Some of these mergers may be detectable today as GW sources in either high- or low-frequency bands. We now examine these mergers in detail before discussing merger rate implications for modern GW detectors.

\subsection{IMBH-BH Mergers}
\label{sec:IMBH_BH mergers}
Table~\hyperref[table:3]{3} lists the BBH mergers involving IMBHs. When the merger occurs in the cluster, the merger remnant usually escapes the cluster after receiving a sufficiently high GW recoil kick. There are three instances where the IMBH is retained after a BBH merger, the first in model 1e, which, as mentioned in Section~\hyperref[sec:methods]{2}, is the model left un-evolved due to computational cost. The second instance is in model 3d, where the IMBH merges with its $27\,\Msun$ BH companion during a binary--binary encounter with another BH binary (with masses $40.5\,\Msun$ and $30.9\,\Msun$). Rather than being ejected, the IMBH remains in a binary with the $30.9\,\Msun$ BH, until ejection during a merger in a binary--single interaction $\sim 170$~Myr later (listed in the table as well). A similar process occurs in model 2m.

Out of 34 total BBH mergers involving an IMBH, 25 are with stellar-mass BHs and 7 with upper-mass gap BHs. The remaining two are IMBH--IMBH mergers. The likelihood of these massive remnants to be ejected from the cluster prevents any further later-generation mergers.

\begin{deluxetable}{c c c c c c c}
\tabletypesize{\scriptsize}
\setlength{\tabcolsep}{0.7\tabcolsep}  
\centering
\tablecaption{Mergers Involving an IMBH}  
\label{table:3}
\tablehead{
	\colhead{Model} &
	\colhead{$\rm t_{merge}$} &
	\colhead{$M_{1}$} &
	\colhead{$M_{2}$} &
	\colhead{Type} &
	\colhead{Outcome} &
	\colhead{$\rm t_{eject}$}\\
	\colhead{} &
	\colhead{(Gyr)} &
	\colhead{($M_{\odot}$)} &
	\colhead{($M_{\odot}$)} &
	\colhead{}&
	\colhead{}&
	\colhead{(Gyr)} 
}
\startdata
1a & 0.77 & 28.0 & 137.4 & Ejected Binary & N/A & 0.05 \\ 
1c & 0.05 & 32.7 & 182.1 & isolat-binary & Ejected & N/A \\ 
1d & 0.53 & 27.6 & 300.9 & Ejected Binary & N/A & 0.5 \\ 
1e & 1.04 & 429.7 & 32.2 & isolat-binary & Retained & N/A \\ 
1h & 5.94 & 32.7 & 125.2 & Ejected Binary & N/A & 0.05 \\ 
1i & 0.08 & 20.9 & 136.2 & isolat-binary & Ejected & N/A \\ 
1k & 0.05 & 19.6 & 145.6 & isolat-binary & Ejected & N/A \\ 
1m & 0.48 & 18.8 & 122.1 & binary--binary & Ejected & N/A \\ 
1m & 0.51 & 366.6 & 107.1 & Ejected Binary & N/A & 0.45 \\ 
\hline
2b & 0.1 & 92.7 & 170.0 & isolat-binary & Ejected & N/A \\ 
2c & 0.05 & 194.2 & 71.0 & binary--single & Ejected & N/A \\ 
2e & 0.04 & 32.4 & 146.0 & binary--binary & Ejected & N/A \\ 
2g & 0.06 & 26.9 & 185.0 & Ejected Binary & N/A & 0.05 \\ 
2m & 0.09 & 21.7 & 288.5 & isolat-binary & Retained & N/A \\ 
2m & 0.51 & 17.8 & 308.3 & binary--binary & Ejected & N/A \\ 
2m & 0.83 & 14.4 & 324.8 & Ejected Binary & N/A & 0.51 \\ 
2t & 0.03 & 32.6 & 192.6 & isolat-binary & Ejected & N/A \\ 
2v & 2.32 & 35.8 & 126.1 & Ejected Binary & N/A & 0.06 \\ 
2w & 0.03 & 36.8 & 196.8 & binary--single & Ejected & N/A \\ 
2w & 0.05 & 28.1 & 151.6 & Ejected Binary & N/A & 0.05 \\ 
2x & 0.05 & 25.9 & 199.1 & isolat-binary & Ejected & N/A \\
\hline
3a & 0.04 & 21.3 & 168.0 & isolat-binary & Ejected & N/A \\ 
3d & 0.32 & 280.0 & 27.0 & binary--binary & Retained & N/A \\ 
3d & 0.49 & 41.3 & 304.5 & binary--single & Ejected & N/A \\ 
3d & 0.48 & 40.5 & 214.5 & Ejected Binary & N/A & 0.35 \\ 
3f & 0.32 & 31.3 & 305.7 & binary--binary & Ejected & N/A \\ 
3f & 3.27 & 334.1 & 32.5 & Ejected Binary & N/A & 0.32 \\ 
3f & 2.38 & 104.3 & 135.4 & Ejected Binary & N/A & 0.34 \\ 
3g & 2.26 & 181.1 & 149.8 & Ejected Binary & N/A & 0.06 \\ 
3h & 0.02 & 36.8 & 314.6 & binary--single & Ejected & N/A \\ 
3h & 1.56 & 122.1 & 26.9 & Ejected Binary & N/A & 0.04 \\ 
3k & 0.1 & 37.3 & 181.3 & Ejected Binary & N/A & 0.06 \\ 
3n & 0.46 & 182.8 & 285.0 & Ejected Binary & N/A & 0.34 \\ 
3o & 0.36 & 474.4 & 40.5 & isolat-binary & Ejected & N/A \\ 
\enddata
\tablecomments{List of all of the BBH mergers with an IMBH component, with the model in which the merger occurs indicated in column 1. Column 2 lists the time at which the merger occurs after cluster formation while columns $3$--$4$ show the component masses. Column 5 lists the mechanism that leads to the merger, with ``Ejected Binary'' indicating escaping binaries that merge after ejection from the cluster. Lastly, columns 6--7 list the merger outcome and time at which the cluster ejects the merger remnant, respectively.}
\end{deluxetable}

The distribution of eccentricities at a GW frequency of $10$ Hz for all in-cluster IMBH-BH mergers has a median value of $10^{-4}$, decreasing to $10^{-7}$ for ejected binaries. These results are consistent with a study of BBH mergers in dense star clusters by \cite{Rodriguez_2018}. However, we find that $12\% $ of IMBH-BH mergers enter the LIGO frequency band with detectable eccentricities (e $ > 0.05$). These high eccentricity mergers occur specifically when two BHs merge during a binary-mediated interaction.

The distribution of spins in our IMBH-BH mergers generally peaks at two different values. Most of the BHs have close to zero spin (as expected, based on our BH natal spin assumption) and the other population shows spins at about 0.7.  See \cite{Rodriguez2019} for a full study of the fate of BHs resulting from repeated mergers in dense star clusters with different assumptions of BH natal spins.

\subsection{Merger Rates}
\label{sec:rates}
We now estimate merger rates for BH binaries that merge inside the cluster as isolated binaries or through dynamical interactions, as well as ejected binaries with inspiral times less than a Hubble time. To calculate the volumetric BBH merger rates as a function of redshift, we use a similar approach as in \cite{O'Leary} and \cite{Rodriguez16}, where the comoving merger rate is defined as:

\begin{equation}\label{eq:1}
    \mathcal{R}(z) = \frac{dN(z)}{dt} \times \rho_{\rm GC} \times f.
\end{equation}

\noindent Here $\frac{dN(z)}{dt}$ is the number of mergers per unit time per cluster at a given redshift, $\rho_{\rm GC}$ is the volumetric density of GCs in the local universe---for which we assume a constant value of $\rho_{\rm GC} = 2.31\,{\rm Mpc}^{-3}$  \citep[e.g.,][]{Rodriguez15,Rodriguez_2018}---and $f$ is a scaling factor that accounts for the cluster mass function.

We compute $\frac{dN(z)}{dt}$ from the list of merger times $t_{\rm merge}$ for all BBHs in (or ejected from) our models that merge within a Hubble time $t_{\rm Hubble}$. We use $t_{\rm Hubble} = 13.7$~Gyr and randomly sample 100 cluster ages, $t_{\rm age}$, for each merger from the distribution in \cite{El_Badry_2018} with the appropriate metallicity ($Z = 0.1\,Z_\odot$). We then define the effective time at which these mergers occur as $t_{\rm eff} = t_{\rm Hubble} - t_{\rm age} + t_{\rm merger}$ \cite[e.g.,][]{Kremer20}. $\frac{dN(z)}{dt}$ then simply follows by converting $t_{\rm eff}$ from units of time to redshift and computing the number of BBH mergers per redshift bin. Finally, the rate is normalized by the number of cluster ages sampled and the number of models in our study.

Previous studies have shown that the number of BBH mergers increases roughly linearly with cluster mass \citep[e.g.,][]{Rodriguez16,Kremer20CMC}---though this relation may not hold well for clusters that form few massive stars or retain few BHs, e.g., clusters born with top-light stellar IMFs \citep[][]{Weatherford21}. To account for the higher-mass end of the cluster initial mass function, we incorporate an uncertain factor $f$ as a scaling factor in the rate calculation. To estimate this number, we use the method described in \cite{Rodriguez15}, where $f$ is the number of mergers obtained by integrating the linear relation between the number of BBH mergers and total cluster mass over a normalized cluster-mass function divided by the average number of BBH mergers per cluster in the models sampled. \cite{Kremer20CMC} applied this calculation to a catalog of \texttt{CMC} simulations that cover the full range of GC properties observed in the Milky Way. The range of initial GC properties we use falls within the range covered by the catalog, so we adapt their best-fit curve for the relationship between the number of BBHs and the cluster mass function for $r_v=1$~pc \citep[Equation 15 in][]{Kremer20CMC}, which yields $f = 4.65$. Our use of this scaling factor is justified with the following reasoning. First, the only difference in the set of initial cluster conditions is the high-mass binary fraction, which is kept at $5\%$ in \cite{Kremer20CMC}. We do not expect the relationship between the total number of BBH mergers and total cluster mass to change with different binary fraction (since both the total cluster mass and number of BHs increase slightly with binary fraction), so the linear relationship would still hold. Second, the average number of BBH mergers per cluster remains fairly similar ($\sim 150$ in our models, while $\sim 100$  in theirs). Furthermore, \cite{Weatherford21} showed that for cluster models with an initial top-heavy IMF, which produce an order of ten times as many BHs as our models, the relationship between the total cluster mass and number of BBHs remains fairly linear (see Equation 2 in \citealt{Weatherford21}). 

\vspace{0.5cm}
Subsequently, the cumulative merger rate is calculated as 
\begin{equation}\label{eq:2}
    \mathcal{R}_{c}(z) = \int_{0}^{z} \mathcal{R}(z')\times \frac{dV_{c}}{dz'} \times (1+z')^{-1}dz',
\end{equation}
\noindent where $\mathcal{R}(z')$ is the comoving merger rate as described in Eq.~\ref{eq:1}, $\frac{dV_{c}}{dz'}$ is the comoving volume at redshift $z'$ and $(1+z')^{-1}$ corrects for time dilation.

\begin{figure}
\begin{center}
\includegraphics[width=1\linewidth]{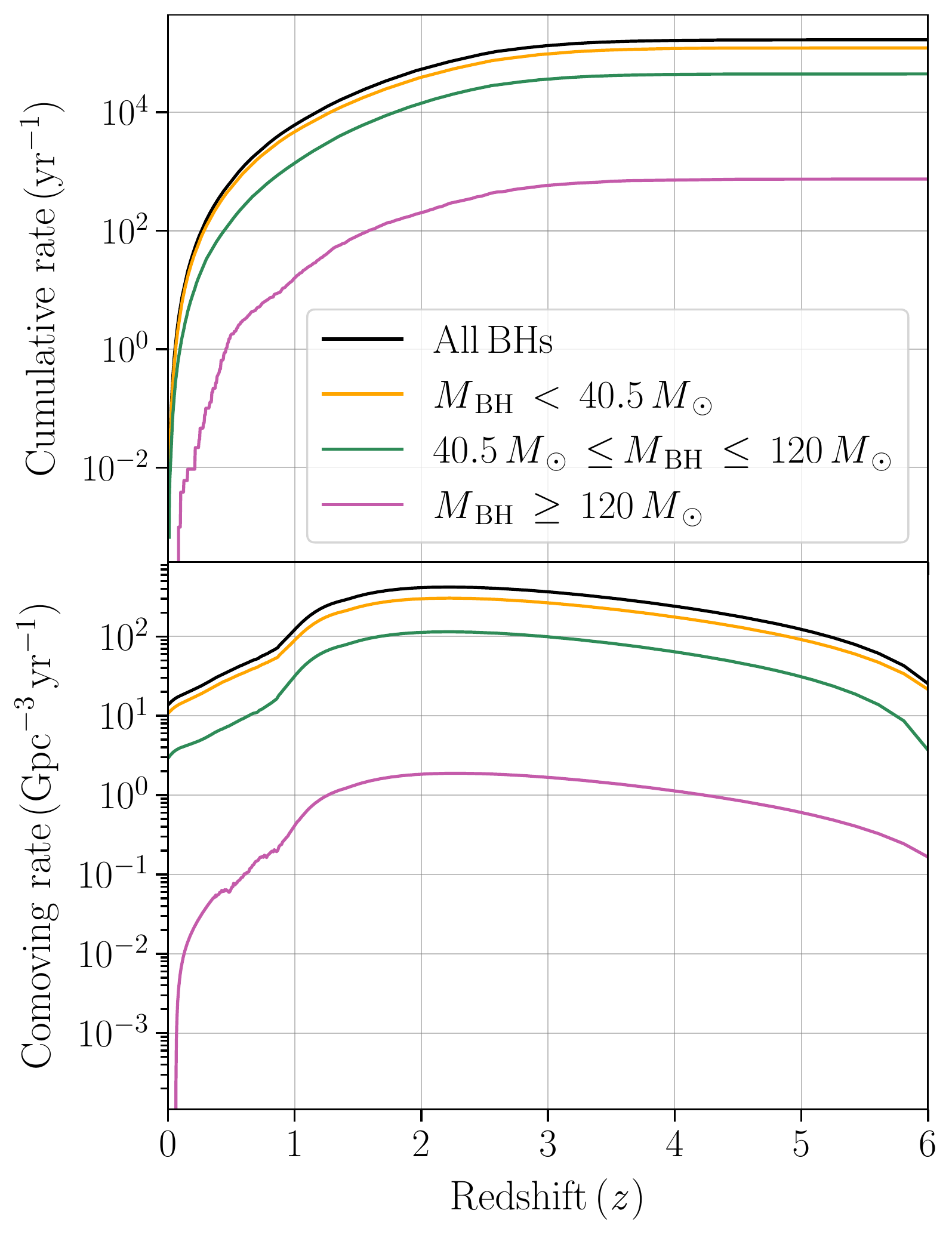}
\caption{\footnotesize \label{fig:rates} Top panel: The cumulative BBH merger rate for BHs, computed using Equation \ref{eq:1}. Bottom panel: the comoving merger rate calculated with Equation \ref{eq:2}. We label as \textit{stellar} mergers cases where both component masses are less than $40.5\,\Msun$. The green and magenta lines show the merger rates for BBHs with at least one component in the ``mass gap'' ($40.5$--$120\,\Msun$) or an IMBH ($M > 120\,\Msun$), respectively. }
\end{center}
\end{figure}

We show in Figure~\hyperref[fig:rates]{10} the cumulative and volumetric rates for different BH mass bins. We estimate a total BBH merger rate in the local universe ($z < 0.2$, to be consistent with the LVK Collaboration) of about \mbox{$20 \,  \rm{Gpc}^{-3}\,\rm{yr}^{-1}$}, which is consistent with the latest rate reported by the LIGO Collaboration, \mbox{$17.9$--$44 \rm\, {Gpc}^{-3}\,\rm{yr}^{-1}$} \citep{LVK21}. Furthermore, in our models, the IMBH merger rate peaks at $z \approx 2$ with a value of about \mbox{$\rm 2\, Gpc^{-3} yr^{-1} $}; the rapid decrease in the IMBH merger rate at lower redshift is due to the dynamical ejection of IMBHs from the cluster. Note that rate calculations in this paper do not include clusters of lower-mass (such as open clusters), which could further increase merger rates \citep[e.g.,][]{DiCarlo19}.

It is not clear whether the scaling factor $f$ used in the method described above remains true for IMBH mergers at the high end of cluster mass. In such clusters, some of these IMBHs could be retained for longer periods of time due to higher escape velocities. Thus, 
we compute an order-of-magnitude merger rate calculation using the general approach as in \cite{Gonzalez_2021}. We define $f_{\rm SF}$ as the fraction of the star formation assumed to occur in star clusters that may produce an IMBH, computed using 

\begin{equation}
    f_{\rm{SF}} \approx \frac{\int_{10^4}^{10^6} M^{-2}\, dM}{\int_{10^2}^{10^6} M^{-2} \,dM} \approx 0.01,
\end{equation}
\noindent where we take $10^{4}\,\Msun $ as the minimum cluster mass to produce an IMBH. We also assume a minimum value of $10^{2}\,\Msun$ for a cluster mass function covering the entire SFR \citep{LadaLada2003}. Finally, the volumetric rate of IMBH mergers at redshift $z$ is defined as 

\begin{equation}
    \label{eq:rate}
    \Gamma(z) \approx  \frac{N_{\rm{IMBH}}}{M_{\rm{cl}}} \,  \rho_{\rm{SF}}(z) \, f_{\rm{SF}}
\end{equation}

\noindent where $\rm N_{\rm IMBH}$ is the average number of IMBHs formed per model, \mbox{$M_{\rm cl} \approx 5 \times 10^{5}\,\Msun$} (the initial cluster mass in our models), and $\rho_{\rm{SF}}(z)$ is the star formation rate density at redshift $z$. We determine a value of \mbox{$0.1\,\Msun\,{\rm Mpc}^{-3}\,{\rm yr}^{-1}$} for $\rho_{\rm{SF}}$ at \mbox{$z \approx 1$} \citep[e.g.,][]{HopkinsBeacom2006}, when metallicities of $0.1\,Z_\odot$ are relevant. This yields an IMBH merger rate for the models with $f_{\rm{b,high}} = 1$ (where \mbox{$N_{\rm IMBH} = 1.27$)} of \mbox{$\Gamma \approx 2.5 \, {\rm Gpc}^{-3}\,\rm{yr}^{-1}$}, a value in agreement with the one calculated above. The IMBH merger rate for the models with \mbox{$f_{\rm{b,high}} = 0.75 $} and \mbox{$f_{\rm{b,high}} = 0.5$} are \mbox{$\Gamma \approx 1.6 \, {\rm Gpc}^{-3}\,\rm{yr}^{-1}$} and \mbox{$\Gamma \approx 1.4\, {\rm Gpc}^{-3}\,\rm{yr}^{-1}$}, respectively.

In a study of the cosmic evolution of BBHs in different star clusters using \texttt{FASTCLUSTER} and \texttt{CosmoRate}, \cite{Mapelli22} calculated a local BBH merger rate of $4$--$\rm 8\,Gpc^{-3}\,yr^{-1}$ in GCs. This is consistent with a previous study on the redshift evolution of the BBH merger rate by \cite{Rodriguez_2018} and \cite{FragioneKocsis2018} that predicted a local rate of $4$--$\rm 18\,Gpc ^{-3}\,yr^{-1}$. We find that the local ($z \approx 0$) merger rate is $\rm 13.9\,Gpc ^{-3}\,yr^{-1}$, in agreement with \cite{Rodriguez_2018}.

\vspace{0.5cm}
\section{Discussion and Conclusions}
\label{sec:discussion}
In this section we review the results of this study and discuss sources of uncertainty that will be the focus of future work. 

\subsection{Summary}
\label{sec:summary}
We have studied the primordial high-mass binary fraction's influence on BH formation in dense star clusters. We observe that at high binary fractions, our models exhibit a tail in the BH mass distribution extending to $500\,\Msun$, beyond the BH upper mass gap and well into the IMBH mass range.

We focus our study on the formation, dynamical evolution, and ejection of IMBHs born with masses in the range $\rm 120$--$500\,\Msun$. A large fraction ($\sim 70\%$) are born from the direct collapse of a massive star formed from consecutive stellar collisions between main-sequence stars and evolved giants. We explore the uncertainties in this formation channel in Section~\hyperref[sec:caveats]{7.3}. The remaining IMBHs are products of BH mergers with massive components.

We have also explored the typical companions of IMBHs in the cluster, noticing a preference towards BHs of masses in the range $\sim 30$--$40\,\Msun$. These binaries are usually short-lived since few-body encounters in the dense cluster core frequently exchange new companions into the IMBH binaries. Eventually, the large number of dynamical interactions will either harden the IMBH--BH binaries to the point of merger, or give the binaries large enough velocity kicks to eject them from the cluster. 

In a study of the dynamical ejection of IMBHs from GCs using the \texttt{MOCCA} code \citep{Giersz14}---which uses a Monte Carlo cluster modeling approach similar to \texttt{CMC}---\cite{Maliszewski_mocca} finds that some IMBHs are retained until a Hubble time. However, the effect of GW recoil kicks due to BBH mergers was not taken into account in the simulations and thus the IMBHs were more likely to be retained. Indeed, we find that all of the IMBHs in our models are ejected from the clusters within the first $\sim 500$~Myr. We explore the mechanisms responsible for these ejections, and discover that most IMBHs are ejected due to the large GW recoil kicks resulting from BH mergers. Some of these IMBHs are ejected in binaries, and we showed that $85\%$ of these binaries merge within a Hubble time and thus are possible sources of GWs detectable by current (LVK) and second-generation (LISA) interferometers.

Lastly, in Section \hyperref[sec:rates]{6.2} we estimate a BBH merger rate of $19.8\,\rm{Gpc}^{-3}\,\rm{yr}^{-1}$ at $z = 0.2$. This is within the range predicted in previous studies of BBH mergers in star clusters \citep[e.g.,][]{Rodriguez_2018}. We also calculate a maximum IMBH merger rate of $\rm \approx 2 \, Gpc^{-3} yr^{-1}$ at 
$z \approx 2$.

\subsection{Ejected IMBHs as seeds of SMBHs}
\label{sec:seeds}
Observations of quasars (QSOs) at high redshifts ($z \gtrsim 6$) indicate the presence of super-massive BHs (SMBHs) in the first billion years of the Universe \citep[e.g.,][]{Fan2006,Decarli2018}. The formation and rapid growth of these objects challenge current theoretical models and have sparked debate on possible formation scenarios \citep[for the most recent review, see][]{Inayoshi2020}. Current formation channels include the collapse of a massive Population III star that forms stellar-mass BH seeds \citep[e.g.,][]{Stacy2012,Hirano2017,Kimura2021}, ``heavy seed scenario'' where  the collapse of a massive gas cloud forms a BH of $M \approx 10^{4}$--$10^{5}\,\Msun$ \citep[e.g.,][]{Oh&Haiman, Mayer2010}, and the formation of BH seeds via repeated stellar mergers \citep[e.g.,][]{Portegies_Zwart_2002,Tagawa_2020} or runaway stellar-mass BH mergers in dense star clusters \citep[e.g.,][]{Davies2011, Kroupa2020}. 

The IMBHs ejected in our models could be promising BH seed candidates for high-redshift SMBHs. This formation channel is explored in \cite{Katz2015}, where metal-poor, high-redshift NSCs are modeled using high-resolution hydro-dynamical cosmological zoom-in simulations as well as direct $N$-body simulations. The study finds that high-redshift NSCs are likely hosts of very massive stars (VMSs) that can collapse into IMBHs. These IMBHs can later grow to masses within the range observed powering high-redshift quasars, if they accrete at the  Eddington rate. However, recent studies made with high-resolution cosmological simulations and direct $N$-body integration of seed BH trajectories found that isolated IMBHs do not have time to sink to the centers of galaxies \citep{Ma21, Pfister19}. A potential solution would be for the IMBHs to remain embedded in the host cluster, and efficiently transported to the Galactic center. Even though all of the IMBHs in our models are ejected from the cluster, this may change for more massive IMBHs and/or in more massive star clusters. More detailed theoretical modelling of massive star evolution at very low metallicities, as well as new observational constraints (e.g., from JWST) on the masses and demographics of young, massive star clusters in the early universe, will be needed 
to better assess this formation channel for high-redshift quasars. 

\subsection{Caveats and Future Work }
\label{sec:caveats}

The most common massive BH formation process in our models is through repeated stellar collisions that form a very massive progenitor. Uncertainties on the properties of the collision products as well as the exact mass-boundaries at which pair instability affects the evolution of the star result in corresponding uncertainties for this formation model. 

As a first step to address these uncertainties, \cite{Ballone22} simulate the collision of two massive stars and follow the structure of the remnant with the smoother-particle hydrodynamics (SPH) code \texttt{StarSmasher} \citep{Gaburov10}. The stars involved in this collision are a core helium-burning (CHeB) star, with mass $M_{\rm CHeB} = 57.6\,\Msun$ and a main-sequence (MS) star with mass $M_{\rm MS} = 41.9\,\Msun$, consistent with a stellar collision in the dynamical simulations of \cite{DiCarlo20} and with typical collisions seen in our models. The study shows that the stellar remnant only loses $12\%$ of the total initial mass during the merger, allowing it to collapse into a massive BH at a later time. 

In a companion paper, \cite{Costa22} carefully model the evolution of the stellar collision product using \texttt{PARSEC} and \texttt{MESA}. They find that the remnant successfully avoids the PI gap and collapses into a BH of mass $\approx 87\,\Msun.$ These results need to be generalized to mergers between different masses and metallicities. Nevertheless, it is the first step in the right direction to study the evolution of massive collision products. 

In this study, we have only considered clusters of a specific metallicity ($Z = 0.1 \, Z_{\odot}$). However, lower metallicities will decrease mass loss in massive stars and thus lead to more massive remnants. Indeed, \cite{Giacobbo18} show that BH masses in merging BHs strongly depend on the progenitor's metallicity. In a study of the impact of metallicity in young star clusters, \cite{DiCarlo20} show that an additional $4\%$ of BHs with masses larger than $60\,\Msun$ form in models with lower metallicities ($ Z \leq 0.002 $ instead of $ Z \leq 0.02$). On the other hand, at very high metallicities, increased stellar wind mass loss may inhibit the growth of massive black holes \citep[e.g.,][]{ShrivastavaKremer2022}. Thus, it would be interesting to expand this study to a wider range of metallicities. Additionally, GCs show evidence of multiple stellar populations, indicating that the properties of primordial binaries may be affected by complex gas-rich environments. Recent studies \citep[e.g.,][]{Rozner2022} have looked into the evolution of these systems and the repercussions on BBH merger rates.

Since the recoil kicks are sensitive to spin magnitudes, our assumption of zero BH natal spin may be influencing the number of ejected BHs. Indeed, \cite{Rodriguez2019} showed that the number of 2G BBH mergers in dense star clusters decreases if a non-zero BH natal spin is assumed. However, since most mergers involving IMBHs already have unequal mass ratios, this property, along with the very low escape speeds of these clusters, should impart GW kicks large enough to eject the IMBH from the cluster. 

Future studies will explore the impact of the IMF shape, coupled with a high binary fraction for massive stars, on the collision rates and massive BH formation. \cite{Weatherford21} showed that a top-heavy stellar IMF increases the number of BHs significantly and  enhances the BBH merger rate. Thus, a grid that encompasses both of these parameters would reveal an interesting spread of clusters with distinct populations of BHs. 

\section{acknowledgements} 

This work was supported by NSF grant AST-2108624 and NASA grant 80NSSC22K0722. G.F.\ and F.A.R.\ acknowledge support from NASA Grant 80NSSC21K1722. K.K.~is supported by an NSF Astronomy and Astrophysics Postdoctoral Fellowship under award AST-2001751. N.C.W.~acknowledges support from the CIERA Riedel Family Graduate Fellowship. Support for M.Z.\ is provided by NASA through the NASA Hubble Fellowship grant HST-HF2-51474.001-A awarded by the Space Telescope Science Institute, which is operated by the Association of Universities for Research in Astronomy, Inc., under NASA contract NAS5-26555. This research was supported in part through the computational resources and staff contributions provided for the Quest high performance computing facility at Northwestern University, which is jointly supported by the Office of the Provost, the Office for Research, and Northwestern University Information Technology.

\bibliographystyle{aasjournal}
\bibliography{main}

\begin{thebibliography}{}
\expandafter\ifx\csname natexlab\endcsname\relax\def\natexlab#1{#1}\fi
\providecommand{\url}[1]{\href{#1}{#1}}
\providecommand{\dodoi}[1]{doi:~\href{http://doi.org/#1}{\nolinkurl{#1}}}
\providecommand{\doeprint}[1]{\href{http://ascl.net/#1}{\nolinkurl{http://ascl.net/#1}}}
\providecommand{\doarXiv}[1]{\href{https://arxiv.org/abs/#1}{\nolinkurl{https://arxiv.org/abs/#1}}}

\bibitem[{{Abbott} {et~al.}(2016){Abbott}, {Abbott}, {Abbott}, {Abernathy},
  {Acernese}, {Ackley}, {Adams}, {Adams}, {Addesso}, {Adhikari}, {Adya},
  {Affeldt}, {Agathos}, {Agatsuma}, {Aggarwal}, {Aguiar}, {Aiello}, {Ain},
  {Ajith}, {Allen}, {Allocca}, {Altin}, {Anderson}, {Anderson}, {Arai},
  {Arain}, {Araya}, {Arceneaux}, {Areeda}, {Arnaud}, {Arun}, {Ascenzi},
  {Ashton}, {Ast}, {Aston}, {Astone}, {Aufmuth}, {Aulbert}, {Babak}, {Bacon},
  {Bader}, {Baker}, {Baldaccini}, {Ballardin}, {Ballmer}, {Barayoga},
  {Barclay}, {Barish}, {Barker}, {Barone}, {Barr}, {Barsotti}, {Barsuglia},
  {Barta}, {Bartlett}, {Barton}, {Bartos}, {Bassiri}, {Basti}, {Batch},
  {Baune}, {Bavigadda}, {Bazzan}, {Behnke}, {Bejger}, {Belczynski}, {Bell},
  {Bell}, {Berger}, {Bergman}, {Bergmann}, {Berry}, {Bersanetti}, {Bertolini},
  {Betzwieser}, {Bhagwat}, {Bhandare}, {Bilenko}, {Billingsley}, {Birch},
  {Birney}, {Birnholtz}, {Biscans}, {Bisht}, {Bitossi}, {Biwer}, {Bizouard},
  {Blackburn}, {Blair}, {Blair}, {Blair}, {Bloemen}, {Bock}, {Bodiya}, {Boer},
  {Bogaert}, {Bogan}, {Bohe}, {Bojtos}, {Bond}, {Bondu}, {Bonnand}, {Boom},
  {Bork}, {Boschi}, {Bose}, {Bouffanais}, {Bozzi}, {Bradaschia}, {Brady},
  {Braginsky}, {Branchesi}, {Brau}, {Briant}, {Brillet}, {Brinkmann},
  {Brisson}, {Brockill}, {Brooks}, {Brown}, {Brown}, {Brown}, {Buchanan},
  {Buikema}, {Bulik}, {Bulten}, {Buonanno}, {Buskulic}, {Buy}, {Byer},
  {Cabero}, {Cadonati}, {Cagnoli}, {Cahillane}, {Bustillo}, {Callister},
  {Calloni}, {Camp}, {Cannon}, {Cao}, {Capano}, {Capocasa}, {Carbognani},
  {Caride}, {Casanueva Diaz}, {Casentini}, {Caudill}, {Cavagli{\`a}},
  {Cavalier}, {Cavalieri}, {Cella}, {Cepeda}, {Baiardi}, {Cerretani},
  {Cesarini}, {Chakraborty}, {Chalermsongsak}, {Chamberlin}, {Chan}, {Chao},
  {Charlton}, {Chassand e-Mottin}, {Chen}, {Chen}, {Cheng}, {Chincarini},
  {Chiummo}, {Cho}, {Cho}, {Chow}, {Christensen}, {Chu}, {Chua}, {Chung},
  {Ciani}, {Clara}, {Clark}, {Cleva}, {Coccia}, {Cohadon}, {Colla}, {Collette},
  {Cominsky}, {Constancio}, {Conte}, {Conti}, {Cook}, {Corbitt}, {Cornish},
  {Corsi}, {Cortese}, {Costa}, {Coughlin}, {Coughlin}, {Coulon}, {Countryman},
  {Couvares}, {Cowan}, {Coward}, {Cowart}, {Coyne}, {Coyne}, {Craig},
  {Creighton}, {Creighton}, {Cripe}, {Crowder}, {Cruise}, {Cumming},
  {Cunningham}, {Cuoco}, {Dal Canton}, {Danilishin}, {D'Antonio}, {Danzmann},
  {Darman}, {Da Silva Costa}, {Dattilo}, {Dave}, {Daveloza}, {Davier},
  {Davies}, {Daw}, {Day}, {De}, {DeBra}, {Debreczeni}, {Degallaix}, {De
  Laurentis}, {Del{\'e}glise}, {Del Pozzo}, {Denker}, {Dent}, {Dereli},
  {Dergachev}, {DeRosa}, {De Rosa}, {DeSalvo}, {Dhurandhar}, {D{\'\i}az}, {Di
  Fiore}, {Di Giovanni}, {Di Lieto}, {Di Pace}, {Di Palma}, {Di Virgilio},
  {Dojcinoski}, {Dolique}, {Donovan}, {Dooley}, {Doravari}, {Douglas},
  {Downes}, {Drago}, {Drever}, {Driggers}, {Du}, {Ducrot}, {Dwyer}, {Edo},
  {Edwards}, {Effler}, {Eggenstein}, {Ehrens}, {Eichholz}, {Eikenberry},
  {Engels}, {Essick}, {Etzel}, {Evans}, {Evans}, {Everett}, {Factourovich},
  {Fafone}, {Fair}, {Fairhurst}, {Fan}, {Fang}, {Farinon}, {Farr}, {Farr},
  {Favata}, {Fays}, {Fehrmann}, {Fejer}, {Feldbaum}, {Ferrante}, {Ferreira},
  {Ferrini}, {Fidecaro}, {Finn}, {Fiori}, {Fiorucci}, {Fisher}, {Flaminio},
  {Fletcher}, {Fong}, {Fournier}, {Franco}, {Frasca}, {Frasconi}, {Frede},
  {Frei}, {Freise}, {Frey}, {Frey}, {Fricke}, {Fritschel}, {Frolov}, {Fulda},
  {Fyffe}, {Gabbard}, {Gair}, {Gammaitoni}, {Gaonkar}, {Garufi}, {Gatto},
  {Gaur}, {Gehrels}, {Gemme}, {Gendre}, {Genin}, {Gennai}, {George}, {Gergely},
  {Germain}, {Ghosh}, {Ghosh}, {Ghosh}, {Giaime}, {Giardina}, {Giazotto},
  {Gill}, {Glaefke}, {Gleason}, {Goetz}, {Goetz}, {Gondan}, {Gonz{\'a}lez},
  {Castro}, {Gopakumar}, {Gordon}, {Gorodetsky}, {Gossan}, {Gosselin},
  {Gouaty}, {Graef}, {Graff}, {Granata}, {Grant}, {Gras}, {Gray}, {Greco},
  {Green}, {Greenhalgh}, {Groot}, {Grote}, {Grunewald}, {Guidi}, {Guo},
  {Gupta}, {Gupta}, {Gushwa}, {Gustafson}, {Gustafson}, {Hacker}, {Hall},
  {Hall}, {Hammond}, {Haney}, {Hanke}, {Hanks}, {Hanna}, {Hannam}, {Hanson},
  {Hardwick}, {Harms}, {Harry}, {Harry}, {Hart}, {Hartman}, {Haster},
  {Haughian}, {Healy}, {Heefner}, {Heidmann}, {Heintze}, {Heinzel}, {Heitmann},
  {Hello}, {Hemming}, {Hendry}, {Heng}, {Hennig}, {Heptonstall}, {Heurs},
  {Hild}, {Hoak}, {Hodge}, {Hofman}, {Hollitt}, {Holt}, {Holz}, {Hopkins},
  {Hosken}, {Hough}, {Houston}, {Howell}, {Hu}, {Huang}, {Huerta}, {Huet},
  {Hughey}, {Husa}, {Huttner}, {Huynh-Dinh}, {Idrisy}, {Indik}, {Ingram},
  {Inta}, {Isa}, {Isac}, {Isi}, {Islas}, {Isogai}, {Iyer}, {Izumi}, {Jacobson},
  {Jacqmin}, {Jang}, {Jani}, {Jaranowski}, {Jawahar}, {Jim{\'e}nez-Forteza},
  {Johnson}, {Johnson-McDaniel}, {Jones}, {Jones}, {Jonker}, {Ju}, {Haris},
  {Kalaghatgi}, {Kalogera}, {Kandhasamy}, {Kang}, {Kanner}, {Karki},
  {Kasprzack}, {Katsavounidis}, {Katzman}, {Kaufer}, {Kaur}, {Kawabe},
  {Kawazoe}, {K{\'e}f{\'e}lian}, {Kehl}, {Keitel}, {Kelley}, {Kells},
  {Kennedy}, {Keppel}, {Key}, {Khalaidovski}, {Khalili}, {Khan}, {Khan},
  {Khan}, {Khazanov}, {Kijbunchoo}, {Kim}, {Kim}, {Kim}, {Kim}, {Kim}, {Kim},
  {King}, {King}, {Kinzel}, {Kissel}, {Kleybolte}, {Klimenko}, {Koehlenbeck},
  {Kokeyama}, {Koley}, {Kondrashov}, {Kontos}, {Koranda}, {Korobko}, {Korth},
  {Kowalska}, {Kozak}, {Kringel}, {Krishnan}, {Kr{\'o}lak}, {Krueger}, {Kuehn},
  {Kumar}, {Kumar}, {Kuo}, {Kutynia}, {Kwee}, {Lackey}, {Landry}, {Lange},
  {Lantz}, {Lasky}, {Lazzarini}, {Lazzaro}, {Leaci}, {Leavey}, {Lebigot},
  {Lee}, {Lee}, {Lee}, {Lee}, {Lenon}, {Leonardi}, {Leong}, {Leroy},
  {Letendre}, {Levin}, {Levine}, {Li}, {Libson}, {Littenberg}, {Lockerbie},
  {Logue}, {Lombardi}, {London}, {Lord}, {Lorenzini}, {Loriette}, {Lormand},
  {Losurdo}, {Lough}, {Lousto}, {Lovelace}, {L{\"u}ck}, {Lundgren}, {Luo},
  {Lynch}, {Ma}, {MacDonald}, {Machenschalk}, {MacInnis}, {Macleod},
  {Maga{\~n}a-Sandoval}, {Magee}, {Mageswaran}, {Majorana}, {Maksimovic},
  {Malvezzi}, {Man}, {Mandel}, {Mandic}, {Mangano}, {Mansell}, {Manske},
  {Mantovani}, {Marchesoni}, {Marion}, {M{\'a}rka}, {M{\'a}rka}, {Markosyan},
  {Maros}, {Martelli}, {Martellini}, {Martin}, {Martin}, {Martynov}, {Marx},
  {Mason}, {Masserot}, {Massinger}, {Masso-Reid}, {Matichard}, {Matone},
  {Mavalvala}, {Mazumder}, {Mazzolo}, {McCarthy}, {McClelland}, {McCormick},
  {McGuire}, {McIntyre}, {McIver}, {McManus}, {McWilliams}, {Meacher},
  {Meadors}, {Meidam}, {Melatos}, {Mendell}, {Mendoza-Gandara}, {Mercer},
  {Merilh}, {Merzougui}, {Meshkov}, {Messenger}, {Messick}, {Meyers},
  {Mezzani}, {Miao}, {Michel}, {Middleton}, {Mikhailov}, {Milano}, {Miller},
  {Millhouse}, {Minenkov}, {Ming}, {Mirshekari}, {Mishra}, {Mitra},
  {Mitrofanov}, {Mitselmakher}, {Mittleman}, {Moggi}, {Mohan}, {Mohapatra},
  {Montani}, {Moore}, {Moore}, {Moraru}, {Moreno}, {Morriss}, {Mossavi},
  {Mours}, {Mow-Lowry}, {Mueller}, {Mueller}, {Muir}, {Mukherjee}, {Mukherjee},
  {Mukherjee}, {Mukund}, {Mullavey}, {Munch}, {Murphy}, {Murray}, {Mytidis},
  {Nardecchia}, {Naticchioni}, {Nayak}, {Necula}, {Nedkova}, {Nelemans},
  {Neri}, {Neunzert}, {Newton}, {Nguyen}, {Nielsen}, {Nissanke}, {Nitz},
  {Nocera}, {Nolting}, {Normandin}, {Nuttall}, {Oberling}, {Ochsner}, {O'Dell},
  {Oelker}, {Ogin}, {Oh}, {Oh}, {Ohme}, {Oliver}, {Oppermann}, {Oram},
  {O'Reilly}, {O'Shaughnessy}, {Ott}, {Ottaway}, {Ottens}, {Overmier}, {Owen},
  {Pai}, {Pai}, {Palamos}, {Palashov}, {Palomba}, {Pal-Singh}, {Pan}, {Pan},
  {Pankow}, {Pannarale}, {Pant}, {Paoletti}, {Paoli}, {Papa}, {Paris},
  {Parker}, {Pascucci}, {Pasqualetti}, {Passaquieti}, {Passuello},
  {Patricelli}, {Patrick}, {Pearlstone}, {Pedraza}, {Pedurand }, {Pekowsky},
  {Pele}, {Penn}, {Perreca}, {Pfeiffer}, {Phelps}, {Piccinni}, {Pichot},
  {Pickenpack}, {Piergiovanni}, {Pierro}, {Pillant}, {Pinard}, {Pinto},
  {Pitkin}, {Poeld}, {Poggiani}, {Popolizio}, {Post}, {Powell}, {Prasad},
  {Predoi}, {Premachandra}, {Prestegard}, {Price}, {Prijatelj}, {Principe},
  {Privitera}, {Prix}, {Prodi}, {Prokhorov}, {Puncken}, {Punturo}, {Puppo},
  {P{\"u}rrer}, {Qi}, {Qin}, {Quetschke}, {Quintero}, {Quitzow-James}, {Raab},
  {Rabeling}, {Radkins}, {Raffai}, {Raja}, {Rakhmanov}, {Ramet}, {Rapagnani},
  {Raymond}, {Razzano}, {Re}, {Read}, {Reed}, {Regimbau}, {Rei}, {Reid},
  {Reitze}, {Rew}, {Reyes}, {Ricci}, {Riles}, {Robertson}, {Robie}, {Robinet},
  {Rocchi}, {Rolland}, {Rollins}, {Roma}, {Romano}, {Romano}, {Romanov},
  {Romie}, {Rosi{\'n}ska}, {Rowan}, {R{\"u}diger}, {Ruggi}, {Ryan}, {Sachdev},
  {Sadecki}, {Sadeghian}, {Salconi}, {Saleem}, {Salemi}, {Samajdar}, {Sammut},
  {Sampson}, {Sanchez}, {Sandberg}, {Sandeen}, {Sand ers}, {Sanders},
  {Sassolas}, {Sathyaprakash}, {Saulson}, {Sauter}, {Savage}, {Sawadsky},
  {Schale}, {Schilling}, {Schmidt}, {Schmidt}, {Schnabel}, {Schofield},
  {Sch{\"o}nbeck}, {Schreiber}, {Schuette}, {Schutz}, {Scott}, {Scott},
  {Sellers}, {Sengupta}, {Sentenac}, {Sequino}, {Sergeev}, {Serna},
  {Setyawati}, {Sevigny}, {Shaddock}, {Shaffer}, {Shah}, {Shahriar}, {Shaltev},
  {Shao}, {Shapiro}, {Shawhan}, {Sheperd}, {Shoemaker}, {Shoemaker}, {Siellez},
  {Siemens}, {Sigg}, {Silva}, {Simakov}, {Singer}, {Singer}, {Singh}, {Singh},
  {Singhal}, {Sintes}, {Slagmolen}, {Smith}, {Smith}, {Smith}, {Smith}, {Son},
  {Sorazu}, {Sorrentino}, {Souradeep}, {Srivastava}, {Staley}, {Steinke},
  {Steinlechner}, {Steinlechner}, {Steinmeyer}, {Stephens}, {Stevenson},
  {Stone}, {Strain}, {Straniero}, {Stratta}, {Strauss}, {Strigin}, {Sturani},
  {Stuver}, {Summerscales}, {Sun}, {Sutton}, {Swinkels}, {Szczepa{\'n}czyk},
  {Tacca}, {Talukder}, {Tanner}, {T{\'a}pai}, {Tarabrin}, {Taracchini},
  {Taylor}, {Theeg}, {Thirugnanasambandam}, {Thomas}, {Thomas}, {Thomas},
  {Thorne}, {Thorne}, {Thrane}, {Tiwari}, {Tiwari}, {Tokmakov}, {Tomlinson},
  {Tonelli}, {Torres}, {Torrie}, {T{\"o}yr{\"a}}, {Travasso}, {Traylor},
  {Trifir{\`o}}, {Tringali}, {Trozzo}, {Tse}, {Turconi}, {Tuyenbayev},
  {Ugolini}, {Unnikrishnan}, {Urban}, {Usman}, {Vahlbruch}, {Vajente},
  {Valdes}, {Vallisneri}, {van Bakel}, {van Beuzekom}, {van den Brand}, {Van
  Den Broeck}, {Vand er-Hyde}, {van der Schaaf}, {van Heijningen}, {van
  Veggel}, {Vardaro}, {Vass}, {Vas{\'u}th}, {Vaulin}, {Vecchio}, {Vedovato},
  {Veitch}, {Veitch}, {Venkateswara}, {Verkindt}, {Vetrano}, {Vicer{\'e}},
  {Vinciguerra}, {Vine}, {Vinet}, {Vitale}, {Vo}, {Vocca}, {Vorvick}, {Voss},
  {Vousden}, {Vyatchanin}, {Wade}, {Wade}, {Wade}, {Waldman}, {Walker},
  {Wallace}, {Walsh}, {Wang}, {Wang}, {Wang}, {Wang}, {Wang}, {Ward}, {Ward},
  {Warner}, {Was}, {Weaver}, {Wei}, {Weinert}, {Weinstein}, {Weiss}, {Welborn},
  {Wen}, {We{\ss}els}, {Westphal}, {Wette}, {Whelan}, {Whitcomb}, {White},
  {Whiting}, {Wiesner}, {Wilkinson}, {Willems}, {Williams}, {Williams},
  {Williamson}, {Willis}, {Willke}, {Wimmer}, {Winkelmann}, {Winkler}, {Wipf},
  {Wiseman}, {Wittel}, {Woan}, {Worden}, {Wright}, {Wu}, {Yablon}, {Yakushin},
  {Yam}, {Yamamoto}, {Yancey}, {Yap}, {Yu}, {Yvert}, {Zadro{\.Z}ny},
  {Zangrando}, {Zanolin}, {Zendri}, {Zevin}, {Zhang}, {Zhang}, {Zhang},
  {Zhang}, {Zhao}, {Zhou}, {Zhou}, {Zhu}, {Zucker}, {Zuraw}, {Zweizig}, {LIGO
  Scientific Collaboration}, \& {Virgo Collaboration}}]{Abbott2016}
{Abbott}, B.~P., {Abbott}, R., {Abbott}, T.~D., {et~al.} 2016, \prl, 116,
  061102, \dodoi{10.1103/PhysRevLett.116.061102}

\bibitem[{{Abbott} {et~al.}(2020){Abbott}, {Abbott}, {Abraham}, {Acernese},
  {Ackley}, {Adams}, {Adams}, {Adhikari}, {Adya}, {Affeldt}, {Agathos},
  {Agatsuma}, {Aggarwal}, {Aguiar}, {Aiello}, {Ain}, {Ajith}, {Akcay}, {Allen},
  {Allocca}, {Altin}, {Amato}, {Anand}, {Ananyeva}, {Anderson}, {Anderson},
  {Angelova}, {Ansoldi}, {Antelis}, {Antier}, {Appert}, {Arai}, {Araya},
  {Areeda}, {Ar{\`e}ne}, {Arnaud}, {Aronson}, {Arun}, {Asali}, {Ascenzi},
  {Ashton}, {Aston}, {Astone}, \& et~al.}]{LIGO2020_O3}
{Abbott}, R., {Abbott}, T.~D., {Abraham}, S., {et~al.} 2020, arXiv e-prints,
  arXiv:2010.14527.
\newblock \doarXiv{2010.14527}

\bibitem[{Abbott {et~al.}(2020)Abbott, Abbott, Abraham, Acernese, Ackley,
  Adams, Adhikari, Adya, Affeldt, Agathos, Agatsuma, Aggarwal, Aguiar, Aich,
  Aiello, Ain, Ajith, Akcay, Allen, Allocca, Altin, Amato, Anand, Ananyeva,
  Anderson, Anderson, Angelova, Ansoldi, Antier, Appert, Arai, Araya, Areeda,
  Ar\`ene, Arnaud, Aronson, Arun, Asali, Ascenzi, Ashton, Aston, Astone, Aubin,
  Aufmuth, AultONeal, Austin, Avendano, Babak, Bacon, Badaracco, Bader, Bae,
  Baer, Baird, Baldaccini, Ballardin, Ballmer, Bals, Balsamo, Baltus, Banagiri,
  Bankar, Bankar, Barayoga, Barbieri, Barish, Barker, Barkett, Barneo, Barone,
  Barr, Barsotti, Barsuglia, Barta, Bartlett, Bartos, Bassiri, Basti, Bawaj,
  Bayley, Bazzan, B\'ecsy, Bejger, Belahcene, Bell, Beniwal, Benjamin, Bentley,
  Bergamin, Berger, Bergmann, Bernuzzi, Berry, Bersanetti, Bertolini,
  Betzwieser, Bhandare, Bhandari, Bidler, Biggs, Bilenko, Billingsley, Birney,
  Birnholtz, Biscans, Bischi, Biscoveanu, Bisht, Bissenbayeva, Bitossi,
  Bizouard, Blackburn, Blackman, Blair, Blair, Blair, Bobba, Bode, Boer,
  Boetzel, Bogaert, Bondu, Bonilla, Bonnand, Booker, Boom, Bork, Boschi, Bose,
  Bossilkov, Bosveld, Bouffanais, Bozzi, Bradaschia, Brady, Bramley, Branchesi,
  Brau, Breschi, Briant, Briggs, Brighenti, Brillet, Brinkmann, Brockill,
  Brooks, Brooks, Brown, Brunett, Bruno, Bruntz, Buikema, Bulik, Bulten,
  Buonanno, Buscicchio, Buskulic, Byer, Cabero, Cadonati, Cagnoli, Cahillane,
  Calder\'on~Bustillo, Callaghan, Callister, Calloni, Camp, Canepa, Cannon,
  Cao, Cao, Carapella, Carbognani, Caride, Carney, Carullo, Casanueva~Diaz,
  Casentini, Casta\~neda, Caudill, Cavagli\`a, Cavalier, Cavalieri, Cella,
  Cerd\'a-Dur\'an, Cesarini, Chaibi, Chakravarti, Chan, Chan, Chandra, Chao,
  Charlton, Chase, Chassande-Mottin, Chatterjee, Chaturvedi, Chatziioannou,
  Chen, Chen, Chen, Cheng, Cheong, Chia, Chiadini, Chierici, Chincarini,
  Chiummo, Cho, Cho, Cho, Christensen, Chu, Chua, Chung, Chung, Ciani,
  Ciecielag, Cie\ifmmode~\acute{s}\else \'{s}\fi{}lar, Ciobanu, Ciolfi,
  Cipriano, Cirone, Clara, Clark, Clearwater, Clesse, Cleva, Coccia, Cohadon,
  Cohen, Colleoni, Collette, Collins, Colpi, Constancio, Conti, Cooper, Corban,
  Corbitt, Cordero-Carri\'on, Corezzi, Corley, Cornish, Corre, Corsi, Cortese,
  Costa, Cotesta, Coughlin, Coughlin, Coulon, Countryman, Couvares, Covas,
  Coward, Cowart, Coyne, Coyne, Creighton, Creighton, Cripe, Croquette,
  Crowder, Cudell, Cullen, Cumming, Cummings, Cunningham, Cuoco, Curylo,
  Canton, D\'alya, Dana, Daneshgaran-Bajastani, D'Angelo, Danilishin,
  D'Antonio, Danzmann, Darsow-Fromm, Dasgupta, Datrier, Dattilo, Dave, Davier,
  Davies, Davis, Daw, DeBra, Deenadayalan, Degallaix, De~Laurentis,
  Del\'eglise, Delfavero, De~Lillo, Del~Pozzo, DeMarchi, D'Emilio, Demos, Dent,
  De~Pietri, De~Rosa, De~Rossi, DeSalvo, de~Varona, Dhurandhar, D\'{\i}az,
  Diaz-Ortiz, Dietrich, Di~Fiore, Di~Fronzo, Di~Giorgio, Di~Giovanni,
  Di~Giovanni, Di~Girolamo, Di~Lieto, Ding, Di~Pace, Di~Palma, Di~Renzo,
  Divakarla, Dmitriev, Doctor, Donovan, Dooley, Doravari, Dorrington, Downes,
  Drago, Driggers, Du, Ducoin, Dupej, Durante, D'Urso, Dwyer, Easter, Eddolls,
  Edelman, Edo, Edy, Effler, Ehrens, Eichholz, Eikenberry, Eisenmann,
  Eisenstein, Ejlli, Errico, Essick, Estelles, Estevez, Etienne, Etzel, Evans,
  Evans, Ewing, Fafone, Fairhurst, Fan, Farinon, Farr, Farr, Fauchon-Jones,
  Favata, Fays, Fazio, Feicht, Fejer, Feng, Fenyvesi, Ferguson,
  Fernandez-Galiana, Ferrante, Ferreira, Ferreira, Fidecaro, Fiori, Fiorucci,
  Fishbach, Fisher, Fittipaldi, Fitz-Axen, Fiumara, Flaminio, Floden, Flynn,
  Fong, Font, Forsyth, Fournier, Frasca, Frasconi, Frei, Freise, Frey, Frey,
  Fritschel, Frolov, Fronz\`e, Fulda, Fyffe, Gabbard, Gadre, Gaebel, Gair,
  Galaudage, Ganapathy, Ganguly, Gaonkar, Garc\'{\i}a-Quir\'os, Garufi,
  Gateley, Gaudio, Gayathri, Gemme, Genin, Gennai, George, George, Gergely,
  Ghonge, Ghosh, Ghosh, Ghosh, Giacomazzo, Giaime, Giardina, Gibson, Gier,
  Gill, Glanzer, Gniesmer, Godwin, Goetz, Goetz, Gohlke, Goncharov, Gonz\'alez,
  Gopakumar, Gossan, Gosselin, Gouaty, Grace, Grado, Granata, Grant, Gras,
  Grassia, Gray, Gray, Greco, Green, Green, Gretarsson, Griggs, Grignani,
  Grimaldi, Grimm, Grote, Grunewald, Gruning, Guidi, Guimaraes, Guix\'e,
  Gulati, Guo, Gupta, Gupta, Gupta, Gustafson, Gustafson, Haegel, Halim, Hall,
  Hamilton, Hammond, Haney, Hanke, Hanks, Hanna, Hannam, Hannuksela, Hansen,
  Hanson, Harder, Hardwick, Haris, Harms, Harry, Harry, Hasskew, Haster,
  Haughian, Hayes, Healy, Heidmann, Heintze, Heinze, Heitmann, Hellman, Hello,
  Hemming, Hendry, Heng, Hennes, Hennig, Heurs, Hild, Hinderer, Hoback,
  Hochheim, Hofgard, Hofman, Holgado, Holland, Holt, Holz, Hopkins, Horst,
  Hough, Howell, Hoy, Huang, H\"ubner, Huerta, Huet, Hughey, Hui, Husa,
  Huttner, Huxford, Huynh-Dinh, Idzkowski, Iess, Inchauspe, Ingram, Intini,
  Isac, Isi, Iyer, Jacqmin, Jadhav, Jadhav, James, Jani, Janthalur, Jaranowski,
  Jariwala, Jaume, Jenkins, Jiang, Johns, Johnson-McDaniel, Jones, Jones,
  Jones, Jones, Jones, Jonker, Ju, Junker, Kalaghatgi, Kalogera, Kamai,
  Kandhasamy, Kang, Kanner, Kapadia, Karki, Kashyap, Kasprzack, Kastaun,
  Katsanevas, Katsavounidis, Katzman, Kaufer, Kawabe, K\'ef\'elian, Keitel,
  Keivani, Kennedy, Key, Khadka, Khalili, Khan, Khan, Khan, Khazanov, Khetan,
  Khursheed, Kijbunchoo, Kim, Kim, Kim, Kim, Kim, Kim, Kim, Kimball, King,
  Kinley-Hanlon, Kirchhoff, Kissel, Kleybolte, Klimenko, Knowles, Knyazev,
  Koch, Koehlenbeck, Koekoek, Koley, Kondrashov, Kontos, Koper, Korobko, Korth,
  Kovalam, Kozak, Kringel, Krishnendu, Kr\'olak, Krupinski, Kuehn, Kumar,
  Kumar, Kumar, Kumar, Kumar, Kuo, Kutynia, Lackey, Laghi, Lalande, Lam,
  Lamberts, Landry, Lane, Lang, Lange, Lantz, Lanza, La~Rosa, Lartaux-Vollard,
  Lasky, Laxen, Lazzarini, Lazzaro, Leaci, Leavey, Lecoeuche, Lee, Lee, Lee,
  Lee, Lee, Lehmann, Leroy, Letendre, Levin, Li, Li, li, Li, Li, Linde, Linker,
  Linley, Littenberg, Liu, Liu, Llorens-Monteagudo, Lo, Lockwood, London,
  Longo, Lorenzini, Loriette, Lormand, Losurdo, Lough, Lousto, Lovelace,
  L\"uck, Lumaca, Lundgren, Ma, Macas, Macfoy, MacInnis, Macleod, MacMillan,
  Macquet, Maga\~na Hernandez, Maga\~na Sandoval, Magee, Majorana, Maksimovic,
  Malik, Man, Mandic, Mangano, Mansell, Manske, Mantovani, Mapelli, Marchesoni,
  Marion, M\'arka, M\'arka, Markakis, Markosyan, Markowitz, Maros, Marquina,
  Marsat, Martelli, Martin, Martin, Martinez, Martynov, Masalehdan, Mason,
  Massera, Masserot, Massinger, Masso-Reid, Mastrogiovanni, Matas, Matichard,
  Mavalvala, Maynard, McCann, McCarthy, McClelland, McCormick, McCuller,
  McGuire, McIsaac, McIver, McManus, McRae, McWilliams, Meacher, Meadors,
  Mehmet, Mehta, Mejuto~Villa, Melatos, Mendell, Mercer, Mereni, Merfeld,
  Merilh, Merritt, Merzougui, Meshkov, Messenger, Messick, Metzdorff, Meyers,
  Meylahn, Mhaske, Miani, Miao, Michaloliakos, Michel, Middleton, Milano,
  Miller, Millhouse, Mills, Milotti, Milovich-Goff, Minazzoli, Minenkov,
  Mishkin, Mishra, Mistry, Mitra, Mitrofanov, Mitselmakher, Mittleman, Mo,
  Mogushi, Mohapatra, Mohite, Molina-Ruiz, Mondin, Montani, Moore, Moraru,
  Morawski, Moreno, Morisaki, Mours, Mow-Lowry, Mozzon, Muciaccia, Mukherjee,
  Mukherjee, Mukherjee, Mukherjee, Mukund, Mullavey, Munch, Mu\~niz, Murray,
  Nagar, Nardecchia, Naticchioni, Nayak, Neil, Neilson, Nelemans, Nelson, Nery,
  Neunzert, Ng, Ng, Nguyen, Nguyen, Nichols, Nichols, Nissanke, Nitz, Nocera,
  Noh, North, Nothard, Nuttall, Oberling, O'Brien, Oganesyan, Ogin, Oh, Oh,
  Ohme, Ohta, Okada, Oliver, Olivetto, Oppermann, Oram, O'Reilly, Ormiston,
  Ortega, O'Shaughnessy, Ossokine, Osthelder, Ottaway, Overmier, Owen, Pace,
  Pagano, Page, Pagliaroli, Pai, Pai, Palamos, Palashov, Palomba, Pan, Panda,
  Pang, Pankow, Pannarale, Pant, Paoletti, Paoli, Parida, Parker, Pascucci,
  Pasqualetti, Passaquieti, Passuello, Patricelli, Payne, Pearlstone, Pechsiri,
  Pedersen, Pedraza, Pele, Penn, Perego, Perez, P\'erigois, Perreca, Perri\`es,
  Petermann, Pfeiffer, Phelps, Phukon, Piccinni, Pichot, Piendibene,
  Piergiovanni, Pierro, Pillant, Pinard, Pinto, Piotrzkowski, Pirello, Pitkin,
  Plastino, Poggiani, Pong, Ponrathnam, Popolizio, Porter, Powell, Prajapati,
  Prasai, Prasanna, Pratten, Prestegard, Principe, Prodi, Prokhorov, Punturo,
  Puppo, P\"urrer, Qi, Quetschke, Quinonez, Raab, Raaijmakers, Radkins,
  Radulesco, Raffai, Rafferty, Raja, Rajan, Rajbhandari, Rakhmanov, Ramirez,
  Ramos-Buades, Rana, Rao, Rapagnani, Raymond, Razzano, Read, Regimbau, Rei,
  Reid, Reitze, Rettegno, Ricci, Richardson, Richardson, Ricker,
  Riemenschneider, Riles, Rizzo, Robertson, Robinet, Rocchi, Rodriguez-Soto,
  Rolland, Rollins, Roma, Romanelli, Romano, Romel, Romero-Shaw, Romie, Rose,
  Rose, Rose, Rosi\ifmmode~\acute{n}\else \'{n}\fi{}ska, Rosofsky, Ross, Rowan,
  Rowlinson, Roy, Roy, Roy, Ruggi, Rutins, Ryan, Sachdev, Sadecki,
  Sakellariadou, Salafia, Salconi, Saleem, Salemi, Samajdar, Sanchez, Sanchez,
  Sanchis-Gual, Sanders, Santiago, Santos, Sarin, Sassolas, Sathyaprakash,
  Sauter, Savage, Savant, Sawant, Sayah, Schaetzl, Schale, Scheel, Scheuer,
  Schmidt, Schnabel, Schofield, Sch\"onbeck, Schreiber, Schulte, Schutz,
  Schwarm, Schwartz, Scott, Scott, Seidel, Sellers, Sengupta, Sennett,
  Sentenac, Sequino, Sergeev, Setyawati, Shaddock, Shaffer, Sharifi, Shahriar,
  Sharma, Sharma, Shawhan, Shen, Shikauchi, Shink, Shoemaker, Shoemaker,
  Shukla, ShyamSundar, Siellez, Sieniawska, Sigg, Singer, Singh, Singh, Singha,
  Singhal, Sintes, Sipala, Skliris, Slagmolen, Slaven-Blair, Smetana, Smith,
  Smith, Somala, Son, Soni, Sorazu, Sordini, Sorrentino, Souradeep, Sowell,
  Spencer, Spera, Srivastava, Srivastava, Staats, Stachie, Standke, Steer,
  Steinke, Steinlechner, Steinlechner, Steinmeyer, Stevenson, Stocks, Stops,
  Stover, Strain, Stratta, Strunk, Sturani, Stuver, Sudhagar, Sudhir,
  Summerscales, Sun, Sunil, Sur, Suresh, Sutton, Swinkels,
  Szczepa\ifmmode~\acute{n}\else \'{n}\fi{}czyk, Tacca, Tait, Talbot,
  Tanasijczuk, Tanner, Tao, T\'apai, Tapia, Tapia San~Martin, Tasson, Taylor,
  Tenorio, Terkowski, Thirugnanasambandam, Thomas, Thomas, Thompson, Thondapu,
  Thorne, Thrane, Tinsman, Saravanan, Tiwari, Tiwari, Tiwari, Toland, Tonelli,
  Tornasi, Torres-Forn\'e, Torrie, Tosta~e Melo, T\"oyr\"a, Travasso, Traylor,
  Tringali, Tripathee, Trovato, Trudeau, Tsang, Tse, Tso, Tsukada, Tsuna,
  Tsutsui, Turconi, Ubhi, Udall, Ueno, Ugolini, Unnikrishnan, Urban, Usman,
  Utina, Vahlbruch, Vajente, Valdes, Valentini, van Bakel, van Beuzekom,
  van~den Brand, Van Den~Broeck, Vander-Hyde, van~der Schaaf, Van~Heijningen,
  van Veggel, Vardaro, Varma, Vass, Vas\'uth, Vecchio, Vedovato, Veitch,
  Veitch, Venkateswara, Venugopalan, Verkindt, Veske, Vetrano, Vicer\'e, Viets,
  Vinciguerra, Vine, Vinet, Vitale, Vivanco, Vo, Vocca, Vorvick, Vyatchanin,
  Wade, Wade, Wade, Walet, Walker, Wallace, Wallace, Walsh, Wang, Wang, Wang,
  Ward, Warden, Warner, Was, Watchi, Weaver, Wei, Weinert, Weinstein, Weiss,
  Wellmann, Wen, We\ss{}els, Westhouse, Wette, Whelan, Whiting, Whittle,
  Wilken, Williams, Williamson, Willis, Willke, Winkler, Wipf, Wittel, Woan,
  Woehler, Wofford, Wong, Wright, Wu, Wysocki, Xiao, Yamamoto, Yang, Yang,
  Yang, Yap, Yazback, Yeeles, Yu, Yu, Yuen, Zadro\ifmmode~\dot{z}\else
  \.{z}\fi{}ny, Zadro\ifmmode~\dot{z}\else \.{z}\fi{}ny, Zanolin, Zelenova,
  Zendri, Zevin, Zhang, Zhang, Zhang, Zhao, Zhao, Zhou, Zhou, Zhu, Zimmerman,
  Zucker, \& Zweizig}]{GW190521}
Abbott, R., Abbott, T.~D., Abraham, S., {et~al.} 2020, Phys. Rev. Lett., 125,
  101102, \dodoi{10.1103/PhysRevLett.125.101102}

\bibitem[{{Antonini} {et~al.}(2019){Antonini}, {Gieles}, \&
  {Gualandris}}]{AntoniniGieles2019}
{Antonini}, F., {Gieles}, M., \& {Gualandris}, A. 2019, \mnras, 486, 5008,
  \dodoi{10.1093/mnras/stz1149}

\bibitem[{{Bacon} {et~al.}(1996){Bacon}, {Sigurdsson}, \& {Davies}}]{Bacon1996}
{Bacon}, D., {Sigurdsson}, S., \& {Davies}, M.~B. 1996, \mnras, 281, 830,
  \dodoi{10.1093/mnras/281.3.830}

\bibitem[{{Ballone} {et~al.}(2022){Ballone}, {Costa}, {Mapelli}, \&
  {MacLeod}}]{Ballone22}
{Ballone}, A., {Costa}, G., {Mapelli}, M., \& {MacLeod}, M. 2022, arXiv
  e-prints, arXiv:2204.03493.
\newblock \doarXiv{2204.03493}

\bibitem[{Banerjee(2020)}]{Banerjee_2020}
Banerjee, S. 2020, Monthly Notices of the Royal Astronomical Society, 500,
  3002–3026, \dodoi{10.1093/mnras/staa2392}

\bibitem[{{Banerjee}(2022)}]{Banerjee2022}
{Banerjee}, S. 2022, \aap, 665, A20, \dodoi{10.1051/0004-6361/202142331}

\bibitem[{{Banerjee, S.} {et~al.}(2020){Banerjee, S.}, {Belczynski, K.},
  {Fryer, C. L.}, {Berczik, P.}, {Hurley, J. R.}, {Spurzem, R.}, \& {Wang,
  L.}}]{Banerjee_2020_b}
{Banerjee, S.}, {Belczynski, K.}, {Fryer, C. L.}, {et~al.} 2020, A\&A, 639,
  A41, \dodoi{10.1051/0004-6361/201935332}

\bibitem[{Belczynski(2020)}]{Belczynski_2020}
Belczynski, K. 2020, The Astrophysical Journal, 905, L15,
  \dodoi{10.3847/2041-8213/abcbf1}

\bibitem[{Belczynski {et~al.}(2016)Belczynski, Heger, Gladysz, Ruiter, Woosley,
  Wiktorowicz, Chen, Bulik, O’Shaughnessy, Holz, Fryer, \&
  Berti}]{Belczynski2016b}
Belczynski, K., Heger, A., Gladysz, W., {et~al.} 2016, Astronomy {\&}
  Astrophysics, 594, A97, \dodoi{10.1051/0004-6361/201628980}

\bibitem[{{Belczynski} {et~al.}(2020){Belczynski}, {Klencki}, {Fields},
  {Olejak}, {Berti}, {Meynet}, {Fryer}, {Holz}, {O'Shaughnessy}, {Brown},
  {Bulik}, {Leung}, {Nomoto}, {Madau}, {Hirschi}, {Kaiser}, {Jones}, {Mondal},
  {Chruslinska}, {Drozda}, {Gerosa}, {Doctor}, {Giersz}, {Ekstrom}, {Georgy},
  {Askar}, {Baibhav}, {Wysocki}, {Natan}, {Farr}, {Wiktorowicz}, {Coleman
  Miller}, {Farr}, \& {Lasota}}]{Bel2020_pisne}
{Belczynski}, K., {Klencki}, J., {Fields}, C.~E., {et~al.} 2020, \aap, 636,
  A104, \dodoi{10.1051/0004-6361/201936528}

\bibitem[{{Bond} {et~al.}(1984){Bond}, {Arnett}, \& {Carr}}]{Bond1984}
{Bond}, J.~R., {Arnett}, W.~D., \& {Carr}, B.~J. 1984, \apj, 280, 825,
  \dodoi{10.1086/162057}

\bibitem[{{Breivik} {et~al.}(2020){Breivik}, {Coughlin}, {Zevin}, {Rodriguez},
  {Kremer}, {Ye}, {Andrews}, {Kurkowski}, {Digman}, {Larson}, \&
  {Rasio}}]{Breivik19}
{Breivik}, K., {Coughlin}, S., {Zevin}, M., {et~al.} 2020, \apj, 898, 71,
  \dodoi{10.3847/1538-4357/ab9d85}

\bibitem[{{Bromm} \& {Larson}(2004)}]{BrommLarson2004}
{Bromm}, V., \& {Larson}, R.~B. 2004, \araa, 42, 79,
  \dodoi{10.1146/annurev.astro.42.053102.134034}

\bibitem[{{Campanelli} {et~al.}(2007){Campanelli}, {Lousto}, {Zlochower}, \&
  {Merritt}}]{Campanelli2007}
{Campanelli}, M., {Lousto}, C., {Zlochower}, Y., \& {Merritt}, D. 2007, \apjl,
  659, L5, \dodoi{10.1086/516712}

\bibitem[{{Carr} {et~al.}(2016){Carr}, {K{\"u}hnel}, \& {Sandstad}}]{Carr2016}
{Carr}, B., {K{\"u}hnel}, F., \& {Sandstad}, M. 2016, \prd, 94, 083504,
  \dodoi{10.1103/PhysRevD.94.083504}

\bibitem[{Chatterjee {et~al.}(2010)Chatterjee, Fregeau, Umbreit, \&
  Rasio}]{Chatterjee_2010}
Chatterjee, S., Fregeau, J.~M., Umbreit, S., \& Rasio, F.~A. 2010, The
  Astrophysical Journal, 719, 915, \dodoi{10.1088/0004-637x/719/1/915}

\bibitem[{{Chatterjee} {et~al.}(2017){Chatterjee}, {Rodriguez}, \&
  {Rasio}}]{Chatterjee2017}
{Chatterjee}, S., {Rodriguez}, C.~L., \& {Rasio}, F.~A. 2017, \apj, 834, 68,
  \dodoi{10.3847/1538-4357/834/1/68}

\bibitem[{{Chatterjee} {et~al.}(2013){Chatterjee}, {Umbreit}, {Fregeau}, \&
  {Rasio}}]{Chatterjee2013}
{Chatterjee}, S., {Umbreit}, S., {Fregeau}, J.~M., \& {Rasio}, F.~A. 2013,
  \mnras, 429, 2881, \dodoi{10.1093/mnras/sts464}

\bibitem[{{Chen} \& {Shen}(2018)}]{Chen_tdes}
{Chen}, J.-H., \& {Shen}, R.-F. 2018, \apj, 867, 20,
  \dodoi{10.3847/1538-4357/aadfda}

\bibitem[{{Costa} {et~al.}(2022){Costa}, {Ballone}, {Mapelli}, \&
  {Bressan}}]{Costa22}
{Costa}, G., {Ballone}, A., {Mapelli}, M., \& {Bressan}, A. 2022, arXiv
  e-prints, arXiv:2204.03492.
\newblock \doarXiv{2204.03492}

\bibitem[{{Costa} {et~al.}(2020){Costa}, {Bressan}, {Mapelli}, {Marigo},
  {Iorio}, \& {Spera}}]{Costa2020}
{Costa}, G., {Bressan}, A., {Mapelli}, M., {et~al.} 2020, arXiv e-prints,
  arXiv:2010.02242.
\newblock \doarXiv{2010.02242}

\bibitem[{{Davies} {et~al.}(2011){Davies}, {Miller}, \&
  {Bellovary}}]{Davies2011}
{Davies}, M.~B., {Miller}, M.~C., \& {Bellovary}, J.~M. 2011, \apjl, 740, L42,
  \dodoi{10.1088/2041-8205/740/2/L42}

\bibitem[{{Decarli} {et~al.}(2018){Decarli}, {Walter}, {Venemans},
  {Ba{\~n}ados}, {Bertoldi}, {Carilli}, {Fan}, {Farina}, {Mazzucchelli},
  {Riechers}, {Rix}, {Strauss}, {Wang}, \& {Yang}}]{Decarli2018}
{Decarli}, R., {Walter}, F., {Venemans}, B.~P., {et~al.} 2018, \apj, 854, 97,
  \dodoi{10.3847/1538-4357/aaa5aa}

\bibitem[{{Di Carlo} {et~al.}(2019){Di Carlo}, {Giacobbo}, {Mapelli},
  {Pasquato}, {Spera}, {Wang}, \& {Haardt}}]{DiCarlo19}
{Di Carlo}, U.~N., {Giacobbo}, N., {Mapelli}, M., {et~al.} 2019, \mnras, 487,
  2947, \dodoi{10.1093/mnras/stz1453}

\bibitem[{{Di Carlo} {et~al.}(2020){Di Carlo}, {Mapelli}, {Giacobbo}, {Spera},
  {Bouffanais}, {Rastello}, {Santoliquido}, {Pasquato}, {Ballone}, {Trani},
  {Torniamenti}, \& {Haardt}}]{DiCarlo20}
{Di Carlo}, U.~N., {Mapelli}, M., {Giacobbo}, N., {et~al.} 2020, \mnras, 498,
  495, \dodoi{10.1093/mnras/staa2286}

\bibitem[{{Di Carlo} {et~al.}(2021){Di Carlo}, {Mapelli}, {Pasquato},
  {Rastello}, {Ballone}, {Dall'Amico}, {Giacobbo}, {Iorio}, {Spera},
  {Torniamenti}, \& {Haardt}}]{DiCarlo21}
{Di Carlo}, U.~N., {Mapelli}, M., {Pasquato}, M., {et~al.} 2021, \mnras, 507,
  5132, \dodoi{10.1093/mnras/stab2390}

\bibitem[{{Duquennoy} \& {Mayor}(1991)}]{Duquennoy&Mayor}
{Duquennoy}, A., \& {Mayor}, M. 1991, \aap, 248, 485

\bibitem[{El-Badry {et~al.}(2018)El-Badry, Quataert, Weisz, Choksi, \&
  Boylan-Kolchin}]{El_Badry_2018}
El-Badry, K., Quataert, E., Weisz, D.~R., Choksi, N., \& Boylan-Kolchin, M.
  2018, Monthly Notices of the Royal Astronomical Society, 482, 4528,
  \dodoi{10.1093/mnras/sty3007}

\bibitem[{{Fan} {et~al.}(2006){Fan}, {Strauss}, {Richards}, {Hennawi},
  {Becker}, {White}, {Diamond-Stanic}, {Donley}, {Jiang}, {Kim}, {Vestergaard},
  {Young}, {Gunn}, {Lupton}, {Knapp}, {Schneider}, {Brandt}, {Bahcall},
  {Barentine}, {Brinkmann}, {Brewington}, {Fukugita}, {Harvanek}, {Kleinman},
  {Krzesinski}, {Long}, {Neilsen}, {Nitta}, {Snedden}, \& {Voges}}]{Fan2006}
{Fan}, X., {Strauss}, M.~A., {Richards}, G.~T., {et~al.} 2006, \aj, 131, 1203,
  \dodoi{10.1086/500296}

\bibitem[{Farmer {et~al.}(2020)Farmer, Renzo, de~Mink, Fishbach, \&
  Justham}]{Farmer_2020}
Farmer, R., Renzo, M., de~Mink, S.~E., Fishbach, M., \& Justham, S. 2020, The
  Astrophysical Journal, 902, L36, \dodoi{10.3847/2041-8213/abbadd}

\bibitem[{{Farmer} {et~al.}(2019){Farmer}, {Renzo}, {de Mink}, {Marchant}, \&
  {Justham}}]{Farmer2019}
{Farmer}, R., {Renzo}, M., {de Mink}, S.~E., {Marchant}, P., \& {Justham}, S.
  2019, \apj, 887, 53, \dodoi{10.3847/1538-4357/ab518b}

\bibitem[{{Fowler} \& {Hoyle}(1964)}]{Fowler1964}
{Fowler}, W.~A., \& {Hoyle}, F. 1964, \apjs, 9, 201, \dodoi{10.1086/190103}

\bibitem[{{Fragione} \& {Kocsis}(2018)}]{FragioneKocsis2018}
{Fragione}, G., \& {Kocsis}, B. 2018, \prl, 121, 161103,
  \dodoi{10.1103/PhysRevLett.121.161103}

\bibitem[{{Fragione} {et~al.}(2022){Fragione}, {Kocsis}, {Rasio}, \&
  {Silk}}]{FragioneKocsis2022}
{Fragione}, G., {Kocsis}, B., {Rasio}, F.~A., \& {Silk}, J. 2022, \apj, 927,
  231, \dodoi{10.3847/1538-4357/ac5026}

\bibitem[{{Fragione} {et~al.}(2020){Fragione}, {Loeb}, \&
  {Rasio}}]{FragioneLoeb2020}
{Fragione}, G., {Loeb}, A., \& {Rasio}, F.~A. 2020, \apjl, 902, L26,
  \dodoi{10.3847/2041-8213/abbc0a}

\bibitem[{{Fragione} \& {Silk}(2020)}]{FragioneSilk2020}
{Fragione}, G., \& {Silk}, J. 2020, \mnras, 498, 4591,
  \dodoi{10.1093/mnras/staa2629}

\bibitem[{{Fregeau} {et~al.}(2009){Fregeau}, {Ivanova}, \&
  {Rasio}}]{Fregeau2009}
{Fregeau}, J.~M., {Ivanova}, N., \& {Rasio}, F.~A. 2009, \apj, 707, 1533,
  \dodoi{10.1088/0004-637X/707/2/1533}

\bibitem[{Fregeau \& Rasio(2007)}]{Fregeau_2007}
Fregeau, J.~M., \& Rasio, F.~A. 2007, The Astrophysical Journal, 658, 1047,
  \dodoi{10.1086/511809}

\bibitem[{{Fuller} \& {Ma}(2019)}]{Fuller&Ma}
{Fuller}, J., \& {Ma}, L. 2019, \apjl, 881, L1,
  \dodoi{10.3847/2041-8213/ab339b}

\bibitem[{{Fuller} {et~al.}(2019){Fuller}, {Piro}, \& {Jermyn}}]{Fuller19}
{Fuller}, J., {Piro}, A.~L., \& {Jermyn}, A.~S. 2019, \mnras, 485, 3661,
  \dodoi{10.1093/mnras/stz514}

\bibitem[{{Gaburov} {et~al.}(2010){Gaburov}, {Lombardi}, \& {Portegies
  Zwart}}]{Gaburov10}
{Gaburov}, E., {Lombardi}, James~C., J., \& {Portegies Zwart}, S. 2010, \mnras,
  402, 105, \dodoi{10.1111/j.1365-2966.2009.15900.x}

\bibitem[{{Gerosa} \& {Berti}(2019)}]{GerosaBerti2019}
{Gerosa}, D., \& {Berti}, E. 2019, \prd, 100, 041301,
  \dodoi{10.1103/PhysRevD.100.041301}

\bibitem[{{Giacobbo} \& {Mapelli}(2018)}]{Giacobbo18}
{Giacobbo}, N., \& {Mapelli}, M. 2018, \mnras, 480, 2011,
  \dodoi{10.1093/mnras/sty1999}

\bibitem[{{Giersz} {et~al.}(2015){Giersz}, {Leigh}, {Hypki}, {L{\"u}tzgendorf},
  \& {Askar}}]{Giersz2015}
{Giersz}, M., {Leigh}, N., {Hypki}, A., {L{\"u}tzgendorf}, N., \& {Askar}, A.
  2015, \mnras, 454, 3150, \dodoi{10.1093/mnras/stv2162}

\bibitem[{{Giersz} {et~al.}(2014){Giersz}, {Leigh}, {Marks}, {Hypki}, \&
  {Askar}}]{Giersz14}
{Giersz}, M., {Leigh}, N., {Marks}, M., {Hypki}, A., \& {Askar}, A. 2014, arXiv
  e-prints, arXiv:1411.7603.
\newblock \doarXiv{1411.7603}

\bibitem[{{Giesers} {et~al.}(2019){Giesers}, {Kamann}, {Dreizler}, {Husser},
  {Askar}, {G{\"o}ttgens}, {Brinchmann}, {Latour}, {Weilbacher}, {Wendt}, \&
  {Roth}}]{Giesers19}
{Giesers}, B., {Kamann}, S., {Dreizler}, S., {et~al.} 2019, \aap, 632, A3,
  \dodoi{10.1051/0004-6361/201936203}

\bibitem[{Gonz\'alez {et~al.}(2021)Gonz\'alez, Kremer, Chatterjee, Fragione,
  Rodriguez, Weatherford, Ye, \& Rasio}]{Gonzalez_2021}
Gonz\'alez, E., Kremer, K., Chatterjee, S., {et~al.} 2021, The Astrophysical
  Journal, 908, L29, \dodoi{10.3847/2041-8213/abdf5b}

\bibitem[{Gürkan {et~al.}(2006)Gürkan, Fregeau, \& Rasio}]{Gurkan}
Gürkan, M.~A., Fregeau, J.~M., \& Rasio, F.~A. 2006, The Astrophysical
  Journal, 640, L39–L42, \dodoi{10.1086/503295}

\bibitem[{{Heger} \& {Woosley}(2002)}]{HegerWoosley2002}
{Heger}, A., \& {Woosley}, S.~E. 2002, \apj, 567, 532, \dodoi{10.1086/338487}

\bibitem[{{Heggie} \& {Hut}(2003)}]{HeggieHut2003}
{Heggie}, D., \& {Hut}, P. 2003, {The Gravitational Million-Body Problem: A
  Multidisciplinary Approach to Star Cluster Dynamics}

\bibitem[{{Heggie}(1975)}]{Heggie1975}
{Heggie}, D.~C. 1975, \mnras, 173, 729, \dodoi{10.1093/mnras/173.3.729}

\bibitem[{{Hirano} \& {Bromm}(2017)}]{Hirano2017}
{Hirano}, S., \& {Bromm}, V. 2017, \mnras, 470, 898,
  \dodoi{10.1093/mnras/stx1220}

\bibitem[{{Hopkins} \& {Beacom}(2006)}]{HopkinsBeacom2006}
{Hopkins}, A.~M., \& {Beacom}, J.~F. 2006, \apj, 651, 142,
  \dodoi{10.1086/506610}

\bibitem[{{Inayoshi} {et~al.}(2020){Inayoshi}, {Visbal}, \&
  {Haiman}}]{Inayoshi2020}
{Inayoshi}, K., {Visbal}, E., \& {Haiman}, Z. 2020, \araa, 58, 27,
  \dodoi{10.1146/annurev-astro-120419-014455}

\bibitem[{Ivanova {et~al.}(2005)Ivanova, Belczynski, Fregeau, \&
  Rasio}]{Ivanova_2005}
Ivanova, N., Belczynski, K., Fregeau, J.~M., \& Rasio, F.~A. 2005, Monthly
  Notices of the Royal Astronomical Society, 358, 572,
  \dodoi{10.1111/j.1365-2966.2005.08804.x}

\bibitem[{{Joshi} {et~al.}(2000){Joshi}, {Rasio}, \& {Portegies
  Zwart}}]{Joshi2000}
{Joshi}, K.~J., {Rasio}, F.~A., \& {Portegies Zwart}, S. 2000, \apj, 540, 969,
  \dodoi{10.1086/309350}

\bibitem[{{Kamann} {et~al.}(2021){Kamann}, {Bastian}, {Usher}, {Cabrera-Ziri},
  \& {Saracino}}]{Kamann21}
{Kamann}, S., {Bastian}, N., {Usher}, C., {Cabrera-Ziri}, I., \& {Saracino}, S.
  2021, \mnras, 508, 2302, \dodoi{10.1093/mnras/stab2643}

\bibitem[{{Katz} {et~al.}(2015){Katz}, {Sijacki}, \& {Haehnelt}}]{Katz2015}
{Katz}, H., {Sijacki}, D., \& {Haehnelt}, M.~G. 2015, \mnras, 451, 2352,
  \dodoi{10.1093/mnras/stv1048}

\bibitem[{{Kimura} {et~al.}(2021){Kimura}, {Hosokawa}, \&
  {Sugimura}}]{Kimura2021}
{Kimura}, K., {Hosokawa}, T., \& {Sugimura}, K. 2021, \apj, 911, 52,
  \dodoi{10.3847/1538-4357/abe866}

\bibitem[{{Kremer} {et~al.}(2022){Kremer}, {Lombardi}, {Lu}, {Piro}, \&
  {Rasio}}]{Kremer22}
{Kremer}, K., {Lombardi}, James~C., J., {Lu}, W., {Piro}, A.~L., \& {Rasio},
  F.~A. 2022, arXiv e-prints, arXiv:2201.12368.
\newblock \doarXiv{2201.12368}

\bibitem[{{Kremer} {et~al.}(2020{\natexlab{a}}){Kremer}, {Spera}, {Becker},
  {Chatterjee}, {Di Carlo}, {Fragione}, {Rodriguez}, {Ye}, \&
  {Rasio}}]{Kremer20}
{Kremer}, K., {Spera}, M., {Becker}, D., {et~al.} 2020{\natexlab{a}}, arXiv
  e-prints, arXiv:2006.10771.
\newblock \doarXiv{2006.10771}

\bibitem[{{Kremer} {et~al.}(2020{\natexlab{b}}){Kremer}, {Ye}, {Rui},
  {Weatherford}, {Chatterjee}, {Fragione}, {Rodriguez}, {Spera}, \&
  {Rasio}}]{Kremer20CMC}
{Kremer}, K., {Ye}, C.~S., {Rui}, N.~Z., {et~al.} 2020{\natexlab{b}}, \apjs,
  247, 48, \dodoi{10.3847/1538-4365/ab7919}

\bibitem[{{Kroupa}(2001)}]{Kroupa2001}
{Kroupa}, P. 2001, \mnras, 322, 231, \dodoi{10.1046/j.1365-8711.2001.04022.x}

\bibitem[{{Kroupa} {et~al.}(2020){Kroupa}, {Subr}, {Jerabkova}, \&
  {Wang}}]{Kroupa2020}
{Kroupa}, P., {Subr}, L., {Jerabkova}, T., \& {Wang}, L. 2020, \mnras, 498,
  5652, \dodoi{10.1093/mnras/staa2276}

\bibitem[{{Lada} \& {Lada}(2003)}]{LadaLada2003}
{Lada}, C.~J., \& {Lada}, E.~A. 2003, \araa, 41, 57,
  \dodoi{10.1146/annurev.astro.41.011802.094844}

\bibitem[{{Limongi} \& {Chieffi}(2018)}]{limongi2018}
{Limongi}, M., \& {Chieffi}, A. 2018, \apjs, 237, 13,
  \dodoi{10.3847/1538-4365/aacb24}

\bibitem[{{Loeb} \& {Rasio}(1994)}]{LoebRasio1994}
{Loeb}, A., \& {Rasio}, F.~A. 1994, \apj, 432, 52, \dodoi{10.1086/174548}

\bibitem[{Lousto \& Zlochower(2008)}]{Lousto2008}
Lousto, C.~O., \& Zlochower, Y. 2008, \prd, 77, 044028,
  \dodoi{10.1103/PhysRevD.77.044028}

\bibitem[{{Ma} {et~al.}(2021){Ma}, {Hopkins}, {Ma}, {Angl{\'e}s-Alc{\'a}zar},
  {Faucher-Gigu{\`e}re}, \& {Kelley}}]{Ma21}
{Ma}, L., {Hopkins}, P.~F., {Ma}, X., {et~al.} 2021, \mnras, 508, 1973,
  \dodoi{10.1093/mnras/stab2713}

\bibitem[{Madau \& Rees(2001)}]{Madau_2001}
Madau, P., \& Rees, M.~J. 2001, The Astrophysical Journal, 551, L27,
  \dodoi{10.1086/319848}

\bibitem[{{Mahapatra} {et~al.}(2021){Mahapatra}, {Gupta}, {Favata}, {Arun}, \&
  {Sathyaprakash}}]{Mahapatra2021}
{Mahapatra}, P., {Gupta}, A., {Favata}, M., {Arun}, K.~G., \& {Sathyaprakash},
  B.~S. 2021, \apjl, 918, L31, \dodoi{10.3847/2041-8213/ac20db}

\bibitem[{{Maliszewski} {et~al.}(2021){Maliszewski}, {Giersz},
  {Gondek-Rosi{\'n}ska}, {Askar}, \& {Hypki}}]{Maliszewski_mocca}
{Maliszewski}, K., {Giersz}, M., {Gondek-Rosi{\'n}ska}, D., {Askar}, A., \&
  {Hypki}, A. 2021, arXiv e-prints, arXiv:2111.09223.
\newblock \doarXiv{2111.09223}

\bibitem[{{Mapelli}(2016)}]{Mapelli2016}
{Mapelli}, M. 2016, \mnras, 459, 3432, \dodoi{10.1093/mnras/stw869}

\bibitem[{{Mapelli} {et~al.}(2022){Mapelli}, {Bouffanais}, {Santoliquido},
  {Arca Sedda}, \& {Artale}}]{Mapelli22}
{Mapelli}, M., {Bouffanais}, Y., {Santoliquido}, F., {Arca Sedda}, M., \&
  {Artale}, M.~C. 2022, \mnras, 511, 5797, \dodoi{10.1093/mnras/stac422}

\bibitem[{{Mapelli} {et~al.}(2020){Mapelli}, {Spera}, {Montanari}, {Limongi},
  {Chieffi}, {Giacobbo}, {Bressan}, \& {Bouffanais}}]{Mapelli2020}
{Mapelli}, M., {Spera}, M., {Montanari}, E., {et~al.} 2020, \apj, 888, 76,
  \dodoi{10.3847/1538-4357/ab584d}

\bibitem[{{Marchant} {et~al.}(2019){Marchant}, {Renzo}, {Farmer}, {Pappas},
  {Taam}, {de Mink}, \& {Kalogera}}]{Marchant2019}
{Marchant}, P., {Renzo}, M., {Farmer}, R., {et~al.} 2019, \apj, 882, 36,
  \dodoi{10.3847/1538-4357/ab3426}

\bibitem[{{Mayer} {et~al.}(2010){Mayer}, {Kazantzidis}, {Escala}, \&
  {Callegari}}]{Mayer2010}
{Mayer}, L., {Kazantzidis}, S., {Escala}, A., \& {Callegari}, S. 2010, \nat,
  466, 1082, \dodoi{10.1038/nature09294}

\bibitem[{{McKernan} {et~al.}(2012){McKernan}, {Ford}, {Lyra}, \&
  {Perets}}]{McKernan2012}
{McKernan}, B., {Ford}, K.~E.~S., {Lyra}, W., \& {Perets}, H.~B. 2012, \mnras,
  425, 460, \dodoi{10.1111/j.1365-2966.2012.21486.x}

\bibitem[{{Merritt} {et~al.}(2004){Merritt}, {Milosavljevi{\'c}}, {Favata},
  {Hughes}, \& {Holz}}]{Merritt2004}
{Merritt}, D., {Milosavljevi{\'c}}, M., {Favata}, M., {Hughes}, S.~A., \&
  {Holz}, D.~E. 2004, \apjl, 607, L9, \dodoi{10.1086/421551}

\bibitem[{{Miller} \& {Hamilton}(2002)}]{MillerHamilton2002}
{Miller}, M.~C., \& {Hamilton}, D.~P. 2002, \mnras, 330, 232,
  \dodoi{10.1046/j.1365-8711.2002.05112.x}

\bibitem[{{Milone} {et~al.}(2012){Milone}, {Piotto}, {Bedin}, {Aparicio},
  {Anderson}, {Sarajedini}, {Marino}, {Moretti}, {Davies}, {Chaboyer},
  {Dotter}, {Hempel}, {Mar{\'\i}n-Franch}, {Majewski}, {Paust}, {Reid},
  {Rosenberg}, \& {Siegel}}]{Milone2012}
{Milone}, A.~P., {Piotto}, G., {Bedin}, L.~R., {et~al.} 2012, \aap, 540, A16,
  \dodoi{10.1051/0004-6361/201016384}

\bibitem[{{Moe} \& {Di Stefano}(2017)}]{MoeDiStefano2017}
{Moe}, M., \& {Di Stefano}, R. 2017, \apjs, 230, 15,
  \dodoi{10.3847/1538-4365/aa6fb6}

\bibitem[{{Ober} {et~al.}(1983){Ober}, {El Eid}, \& {Fricke}}]{Ober1983}
{Ober}, W.~W., {El Eid}, M.~F., \& {Fricke}, K.~J. 1983, \aap, 119, 61

\bibitem[{{Oh} \& {Haiman}(2002)}]{Oh&Haiman}
{Oh}, S.~P., \& {Haiman}, Z. 2002, \apj, 569, 558, \dodoi{10.1086/339393}

\bibitem[{{O'Leary} {et~al.}(2009){O'Leary}, {Kocsis}, \& {Loeb}}]{O'Leary}
{O'Leary}, R.~M., {Kocsis}, B., \& {Loeb}, A. 2009, \mnras, 395, 2127,
  \dodoi{10.1111/j.1365-2966.2009.14653.x}

\bibitem[{{Pattabiraman} {et~al.}(2013){Pattabiraman}, {Umbreit}, {Liao},
  {Choudhary}, {Kalogera}, {Memik}, \& {Rasio}}]{Pattabiraman2013}
{Pattabiraman}, B., {Umbreit}, S., {Liao}, W.-k., {et~al.} 2013, \apjs, 204,
  15, \dodoi{10.1088/0067-0049/204/2/15}

\bibitem[{Peters(1964)}]{Peters}
Peters, P.~C. 1964, Phys. Rev., 136, B1224, \dodoi{10.1103/PhysRev.136.B1224}

\bibitem[{Pfister {et~al.}(2019)Pfister, Volonteri, Dubois, Dotti, \&
  Colpi}]{Pfister19}
Pfister, H., Volonteri, M., Dubois, Y., Dotti, M., \& Colpi, M. 2019, Monthly
  Notices of the Royal Astronomical Society, 486, 101,
  \dodoi{10.1093/mnras/stz822}

\bibitem[{{Portegies Zwart} {et~al.}(2004){Portegies Zwart}, {Baumgardt},
  {Hut}, {Makino}, \& {McMillan}}]{Zwart}
{Portegies Zwart}, S.~F., {Baumgardt}, H., {Hut}, P., {Makino}, J., \&
  {McMillan}, S. L.~W. 2004, \nat, 428, 724, \dodoi{10.1038/nature02448}

\bibitem[{{Portegies Zwart} \& {McMillan}(2002)}]{PortegiesZwartMcMillan2002}
{Portegies Zwart}, S.~F., \& {McMillan}, S. L.~W. 2002, \apj, 576, 899,
  \dodoi{10.1086/341798}

\bibitem[{{Ramirez-Ruiz} \& {Rosswog}(2009)}]{Ramirez_tdes}
{Ramirez-Ruiz}, E., \& {Rosswog}, S. 2009, \apjl, 697, L77,
  \dodoi{10.1088/0004-637X/697/2/L77}

\bibitem[{Renzo {et~al.}(2020)Renzo, Farmer, Justham, de Mink, Götberg, \&
  Marchant}]{Renzo2020}
Renzo, M., Farmer, R.~J., Justham, S., {et~al.} 2020, \mnras, 493, 4333,
  \dodoi{10.1093/mnras/staa549}

\bibitem[{{Rizzuto} {et~al.}(2022){Rizzuto}, {Naab}, {Spurzem}, {Arca-Sedda},
  {Giersz}, {Ostriker}, \& {Banerjee}}]{Rizzuto2022}
{Rizzuto}, F.~P., {Naab}, T., {Spurzem}, R., {et~al.} 2022, \mnras, 512, 884,
  \dodoi{10.1093/mnras/stac231}

\bibitem[{{Rizzuto} {et~al.}(2020){Rizzuto}, {Naab}, {Spurzem}, {Giersz},
  {Ostriker}, {Stone}, {Wang}, {Berczik}, \& {Rampp}}]{Rizzuto}
---. 2020, arXiv e-prints, arXiv:2008.09571.
\newblock \doarXiv{2008.09571}

\bibitem[{{Rodriguez} {et~al.}(2018){Rodriguez}, {Amaro-Seoane}, {Chatterjee},
  {Kremer}, {Rasio}, {Samsing}, {Ye}, \& {Zevin}}]{Rodriguez2018b}
{Rodriguez}, C.~L., {Amaro-Seoane}, P., {Chatterjee}, S., {et~al.} 2018, \prd,
  98, 123005, \dodoi{10.1103/PhysRevD.98.123005}

\bibitem[{Rodriguez {et~al.}(2016)Rodriguez, Chatterjee, \&
  Rasio}]{Rodriguez16}
Rodriguez, C.~L., Chatterjee, S., \& Rasio, F.~A. 2016, Phys. Rev. D, 93,
  084029, \dodoi{10.1103/PhysRevD.93.084029}

\bibitem[{Rodriguez \& Loeb(2018)}]{Rodriguez_2018}
Rodriguez, C.~L., \& Loeb, A. 2018, The Astrophysical Journal, 866, L5,
  \dodoi{10.3847/2041-8213/aae377}

\bibitem[{Rodriguez {et~al.}(2015)Rodriguez, Morscher, Pattabiraman,
  Chatterjee, Haster, \& Rasio}]{Rodriguez15}
Rodriguez, C.~L., Morscher, M., Pattabiraman, B., {et~al.} 2015, Phys. Rev.
  Lett., 115, 051101, \dodoi{10.1103/PhysRevLett.115.051101}

\bibitem[{{Rodriguez} {et~al.}(2019){Rodriguez}, {Zevin}, {Amaro-Seoane},
  {Chatterjee}, {Kremer}, {Rasio}, \& {Ye}}]{Rodriguez2019}
{Rodriguez}, C.~L., {Zevin}, M., {Amaro-Seoane}, P., {et~al.} 2019, \prd, 100,
  043027, \dodoi{10.1103/PhysRevD.100.043027}

\bibitem[{Rodriguez {et~al.}(2022)Rodriguez, Weatherford, Coughlin,
  Amaro-Seoane, Breivik, Chatterjee, Fragione, K{\i}ro{\u{g} }lu, Kremer, Rui,
  Ye, Zevin, \& Rasio}]{Rodriguez_2022}
Rodriguez, C.~L., Weatherford, N.~C., Coughlin, S.~C., {et~al.} 2022, The
  Astrophysical Journal Supplement Series, 258, 22,
  \dodoi{10.3847/1538-4365/ac2edf}

\bibitem[{{Rose} {et~al.}(2022){Rose}, {Naoz}, {Sari}, \& {Linial}}]{Rose2022}
{Rose}, S.~C., {Naoz}, S., {Sari}, R., \& {Linial}, I. 2022, \apjl, 929, L22,
  \dodoi{10.3847/2041-8213/ac6426}

\bibitem[{{Roupas} \& {Kazanas}(2019)}]{Roupas2019}
{Roupas}, Z., \& {Kazanas}, D. 2019, \aap, 632, L8,
  \dodoi{10.1051/0004-6361/201937002}

\bibitem[{{Rozner} \& {Perets}(2022)}]{Rozner2022}
{Rozner}, M., \& {Perets}, H.~B. 2022, \apj, 931, 149,
  \dodoi{10.3847/1538-4357/ac6d55}

\bibitem[{{Sana} {et~al.}(2009){Sana}, {Gosset}, \& {Evans}}]{Sana2009}
{Sana}, H., {Gosset}, E., \& {Evans}, C.~J. 2009, \mnras, 400, 1479,
  \dodoi{10.1111/j.1365-2966.2009.15545.x}

\bibitem[{{Sana} {et~al.}(2012){Sana}, {de Mink}, {de Koter}, {Langer},
  {Evans}, {Gieles}, {Gosset}, {Izzard}, {Le Bouquin}, \&
  {Schneider}}]{Sana2012}
{Sana}, H., {de Mink}, S.~E., {de Koter}, A., {et~al.} 2012, Science, 337, 444,
  \dodoi{10.1126/science.1223344}

\bibitem[{{Shrivastava} \& {Kremer}(2022)}]{ShrivastavaKremer2022}
{Shrivastava}, R., \& {Kremer}, K. 2022, Research Notes of the American
  Astronomical Society, 6, 157, \dodoi{10.3847/2515-5172/ac87b1}

\bibitem[{{Spera} \& {Mapelli}(2017)}]{Spera17}
{Spera}, M., \& {Mapelli}, M. 2017, \mnras, 470, 4739,
  \dodoi{10.1093/mnras/stx1576}

\bibitem[{Spera {et~al.}(2019)Spera, Mapelli, Giacobbo, Trani, Bressan, \&
  Costa}]{Spera19}
Spera, M., Mapelli, M., Giacobbo, N., {et~al.} 2019, Monthly Notices of the
  Royal Astronomical Society, 485, 889, \dodoi{10.1093/mnras/stz359}

\bibitem[{{Stacy} {et~al.}(2012){Stacy}, {Greif}, \& {Bromm}}]{Stacy2012}
{Stacy}, A., {Greif}, T.~H., \& {Bromm}, V. 2012, \mnras, 422, 290,
  \dodoi{10.1111/j.1365-2966.2012.20605.x}

\bibitem[{{Stevenson} {et~al.}(2019){Stevenson}, {Sampson}, {Powell},
  {Vigna-G{\'o}mez}, {Neijssel}, {Sz{\'e}csi}, \& {Mandel}}]{Stevenson2019}
{Stevenson}, S., {Sampson}, M., {Powell}, J., {et~al.} 2019, \apj, 882, 121,
  \dodoi{10.3847/1538-4357/ab3981}

\bibitem[{Tagawa {et~al.}(2020)Tagawa, Haiman, \& Kocsis}]{Tagawa_2020}
Tagawa, H., Haiman, Z., \& Kocsis, B. 2020, The Astrophysical Journal, 892, 36,
  \dodoi{10.3847/1538-4357/ab7922}

\bibitem[{{Takahashi} {et~al.}(2018){Takahashi}, {Yoshida}, \&
  {Umeda}}]{Takahashi2018}
{Takahashi}, K., {Yoshida}, T., \& {Umeda}, H. 2018, \apj, 857, 111,
  \dodoi{10.3847/1538-4357/aab95f}

\bibitem[{{The LIGO Scientific Collaboration} {et~al.}(2020{\natexlab{a}}){The
  LIGO Scientific Collaboration}, {the Virgo Collaboration}, {Abbott},
  {Abbott}, {Abraham}, {Acernese}, {Ackley}, {Adams}, {Adams}, {Adhikari},
  {Adya}, {Affeldt}, {Agathos}, {Agatsuma}, {Aggarwal}, {Aguiar}, {Aiello},
  {Ain}, {Ajith}, {Allen}, {Allocca}, {Altin}, {Amato}, {Anand}, {Ananyeva},
  {Anderson}, {Anderson}, {Angelova}, {Ansoldi}, {Antelis}, {Antier}, {Appert},
  {Arai}, {Araya}, {Areeda}, {Ar{\`e}ne}, {Arnaud}, {Aronson}, {Arun}, {Asali},
  {Ascenzi}, {Ashton}, {Aston}, {Astone}, {Aubin}, {Aufmuth}, {AultONeal},
  {Austin}, {Avendano}, {Babak}, {Badaracco}, {Bader}, {Bae}, {Baer},
  {Bagnasco}, {Baird}, {Ball}, {Ballardin}, {Ballmer}, {Bals}, {Balsamo},
  {Baltus}, {Banagiri}, {Bankar}, {Bankar}, {Barayoga}, {Barbieri}, {Barish},
  {Barker}, {Barneo}, {Barnum}, {Barone}, {Barr}, {Barsotti}, {Barsuglia},
  {Barta}, {Bartlett}, {Bartos}, {Bassiri}, {Basti}, {Bawaj}, {Bayley},
  {Bazzan}, {Becher}, {B{\'e}csy}, {Bedakihale}, {Bejger}, {Belahcene},
  {Beniwal}, {Benjamin}, {Benkel}, {Bennett}, {Bentley}, {Bergamin}, {Berger},
  {Bergmann}, {Bernuzzi}, {Berry}, {Bersanetti}, {Bertolini}, {Betzwieser},
  {Bhandare}, {Bhandari}, {Bhattacharjee}, {Bidler}, {Bilenko}, {Billingsley},
  {Birney}, {Birnholtz}, {Biscans}, {Bischi}, {Biscoveanu}, {Bisht}, {Bitossi},
  {Bizouard}, {Blackburn}, {Blackman}, {Blair}, {Blair}, {Blair}, {Blanch},
  {Bobba}, {Bode}, {Boer}, {Boetzel}, {Bogaert}, {Boldrini}, {Bondu},
  {Bonilla}, {Bonnand}, {Booker}, {Boom}, {Borhanian}, {Bork}, {Boschi},
  {Bose}, {Bose}, {Bossilkov}, {Boudart}, {Bouffanais}, {Bozzi}, {Bradaschia},
  {Brady}, {Bramley}, {Branchesi}, {Brau}, {Breschi}, {Briant}, {Briggs},
  {Brighenti}, {Brillet}, {Brinkmann}, {Brockill}, {Brooks}, {Brooks}, {Brown},
  {Brunett}, {Bruno}, {Bruntz}, {Buikema}, {Bulik}, {Bulten}, {Buonanno},
  {Buskulic}, {Byer}, {Cabero}, {Cadonati}, {Caesar}, {Cagnoli}, {Cahillane},
  {Calder{\'o}n Bustillo}, {Callaghan}, {Callister}, {Calloni}, {Camp},
  {Canepa}, {Cannon}, {Cao}, {Cao}, {Carapella}, {Carbognani}, {Carney},
  {Carpinelli}, {Carullo}, {Carver}, {Casanueva Diaz}, {Casentini}, {Caudill},
  {Cavagli{\`a}}, {Cavalier}, {Cavalieri}, {Cella}, {Cerd{\'a}-Dur{\'a}n},
  {Cesarini}, {Chaibi}, {Chakravarti}, {Chan}, {Chan}, {Chandra}, {Chanial},
  {Chao}, {Charlton}, {Chase}, {Chassande-Mottin}, {Chatterjee}, {Chaturvedi},
  {Chatziioannou}, {Chen}, {Chen}, {Chen}, {Chen}, {Cheng}, {Cheong}, {Chia},
  {Chiadini}, {Chierici}, {Chincarini}, {Chiummo}, {Cho}, {Cho}, {Cho},
  {Choate}, {Christensen}, {Chu}, {Chua}, {Chung}, {Chung}, {Ciani},
  {Ciecielag}, {Cie{\'s}lar}, {Cifaldi}, {Ciobanu}, {Ciolfi}, {Cipriano},
  {Cirone}, {Clara}, {Clark}, {Clark}, {Clarke}, {Clearwater}, {Clesse},
  {Cleva}, {Coccia}, {Cohadon}, {Cohen}, {Colleoni}, {Collette}, {Collins},
  {Colpi}, {Constancio}, {Conti}, {Cooper}, {Corban}, {Corbitt},
  {Cordero-Carri{\'o}n}, {Corezzi}, {Corley}, {Cornish}, {Corre}, {Corsi},
  {Cortese}, {Costa}, {Cotesta}, {Coughlin}, {Coughlin}, {Coulon},
  {Countryman}, {Couvares}, {Covas}, {Coward}, {Cowart}, {Coyne}, {Coyne},
  {Creighton}, {Creighton}, {Croquette}, {Crowder}, {Cudell}, {Cullen},
  {Cumming}, {Cummings}, {Cunningham}, {Cuoco}, {Cury\{l\}o}, {Dal Canton},
  {D{\'a}lya}, {Dana}, {DaneshgaranBajastani}, {D'Angelo}, {Danilishin},
  {D'Antonio}, {Danzmann}, {Darsow-Fromm}, {Dasgupta}, {Datrier}, {Dattilo},
  {Dave}, {Davier}, {Davies}, {Davis}, {Daw}, {Dean}, {DeBra}, {Deenadayalan},
  {Degallaix}, {De Laurentis}, {Del{\'e}glise}, {Del Favero}, {De Lillo}, {De
  Lillo}, {Del Pozzo}, {DeMarchi}, {De Matteis}, {D'Emilio}, {Demos}, {Denker},
  {Dent}, {Depasse}, {De Pietri}, {De Rosa}, {De Rossi}, {DeSalvo}, {de
  Varona}, {Dhani}, {Dhurandhar}, {D{\'\i}az}, {Diaz-Ortiz}, {Didio},
  {Dietrich}, {Di Fiore}, {DiFronzo}, {Di Giorgio}, {Di Giovanni}, {Di
  Giovanni}, {Di Girolamo}, {Di Lieto}, {Ding}, {Di Pace}, {Di Palma}, {Di
  Renzo}, {Divakarla}, {Dmitriev}, {Doctor}, {D'Onofrio}, {Donovan}, {Dooley},
  {Doravari}, {Dorrington}, {Downes}, {Drago}, {Driggers}, {Du}, {Ducoin},
  {Dudi}, {Dupej}, {Durante}, {D'Urso}, {Duverne}, {Dwyer}, {Easter},
  {Eddolls}, {Edelman}, {Edo}, {Edy}, {Effler}, {Eichholz}, {Eikenberry},
  {Eisenmann}, {Eisenstein}, {Ejlli}, {Errico}, {Essick}, {Estell{\'e}s},
  {Estevez}, {Etienne}, {Etzel}, {Evans}, {Evans}, {Ewing}, {Fafone}, {Fair},
  {Fairhurst}, {Fan}, {Farah}, {Farinon}, {Farr}, {Farr}, {Fauchon-Jones},
  {Favata}, {Fays}, {Fazio}, {Feicht}, {Fejer}, {Feng}, {Fenyvesi}, {Ferguson},
  {Fernandez-Galiana}, {Ferrante}, {Ferreira}, {Fidecaro}, {Figura}, {Fiori},
  {Fiorucci}, {Fishbach}, {Fisher}, {Fishner}, {Fittipaldi}, {Fitz-Axen},
  {Fiumara}, {Flaminio}, {Floden}, {Flynn}, {Fong}, {Font}, {Forsyth},
  {Fournier}, {Frasca}, {Frasconi}, {Frei}, {Freise}, {Frey}, {Frey},
  {Fritschel}, {Frolov}, {Fronz{\'e}}, {Fulda}, {Fyffe}, {Gabbard}, {Gadre},
  {Gaebel}, {Gair}, {Gais}, {Galaudage}, {Gamba}, {Ganapathy}, {Ganguly},
  {Gaonkar}, {Garaventa}, {Garc{\'\i}a-Quir{\'o}s}, {Garufi}, {Gateley},
  {Gaudio}, {Gayathri}, {Gemme}, {Gennai}, {George}, {George}, {George},
  {Gergely}, {Ghonge}, {Ghosh}, {Ghosh}, {Ghosh}, {Giacomazzo}, {Giacoppo},
  {Giaime}, {Giardina}, {Gibson}, {Gier}, {Gill}, {Giri}, {Glanzer}, {Gleckl},
  {Godwin}, {Goetz}, {Goetz}, {Gohlke}, {Goncharov}, {Gonz{\'a}lez},
  {Gopakumar}, {Gossan}, {Gosselin}, {Gouaty}, {Grace}, {Grado}, {Granata},
  {Granata}, {Grant}, {Gras}, {Grassia}, {Gray}, {Gray}, {Greco}, {Green},
  {Green}, {Gretarsson}, {Griggs}, {Grignani}, {Grimaldi}, {Grimes}, {Grimm},
  {Grote}, {Grunewald}, {Gruning}, {Guerrero}, {Guidi}, {Guimaraes},
  {Guix{\'e}}, {Gulati}, {Guo}, {Gupta}, {Gupta}, {Gupta}, {Gustafson},
  {Gustafson}, {Guzman}, {Haegel}, {Halim}, {Hall}, {Hamilton}, {Hammond},
  {Haney}, {Hanke}, {Hanks}, {Hanna}, {Hannam}, {Hannuksela}, {Hannuksela},
  {Hansen}, {Hansen}, {Hanson}, {Harder}, {Hardwick}, {Haris}, {Harms},
  {Harry}, {Harry}, {Hartwig}, {Hasskew}, {Haster}, {Haughian}, {Hayes},
  {Healy}, {Heidmann}, {Heintze}, {Heinze}, {Heinzel}, {Heitmann}, {Hellman},
  {Hello}, {Helmling-Cornell}, {Hemming}, {Hendry}, {Heng}, {Hennes}, {Hennig},
  {Hennig}, {Hernandez Vivanco}, {Heurs}, {Hild}, {Hill}, {Hines}, {Hochheim},
  {Hofgard}, {Hofman}, {Hohmann}, {Holgado}, {Holland}, {Hollows}, {Holmes},
  {Holt}, {Holz}, {Hopkins}, {Horst}, {Hough}, {Howell}, {Hoy}, {Hoyland},
  {Huang}, {H{\"u}bner}, {Huddart}, {Huerta}, {Hughey}, {Hui}, {Husa},
  {Huttner}, {Hutzler}, {Huxford}, {Huynh-Dinh}, {Idzkowski}, {Iess},
  {Imperato}, {Inchauspe}, {Ingram}, {Intini}, {Isi}, {Iyer},
  {JaberianHamedan}, {Jacqmin}, {Jadhav}, {Jadhav}, {James}, {Jani},
  {Janssens}, {Janthalur}, {Jaranowski}, {Jariwala}, {Jaume}, {Jenkins},
  {Jeunon}, {Jiang}, {Johns}, {Johnson-McDaniel}, {Jones}, {Jones}, {Jones},
  {Jones}, {Jones}, {Jonker}, {Ju}, {Junker}, {Kalaghatgi}, {Kalogera},
  {Kamai}, {Kandhasamy}, {Kang}, {Kanner}, {Kapadia}, {Kapasi}, {Karathanasis},
  {Karki}, {Kashyap}, {Kasprzack}, {Kastaun}, {Katsanevas}, {Katsavounidis},
  {Katzman}, {Kawabe}, {K{\'e}f{\'e}lian}, {Keitel}, {Key}, {Khadka},
  {Khalili}, {Khan}, {Khan}, {Khazanov}, {Khetan}, {Khursheed}, {Kijbunchoo},
  {Kim}, {Kim}, {Kim}, {Kim}, {Kim}, {Kim}, {Kimball}, {King}, {Kinley-Hanlon},
  {Kirchhoff}, {Kissel}, {Kleybolte}, {Klimenko}, {Knowles}, {Knyazev}, {Koch},
  {Koehlenbeck}, {Koekoek}, {Koley}, {Kolstein}, {Komori}, {Kondrashov},
  {Kontos}, {Koper}, {Korobko}, {Korth}, {Kovalam}, {Kozak}, {Kr{\"a}mer},
  {Kringel}, {Krishnendu}, {Kr{\'o}lak}, {Kuehn}, {Kumar}, {Kumar}, {Kumar},
  {Kumar}, {Kuns}, {Kwang}, {Lackey}, {Laghi}, {Lalande}, {Lam}, {Lamberts},
  {Landry}, {Lane}, {Lang}, {Lange}, {Lantz}, {Lanza}, {La Rosa},
  {Lartaux-Vollard}, {Lasky}, {Laxen}, {Lazzarini}, {Lazzaro}, {Leaci},
  {Leavey}, {Lecoeuche}, {Lee}, {Lee}, {Lee}, {Lee}, {Lehmann}, {Leon},
  {Leroy}, {Letendre}, {Levin}, {Li}, {Li}, {Li}, {Li}, {Li}, {Linde},
  {Linker}, {Linley}, {Littenberg}, {Liu}, {Liu}, {Llorens-Monteagudo}, {Lo},
  {Lockwood}, {London}, {Longo}, {Lorenzini}, {Loriette}, {Lormand}, {Losurdo},
  {Lough}, {Lousto}, {Lovelace}, {L{\"u}ck}, {Lumaca}, {Lundgren}, {Ma},
  {Macas}, {MacInnis}, {Macleod}, {MacMillan}, {Macquet}, {Maga{\~n}a
  Hernandez}, {Maga{\~n}a-Sandoval}, {Magazz{\`u}}, {Magee}, {Majorana},
  {Maksimovic}, {Maliakal}, {Malik}, {Man}, {Mandic}, {Mangano}, {Mansell},
  {Manske}, {Mantovani}, {Mapelli}, {Marchesoni}, {Marion}, {M{\'a}rka},
  {M{\'a}rka}, {Markakis}, {Markosyan}, {Markowitz}, {Maros}, {Marquina},
  {Marsat}, {Martelli}, {Martin}, {Martin}, {Martinez}, {Martinez}, {Martynov},
  {Masalehdan}, {Mason}, {Massera}, {Masserot}, {Massinger}, {Masso-Reid},
  {Mastrogiovanni}, {Matas}, {Mateu-Lucena}, {Matichard}, {Matiushechkina},
  {Mavalvala}, {Maynard}, {McCann}, {McCarthy}, {McClelland}, {McCormick},
  {McCuller}, {McGuire}, {McIsaac}, {McIver}, {McManus}, {McRae}, {McWilliams},
  {Meacher}, {Meadors}, {Mehmet}, {Mehta}, {Melatos}, {Melchor}, {Mendell},
  {Menendez-Vazquez}, {Mercer}, {Mereni}, {Merfeld}, {Merilh}, {Merritt},
  {Merzougui}, {Meshkov}, {Messenger}, {Messick}, {Metzdorff}, {Meyers},
  {Meylahn}, {Mhaske}, {Miani}, {Miao}, {Michaloliakos}, {Michel}, {Middleton},
  {Milano}, {Miller}, {Millhouse}, {Mills}, {Milotti}, {Milovich-Goff},
  {Minazzoli}, {Minenkov}, {Mir}, {Mishkin}, {Mishra}, {Mistry}, {Mitra},
  {Mitrofanov}, {Mitselmakher}, {Mittleman}, {Mo}, {Mogushi}, {Mohapatra},
  {Mohite}, {Molina}, {Molina-Ruiz}, {Mondin}, {Montani}, {Moore}, {Moraru},
  {Morawski}, {Moreno}, {Morisaki}, {Mours}, {Mow-Lowry}, {Mozzon},
  {Muciaccia}, {Mukherjee}, {Mukherjee}, {Mukherjee}, {Mukherjee}, {Mukund},
  {Mullavey}, {Munch}, {Mu{\~n}iz}, {Murray}, {Nadji}, {Nagar}, {Nardecchia},
  {Naticchioni}, {Nayak}, {Neil}, {Neilson}, {Nelemans}, {Nelson}, {Nery},
  {Neunzert}, {Ng}, {Ng}, {Nguyen}, {Nguyen}, {Nguyen}, {Nichols}, {Nissanke},
  {Nocera}, {Noh}, {North}, {Nothard}, {Nuttall}, {Oberling}, {O'Brien},
  {O'Dell}, {Oganesyan}, {Ogin}, {Oh}, {Oh}, {Ohme}, {Ohta}, {Okada},
  {Olivetto}, {Oppermann}, {Oram}, {O'Reilly}, {Ormiston}, {Ormsby}, {Ortega},
  {O'Shaughnessy}, {Ossokine}, {Osthelder}, {Ottaway}, {Overmier}, {Owen},
  {Pace}, {Pagano}, {Page}, {Pagliaroli}, {Pai}, {Pai}, {Palamos}, {Palashov},
  {Palomba}, {Pan}, {Panda}, {Pang}, {Pankow}, {Pannarale}, {Pant}, {Paoletti},
  {Paoli}, {Paolone}, {Parker}, {Pascucci}, {Pasqualetti}, {Passaquieti},
  {Passuello}, {Patel}, {Patricelli}, {Payne}, {Pechsiri}, {Pedraza},
  {Pegoraro}, {Pele}, {Penn}, {Perego}, {Perez}, {P{\'e}rigois}, {Perreca},
  {Perri{\`e}s}, {Petermann}, {Petterson}, {Pfeiffer}, {Pham}, {Phukon},
  {Piccinni}, {Pichot}, {Piendibene}, {Piergiovanni}, {Pierini}, {Pierro},
  {Pillant}, {Pilo}, {Pinard}, {Pinto}, {Piotrzkowski}, {Pirello}, {Pitkin},
  {Placidi}, {Plastino}, {Pluchar}, {Poggiani}, {Polini}, {Pong}, {Ponrathnam},
  {Popolizio}, {Porter}, {Poverman}, {Powell}, {Pracchia}, {Prajapati},
  {Prasai}, {Prasanna}, {Pratten}, {Prestegard}, {Principe}, {Prodi},
  {Prokhorov}, {Prosposito}, {Puecher}, {Punturo}, {Puosi}, {Puppo},
  {P{\"u}rrer}, {Qi}, {Quetschke}, {Quinonez}, {Quitzow-James}, {Raab},
  {Raaijmakers}, {Radkins}, {Radulesco}, {Raffai}, {Rafferty}, {Rail}, {Raja},
  {Rajan}, {Rajbhandari}, {Rakhmanov}, {Ramirez}, {Ramirez}, {Ramos-Buades},
  {Rana}, {Rao}, {Rapagnani}, {Rapol}, {Ratto}, {Raymond}, {Razzano}, {Read},
  {Regimbau}, {Rei}, {Reid}, {Reitze}, {Rettegno}, {Ricci}, {Richardson},
  {Richardson}, {Richardson}, {Ricker}, {Riemenschneider}, {Riles}, {Rizzo},
  {Robertson}, {Robinet}, {Rocchi}, {Rocha}, {Rodriguez}, {Rodriguez-Soto},
  {Rolland}, {Rollins}, {Roma}, {Romanelli}, {Romano}, {Romel}, {Romero},
  {Romero-Shaw}, {Romie}, {Ronchini}, {Rose}, {Rose}, {Rose}, {Rosi{\'n}ska},
  {Rosofsky}, {Ross}, {Rowan}, {Rowlinson}, {Roy}, {Roy}, {Ruggi}, {Ryan},
  {Sachdev}, {Sadecki}, {Sakellariadou}, {Salafia}, {Salconi}, {Saleem},
  {Samajdar}, {Sanchez}, {Sanchez}, {Sanchez}, {Sanchis-Gual}, {Sanders},
  {Santiago}, {Santos}, {Saravanan}, {Sarin}, {Sassolas}, {Sathyaprakash},
  {Sauter}, {Savage}, {Savant}, {Sawant}, {Sayah}, {Schaetzl}, {Schale},
  {Scheel}, {Scheuer}, {Schindler-Tyka}, {Schmidt}, {Schnabel}, {Schofield},
  {Sch{\"o}nbeck}, {Schreiber}, {Schulte}, {Schutz}, {Schwarm}, {Schwartz},
  {Scott}, {Scott}, {Seglar-Arroyo}, {Seidel}, {Sellers}, {Sengupta},
  {Sennett}, {Sentenac}, {Sequino}, {Sergeev}, {Setyawati}, {Shaffer},
  {Shahriar}, {Sharifi}, {Sharma}, {Sharma}, {Shawhan}, {Shen}, {Shikauchi},
  {Shink}, {Shoemaker}, {Shoemaker}, {Shukla}, {ShyamSundar}, {Sieniawska},
  {Sigg}, {Singer}, {Singh}, {Singh}, {Singha}, {Singhal}, {Sintes}, {Sipala},
  {Skliris}, {Slagmolen}, {Slaven-Blair}, {Smetana}, {Smith}, {Smith},
  {Somala}, {Son}, {Soni}, {Sorazu}, {Sordini}, {Sorrentino}, {Sorrentino},
  {Soulard}, {Souradeep}, {Sowell}, {Spencer}, {Spera}, {Srivastava},
  {Srivastava}, {Staats}, {Stachie}, {Steer}, {Steinhoff}, {Steinke},
  {Steinlechner}, {Steinlechner}, {Steinmeyer}, {Stolle-McAllister}, {Stops},
  {Stover}, {Strain}, {Stratta}, {Strunk}, {Sturani}, {Stuver}, {S{\"u}dbeck},
  {Sudhagar}, {Sudhir}, {Suh}, {Summerscales}, {Sun}, {Sun}, {Sunil}, {Sur},
  {Suresh}, {Sutton}, {Swinkels}, {Szczepa{\'n}czyk}, {Tacca}, {Tait},
  {Talbot}, {Tanasijczuk}, {Tanner}, {Tao}, {Tapia}, {Tapia San Martin},
  {Tasson}, {Taylor}, {Tenorio}, {Terkowski}, {Thirugnanasambandam}, {Thomas},
  {Thomas}, {Thomas}, {Thompson}, {Thondapu}, {Thorne}, {Thrane}, {Tiwari},
  {Tiwari}, {Tiwari}, {Toland}, {Tolley}, {Tonelli}, {Tornasi},
  {Torres-Forn{\'e}}, {Torrie}, {Melo}, {T{\"o}yr{\"a}}, {Tran}, {Trapananti},
  {Travasso}, {Traylor}, {Tringali}, {Tripathee}, {Trovato}, {Trudeau}, {Tsai},
  {Tsang}, {Tse}, {Tso}, {Tsukada}, {Tsuna}, {Tsutsui}, {Turconi}, {Ubhi},
  {Udall}, {Ueno}, {Ugolini}, {Unnikrishnan}, {Urban}, {Usman}, {Utina},
  {Vahlbruch}, {Vajente}, {Vajpeyi}, {Valdes}, {Valentini}, {Valsan}, {van
  Bakel}, {van Beuzekom}, {van den Brand}, {Van Den Broeck}, {Vander-Hyde},
  {van der Schaaf}, {van Heijningen}, {Vardaro}, {Vargas}, {Varma}, {Vass},
  {Vas{\'u}th}, {Vecchio}, {Vedovato}, {Veitch}, {Veitch}, {Venkateswara},
  {Venneberg}, {Venugopalan}, {Verkindt}, {Verma}, {Veske}, {Vetrano},
  {Vicer{\'e}}, {Viets}, {Vijaykumar}, {Villa-Ortega}, {Vinet}, {Vitale}, {Vo},
  {Vocca}, {Vorvick}, {Vyatchanin}, {Wade}, {Wade}, {Wade}, {Wald}, {Walet},
  {Walker}, {Wallace}, {Wallace}, {Walsh}, {Wang}, {Wang}, {Wang}, {Wang},
  {Ward}, {Warner}, {Was}, {Washington}, {Watchi}, {Weaver}, {Wei}, {Weinert},
  {Weinstein}, {Weiss}, {Wellmann}, {Wen}, {We{\ss}els}, {Westhouse}, {Wette},
  {Whelan}, {White}, {White}, {Whiting}, {Whittle}, {Wilken}, {Williams},
  {Williams}, {Williamson}, {Willis}, {Willke}, {Wilson}, {Wimmer}, {Winkler},
  {Wipf}, {Woan}, {Woehler}, {Wofford}, {Wong}, {Wrangel}, {Wright}, {Wu},
  {Wysocki}, {Xiao}, {Yamamoto}, {Yang}, {Yang}, {Yang}, {Yap}, {Yeeles},
  {Yoon}, {Yu}, {Yu}, {Yuen}, {Zadro{\.z}ny}, {Zanolin}, {Zelenova}, {Zendri},
  {Zevin}, {Zhang}, {Zhang}, {Zhang}, {Zhang}, {Zhao}, {Zhao}, {Zhou}, {Zhou},
  {Zhu}, {Zimmerman}, {Zucker}, \& {Zweizig}}]{GWTC2_gr2020}
{The LIGO Scientific Collaboration}, {the Virgo Collaboration}, {Abbott}, R.,
  {et~al.} 2020{\natexlab{a}}, arXiv e-prints, arXiv:2010.14529.
\newblock \doarXiv{2010.14529}

\bibitem[{{The LIGO Scientific Collaboration} {et~al.}(2020{\natexlab{b}}){The
  LIGO Scientific Collaboration}, {the Virgo Collaboration}, {Abbott},
  {Abbott}, {Abraham}, {Acernese}, {Ackley}, {Adams}, {Adhikari}, {Adya},
  {Affeldt}, {Agathos}, {Agatsuma}, {Aggarwal}, {Aguiar}, {Aich}, {Aiello},
  {Ain}, {Ajith}, {Allen}, {Allocca}, {Altin}, {Amato}, {Anand}, {Ananyeva},
  {Anderson}, {Anderson}, {Angelova}, {Ansoldi}, {Antier}, {Appert}, {Arai},
  {Araya}, {Areeda}, {Ar{\`e}ne}, {Arnaud}, {Aronson}, {Asali}, {Ascenzi},
  {Ashton}, {Assiduo}, {Aston}, {Astone}, {Aubin}, {Aufmuth}, {AultONeal},
  {Austin}, {Avendano}, {Babak}, {Bacon}, {Badaracco}, {Bader}, {Bae}, {Baer},
  {Baird}, {Baldaccini}, {Ballardin}, {Ballmer}, {Bals}, {Balsamo}, {Baltus},
  {Banagiri}, {Bankar}, {Bankar}, {Barayoga}, {Barbieri}, {Barish}, {Barker},
  {Barkett}, {Barneo}, {Barone}, {Barr}, {Barsotti}, {Barsuglia}, {Barta},
  {Bartlett}, {Bartos}, {Bassiri}, {Basti}, {Bawaj}, {Bayley}, {Bazzan},
  {B{\'e}csy}, {Bejger}, {Belahcene}, {Bell}, {Beniwal}, {Benjamin}, {Bentley},
  {Bergamin}, {Berger}, {Bergmann}, {Bernuzzi}, {Berry}, {Bersanetti},
  {Bertolini}, {Betzwieser}, {Bhandare}, {Bhandari}, {Bianchi}, {Bidler},
  {Biggs}, {Bilenko}, {Billingsley}, {Birney}, {Birnholtz}, {Biscans},
  {Bischi}, {Biscoveanu}, {Bisht}, {Bissenbayeva}, {Bitossi}, {Bizouard},
  {Blackburn}, {Blackman}, {Blair}, {Blair}, {Blair}, {Bobba}, {Bode}, {Boer},
  {Boetzel}, {Bogaert}, {Bondu}, {Bonilla}, {Bonnand}, {Booker}, {Boom},
  {Bork}, {Boschi}, {Bose}, {Bossilkov}, {Bosveld}, {Bouffanais}, {Bozzi},
  {Bradaschia}, {Brady}, {Bramley}, {Branchesi}, {Brau}, {Breschi}, {Briant},
  {Briggs}, {Brighenti}, {Brillet}, {Brinkmann}, {Brockill}, {Brooks},
  {Brooks}, {Brown}, {Brunett}, {Bruno}, {Bruntz}, {Buikema}, {Bulik},
  {Bulten}, {Buonanno}, {Buskulic}, {Byer}, {Cabero}, {Cadonati}, {Cagnoli},
  {Cahillane}, {Calder{\'o}n Bustillo}, {Callaghan}, {Callister}, {Calloni},
  {Camp}, {Canepa}, {Caneva Santoro}, {Cannon}, {Cao}, {Cao}, {Carapella},
  {Carbognani}, {Caride}, {Carney}, {Carullo}, {Carver}, {Casanueva Diaz},
  {Casentini}, {Casta{\~n}eda}, {Caudill}, {Cavagli{\`a}}, {Cavalier},
  {Cavalieri}, {Cella}, {Cerd{\'a}-Dur{\'a}n}, {Cesarini}, {Chaibi},
  {Chakravarti}, {Chan}, {Chan}, {Chao}, {Charlton}, {Chase},
  {Chassande-Mottin}, {Chatterjee}, {Chaturvedi}, {Chen}, {Chen}, {Chen},
  {Cheng}, {Cheong}, {Chia}, {Chiadini}, {Chierici}, {Chincarini}, {Chiummo},
  {Cho}, {Cho}, {Cho}, {Christensen}, {Chu}, {Chua}, {Chung}, {Chung}, {Ciani},
  {Ciecielag}, {Cie\{{\'s}\}lar}, {Ciobanu}, {Ciolfi}, {Cipriano}, {Cirone},
  {Clara}, {Clark}, {Clearwater}, {Clesse}, {Cleva}, {Coccia}, {Cohadon},
  {Cohen}, {Colleoni}, {Collette}, {Collins}, {Colpi}, {Constancio}, {Conti},
  {Cooper}, {Corban}, {Corbitt}, {Cordero-Carri{\'o}n}, {Corezzi}, {Corley},
  {Cornish}, {Corre}, {Corsi}, {Cortese}, {Costa}, {Cotesta}, {Coughlin},
  {Coughlin}, {Coulon}, {Countryman}, {Couvares}, {Covas}, {Coward}, {Cowart},
  {Coyne}, {Coyne}, {Creighton}, {Creighton}, {Cripe}, {Croquette}, {Crowder},
  {Cudell}, {Cullen}, {Cumming}, {Cummings}, {Cunningham}, {Cuoco}, {Curylo},
  {Dal Canton}, {D{\'a}lya}, {Dana}, {Daneshgaran-Bajastani}, {D'Angelo},
  {Danilishin}, {D'Antonio}, {Danzmann}, {Darsow-Fromm}, {Dasgupta}, {Datrier},
  {Dattilo}, {Dave}, {Davier}, {Davies}, {Davis}, {Daw}, {DeBra},
  {Deenadayalan}, {Degallaix}, {De Laurentis}, {Del{\'e}glise}, {Delfavero},
  {De Lillo}, {Del Pozzo}, {DeMarchi}, {D'Emilio}, {Demos}, {Dent}, {De
  Pietri}, {De Rosa}, {De Rossi}, {DeSalvo}, {de Varona}, {Dhurandhar},
  {D{\'\i}az}, {Diaz-Ortiz}, {Dietrich}, {Di Fiore}, {Di Fronzo}, {Di Giorgio},
  {Di Giovanni}, {Di Giovanni}, {Di Girolamo}, {Di Lieto}, {Ding}, {Di Pace},
  {Di Palma}, {Di Renzo}, {Divakarla}, {Dmitriev}, {Doctor}, {Donovan},
  {Dooley}, {Doravari}, {Dorrington}, {Downes}, {Drago}, {Driggers}, {Du},
  {Ducoin}, {Dupej}, {Durante}, {D'Urso}, {Dwyer}, {Easter}, {Eddolls},
  {Edelman}, {Edo}, {Edy}, {Effler}, {Ehrens}, {Eichholz}, {Eikenberry},
  {Eisenmann}, {Eisenstein}, {Ejlli}, {Errico}, {Essick}, {Estelles},
  {Estevez}, {Etienne}, {Etzel}, {Evans}, {Evans}, {Ewing}, {Fafone},
  {Fairhurst}, {Fan}, {Farinon}, {Farr}, {Farr}, {Fauchon-Jones}, {Favata},
  {Fays}, {Fazio}, {Feicht}, {Fejer}, {Feng}, {Fenyvesi}, {Ferguson},
  {Fernandez-Galiana}, {Ferrante}, {Ferreira}, {Ferreira}, {Fidecaro}, {Fiori},
  {Fiorucci}, {Fishbach}, {Fisher}, {Fittipaldi}, {Fitz-Axen}, {Fiumara},
  {Flaminio}, {Floden}, {Flynn}, {Fong}, {Font}, {Forsyth}, {Fournier},
  {Frasca}, {Frasconi}, {Frei}, {Freise}, {Frey}, {Frey}, {Fritschel},
  {Frolov}, {Fronz{\`e}}, {Fulda}, {Fyffe}, {Gabbard}, {Gadre}, {Gaebel},
  {Gair}, {Galaudage}, {Ganapathy}, {Gaonkar}, {Garc{\'\i}a-Quir{\'o}s},
  {Garufi}, {Gateley}, {Gaudio}, {Gayathri}, {Gemme}, {Genin}, {Gennai},
  {George}, {George}, {Gergely}, {Ghonge}, {Ghosh}, {Ghosh}, {Ghosh},
  {Giacomazzo}, {Giaime}, {Giardina}, {Gibson}, {Gier}, {Gill}, {Glanzer},
  {Gniesmer}, {Godwin}, {Goetz}, {Goetz}, {Gohlke}, {Goncharov},
  {Gonz{\'a}lez}, {Gopakumar}, {Gossan}, {Gosselin}, {Gouaty}, {Grace},
  {Grado}, {Granata}, {Grant}, {Gras}, {Grassia}, {Gray}, {Gray}, {Greco},
  {Green}, {Green}, {Gretarsson}, {Griggs}, {Grignani}, {Grimaldi}, {Grimm},
  {Grote}, {Grunewald}, {Gruning}, {Guidi}, {Guimaraes}, {Guix{\'e}}, {Gulati},
  {Guo}, {Gupta}, {Gupta}, {Gupta}, {Gustafson}, {Gustafson}, {Haegel},
  {Halim}, {Hall}, {Hamilton}, {Hammond}, {Haney}, {Hanke}, {Hanks}, {Hanna},
  {Hannam}, {Hannuksela}, {Hansen}, {Hanson}, {Harder}, {Hardwick}, {Haris},
  {Harms}, {Harry}, {Harry}, {Hasskew}, {Haster}, {Haughian}, {Hayes}, {Healy},
  {Heidmann}, {Heintze}, {Heinze}, {Heitmann}, {Hellman}, {Hello}, {Hemming},
  {Hendry}, {Heng}, {Hennes}, {Hennig}, {Heurs}, {Hild}, {Hinderer}, {Hoback},
  {Hochheim}, {Hofgard}, {Hofman}, {Holgado}, {Holland}, {Holt}, {Holz},
  {Hopkins}, {Horst}, {Hough}, {Howell}, {Hoy}, {Huang}, {H{\"u}bner},
  {Huerta}, {Huet}, {Hughey}, {Hui}, {Husa}, {Huttner}, {Huxford},
  {Huynh-Dinh}, {Idzkowski}, {Iess}, {Inchauspe}, {Ingram}, {Intini}, {Isac},
  {Isi}, {Iyer}, {Jacqmin}, {Jadhav}, {Jadhav}, {James}, {Jani}, {Janthalur},
  {Jaranowski}, {Jariwala}, {Jaume}, {Jenkins}, {Jiang}, {Johns}, {Jones},
  {Jones}, {Jones}, {Jones}, {Jones}, {Jonker}, {Ju}, {Junker}, {Kalaghatgi},
  {Kalogera}, {Kamai}, {Kandhasamy}, {Kang}, {Kanner}, {Kapadia}, {Karki},
  {Kashyap}, {Kasprzack}, {Kastaun}, {Katsanevas}, {Katsavounidis}, {Katzman},
  {Kaufer}, {Kawabe}, {K{\'e}f{\'e}lian}, {Keitel}, {Keivani}, {Kennedy},
  {Key}, {Khadka}, {Khalili}, {Khan}, {Khan}, {Khan}, {Khazanov}, {Khetan},
  {Khursheed}, {Kijbunchoo}, {Kim}, {Kim}, {Kim}, {Kim}, {Kim}, {Kim}, {Kim},
  {Kimball}, {King}, {Kinley-Hanlon}, {Kirchhoff}, {Kissel}, {Kleybolte},
  {Klimenko}, {Knowles}, {Knyazev}, {Koch}, {Koehlenbeck}, {Koekoek}, {Koley},
  {Kondrashov}, {Kontos}, {Koper}, {Korobko}, {Korth}, {Kovalam}, {Kozak},
  {Kringel}, {Krishnendu}, {Kr{\'o}lak}, {Krupinski}, {Kuehn}, {Kumar},
  {Kumar}, {Kumar}, {Kumar}, {Kumar}, {Kuo}, {Kutynia}, {Lackey}, {Laghi},
  {Lalande}, {Lam}, {Lamberts}, {Landry}, {Lane}, {Lang}, {Lange}, {Lantz},
  {Lanza}, {La Rosa}, {Lartaux-Vollard}, {Lasky}, {Laxen}, {Lazzarini},
  {Lazzaro}, {Leaci}, {Leavey}, {Lecoeuche}, {Lee}, {Lee}, {Lee}, {Lee}, {Lee},
  {Lehmann}, {Leroy}, {Letendre}, {Levin}, {Li}, {Li}, {li}, {Li}, {Li},
  {Linde}, {Linker}, {Linley}, {Littenberg}, {Liu}, {Liu},
  {Llorens-Monteagudo}, {Lo}, {Lockwood}, {London}, {Longo}, {Lorenzini},
  {Loriette}, {Lormand}, {Losurdo}, {Lough}, {Lousto}, {Lovelace}, {L{\"u}ck},
  {Lumaca}, {Lundgren}, {Ma}, {Macas}, {Macfoy}, {MacInnis}, {Macleod},
  {MacMillan}, {Macquet}, {Maga{\~n}a Hernandez}, {Maga{\~n}a-Sandoval},
  {Magee}, {Majorana}, {Maksimovic}, {Malik}, {Man}, {Mandic}, {Mangano},
  {Mansell}, {Manske}, {Mantovani}, {Mapelli}, {Marchesoni}, {Marion},
  {M{\'a}rka}, {M{\'a}rka}, {Markakis}, {Markosyan}, {Markowitz}, {Maros},
  {Marquina}, {Marsat}, {Martelli}, {Martin}, {Martin}, {Martinez}, {Martynov},
  {Masalehdan}, {Mason}, {Massera}, {Masserot}, {Massinger}, {Masso-Reid},
  {Mastrogiovanni}, {Matas}, {Matichard}, {Mavalvala}, {Maynard}, {McCann},
  {McCarthy}, {McClelland}, {McCormick}, {McCuller}, {McGuire}, {McIsaac},
  {McIver}, {McManus}, {McRae}, {McWilliams}, {Meacher}, {Meadors}, {Mehmet},
  {Mehta}, {Mejuto Villa}, {Melatos}, {Mendell}, {Mercer}, {Mereni}, {Merfeld},
  {Merilh}, {Merritt}, {Merzougui}, {Meshkov}, {Messenger}, {Messick},
  {Metzdorff}, {Meyers}, {Meylahn}, {Mhaske}, {Miani}, {Miao}, {Michaloliakos},
  {Michel}, {Middleton}, {Milano}, {Miller}, {Millhouse}, {Mills}, {Milotti},
  {Milovich-Goff}, {Minazzoli}, {Minenkov}, {Mishkin}, {Mishra}, {Mistry},
  {Mitra}, {Mitrofanov}, {Mitselmakher}, {Mittleman}, {Mo}, {Mogushi},
  {Mohapatra}, {Mohite}, {Molina-Ruiz}, {Mondin}, {Montani}, {Moore}, {Moraru},
  {Morawski}, {Moreno}, {Morisaki}, {Mours}, {Mow-Lowry}, {Mozzon},
  {Muciaccia}, {Mukherjee}, {Mukherjee}, {Mukherjee}, {Mukherjee}, {Mukund},
  {Mullavey}, {Munch}, {Mu{\~n}iz}, {Murray}, {Nagar}, {Nardecchia},
  {Naticchioni}, {Nayak}, {Neil}, {Neilson}, {Nelemans}, {Nelson}, {Nery},
  {Neunzert}, {Ng}, {Ng}, {Nguyen}, {Nguyen}, {Nichols}, {Nichols}, {Nissanke},
  {Nocera}, {Noh}, {North}, {Nothard}, {Nuttall}, {Oberling}, {O'Brien},
  {Oganesyan}, {Ogin}, {Oh}, {Oh}, {Ohme}, {Ohta}, {Okada}, {Oliver},
  {Olivetto}, {Oppermann}, {Oram}, {O'Reilly}, {Ormiston}, {Ormsby}, {Ortega},
  {O'Shaughnessy}, {Ossokine}, {Osthelder}, {Ottaway}, {Overmier}, {Owen},
  {Pace}, {Pagano}, {Page}, {Pagliaroli}, {Pai}, {Pai}, {Palamos}, {Palashov},
  {Palomba}, {Pan}, {Panda}, {Pang}, {Pankow}, {Pannarale}, {Pant}, {Paoletti},
  {Paoli}, {Parida}, {Parker}, {Pascucci}, {Pasqualetti}, {Passaquieti},
  {Passuello}, {Patel}, {Patricelli}, {Payne}, {Pearlstone}, {Pechsiri},
  {Pedersen}, {Pedraza}, {Pele}, {Penn}, {Perego}, {Perez}, {P{\'e}rigois},
  {Perreca}, {Perri{\'e}s}, {Petermann}, {Pfeiffer}, {Phelps}, {Phukon},
  {Piccinni}, {Pichot}, {Piendibene}, {Piergiovanni}, {Pierro}, {Pillant},
  {Pinard}, {Pinto}, {Piotrzkowski}, {Pirello}, {Pitkin}, {Plastino},
  {Poggiani}, {Pong}, {Ponrathnam}, {Popolizio}, {Porter}, {Powell},
  {Prajapati}, {Prasai}, {Prasanna}, {Pratten}, {Prestegard}, {Principe},
  {Prodi}, {Prokhorov}, {Punturo}, {Puppo}, {P{\"u}rrer}, {Qi}, {Quetschke},
  {Quinonez}, {Raab}, {Raaijmakers}, {Radkins}, {Radulesco}, {Raffai},
  {Rafferty}, {Raja}, {Rajan}, {Rajbhandari}, {Rakhmanov}, {Ramirez},
  {Ramos-Buades}, {Rana}, {Rao}, {Rapagnani}, {Raymond}, {Razzano}, {Read},
  {Regimbau}, {Rei}, {Reid}, {Reitze}, {Rettegno}, {Ricci}, {Richardson},
  {Richardson}, {Ricker}, {Riemenschneider}, {Riles}, {Rizzo}, {Robertson},
  {Robinet}, {Rocchi}, {Rodriguez-Soto}, {Rolland}, {Rollins}, {Roma},
  {Romanelli}, {Romano}, {Romel}, {Romero-Shaw}, {Romie}, {Rose}, {Rose},
  {Rose}, {Rosi{\'n}ska}, {Rosofsky}, {Ross}, {Rowan}, {Rowlinson}, {Roy},
  {Roy}, {Roy}, {Ruggi}, {Rutins}, {Ryan}, {Sachdev}, {Sadecki},
  {Sakellariadou}, {Salafia}, {Salconi}, {Saleem}, {Samajdar}, {Sanchez},
  {Sanchez}, {Sanchis-Gual}, {Sanders}, {Santiago}, {Santos}, {Sarin},
  {Sassolas}, {Sathyaprakash}, {Sauter}, {Savage}, {Savant}, {Sawant}, {Sayah},
  {Schaetzl}, {Schale}, {Scheel}, {Scheuer}, {Schmidt}, {Schnabel},
  {Schofield}, {Sch{\"o}nbeck}, {Schreiber}, {Schulte}, {Schutz}, {Schwarm},
  {Schwartz}, {Scott}, {Scott}, {Seidel}, {Sellers}, {Sengupta}, {Sennett},
  {Sentenac}, {Sequino}, {Sergeev}, {Setyawati}, {Shaddock}, {Shaffer},
  {Shahriar}, {Sharifi}, {Sharma}, {Sharma}, {Shawhan}, {Shen}, {Shikauchi},
  {Shink}, {Shoemaker}, {Shoemaker}, {Shukla}, {ShyamSundar}, {Siellez},
  {Sieniawska}, {Sigg}, {Singer}, {Singh}, {Singh}, {Singha}, {Singhal},
  {Sintes}, {Sipala}, {Skliris}, {Slagmolen}, {Slaven-Blair}, {Smetana},
  {Smith}, {Smith}, {Somala}, {Son}, {Soni}, {Sorazu}, {Sordini}, {Sorrentino},
  {Souradeep}, {Sowell}, {Spencer}, {Spera}, {Srivastava}, {Srivastava},
  {Staats}, {Stachie}, {Standke}, {Steer}, {Steinke}, {Steinlechner},
  {Steinlechner}, {Steinmeyer}, {Stocks}, {Stops}, {Stover}, {Strain},
  {Stratta}, {Strunk}, {Sturani}, {Stuver}, {Sudhagar}, {Sudhir},
  {Summerscales}, {Sun}, {Sunil}, {Sur}, {Suresh}, {Sutton}, {Swinkels},
  {Szczepa{\'n}czyk}, {Tacca}, {Tait}, {Talbot}, {Tanasijczuk}, {Tanner},
  {Tao}, {T{\'a}pai}, {Tapia}, {Tapia San Martin}, {Tasson}, {Taylor},
  {Tenorio}, {Terkowski}, {Thirugnanasambandam}, {Thomas}, {Thomas},
  {Thompson}, {Thondapu}, {Thorne}, {Thrane}, {Tinsman}, {Saravanan}, {Tiwari},
  {Tiwari}, {Tiwari}, {Toland}, {Tonelli}, {Tornasi}, {Torres-Forn{\'e}},
  {Torrie}, {Melo}, {T{\"o}yr{\"a}}, {Travasso}, {Traylor}, {Tringali},
  {Tripathee}, {Trovato}, {Trudeau}, {Tsang}, {Tse}, {Tso}, {Tsukada}, {Tsuna},
  {Tsutsui}, {Turconi}, {Ubhi}, {Ueno}, {Ugolini}, {Unnikrishnan}, {Urban},
  {Usman}, {Utina}, {Vahlbruch}, {Vajente}, {Valdes}, {Valentini}, {van Bakel},
  {van Beuzekom}, {van den Brand}, {Van Den Broeck}, {Vander-Hyde}, {van der
  Schaaf}, {Van Heijningen}, {van Veggel}, {Vardaro}, {Varma}, {Vass},
  {Vas{\'u}th}, {Vecchio}, {Vedovato}, {Veitch}, {Veitch}, {Venkateswara},
  {Venugopalan}, {Verkindt}, {Veske}, {Vetrano}, {Vicer{\'e}}, {Viets},
  {Vinciguerra}, {Vine}, {Vinet}, {Vitale}, {Hernandez Vivanco}, {Vo}, {Vocca},
  {Vorvick}, {Vyatchanin}, {Wade}, {Wade}, {Wade}, {Walet}, {Walker},
  {Wallace}, {Wallace}, {Walsh}, {Wang}, {Wang}, {Wang}, {Ward}, {Warden},
  {Warner}, {Was}, {Watchi}, {Weaver}, {Wei}, {Weinert}, {Weinstein}, {Weiss},
  {Wellmann}, {Wen}, {We{\ss}els}, {Westhouse}, {Wette}, {Whelan}, {Whiting},
  {Whittle}, {Wilken}, {Williams}, {Williamson}, {Willis}, {Willke}, {Winkler},
  {Wipf}, {Wittel}, {Woan}, {Woehler}, {Wofford}, {Wong}, {Wright}, {Wu},
  {Wysocki}, {Xiao}, {Yamamoto}, {Yang}, {Yang}, {Yang}, {Yap}, {Yazback},
  {Yeeles}, {Yu}, {Yu}, {Yuen}, {Zadro{\.z}ny}, {Zadro{\.z}ny}, {Zanolin},
  {Zelenova}, {Zendri}, {Zevin}, {Zhang}, {Zhang}, {Zhang}, {Zhao}, {Zhao},
  {Zheng}, {Zhou}, {Zhou}, {Zhu}, {Zucker}, \& {Zweizig}}]{GWTC2_grb2020}
---. 2020{\natexlab{b}}, arXiv e-prints, arXiv:2010.14550.
\newblock \doarXiv{2010.14550}

\bibitem[{{The LIGO Scientific Collaboration} {et~al.}(2020{\natexlab{c}}){The
  LIGO Scientific Collaboration}, {the Virgo Collaboration}, {Abbott},
  {Abbott}, {Abraham}, {Acernese}, {Ackley}, {Adams}, {Adams}, {Adhikari},
  {Adya}, {Affeldt}, {Agathos}, {Agatsuma}, {Aggarwal}, {Aguiar}, {Aiello},
  {Ain}, {Ajith}, {Allen}, {Allocca}, {Altin}, {Amato}, {Anand}, {Ananyeva},
  {Anderson}, {Anderson}, {Angelova}, {Ansoldi}, {Antelis}, {Antier}, {Appert},
  {Arai}, {Araya}, {Areeda}, {Ar{\`e}ne}, {Arnaud}, {Aronson}, {Arun}, {Asali},
  {Ascenzi}, {Ashton}, {Aston}, {Astone}, {Aubin}, {Aufmuth}, {AultONeal},
  {Austin}, {Avendano}, {Babak}, {Badaracco}, {Bader}, {Bae}, {Baer},
  {Bagnasco}, {Baird}, {Ball}, {Ballardin}, {Ballmer}, {Bals}, {Balsamo},
  {Baltus}, {Banagiri}, {Bankar}, {Bankar}, {Barayoga}, {Barbieri}, {Barish},
  \& et~al.}]{LIGO2020_O3populations}
---. 2020{\natexlab{c}}, arXiv e-prints, arXiv:2010.14533.
\newblock \doarXiv{2010.14533}

\bibitem[{{The LIGO Scientific Collaboration} {et~al.}(2021){The LIGO
  Scientific Collaboration}, {the Virgo Collaboration}, {the KAGRA
  Collaboration}, {Abbott}, {Abbott}, {Acernese}, {Ackley}, {Adams},
  {Adhikari}, {Adhikari}, {Adya}, {Affeldt}, {Agarwal}, {Agathos}, {Agatsuma},
  {Aggarwal}, {Aguiar}, {Aiello}, {Ain}, {Ajith}, {Akutsu}, {Albanesi},
  {Allocca}, {Altin}, {Amato}, {Anand}, {Anand}, {Ananyeva}, {Anderson},
  {Anderson}, {Ando}, {Andrade}, {Andres}, {Andri{\'c}}, {Angelova}, {Ansoldi},
  {Antelis}, {Antier}, {Antonini}, {Appert}, {Arai}, {Arai}, {Arai}, {Araki},
  {Araya}, {Araya}, {Areeda}, {Ar{\`e}ne}, {Aritomi}, {Arnaud}, {Aronson},
  {Arun}, {Asada}, {Asali}, {Ashton}, {Aso}, {Assiduo}, {Aston}, {Astone},
  {Aubin}, {Austin}, {Babak}, {Badaracco}, {Bader}, {Badger}, {Bae}, {Bae},
  {Baer}, {Bagnasco}, {Bai}, {Baiotti}, {Baird}, {Bajpai}, {Ball}, {Ballardin},
  {Ballmer}, {Balsamo}, {Baltus}, {Banagiri}, {Bankar}, {Barayoga}, {Barbieri},
  {Barish}, {Barker}, {Barneo}, {Barone}, {Barr}, {Barsotti}, {Barsuglia},
  {Barta}, {Bartlett}, {Barton}, {Bartos}, {Bassiri}, {Basti}, {Bawaj},
  {Bayley}, {Baylor}, {Bazzan}, {B{\'e}csy}, {Bedakihale}, {Bejger},
  {Belahcene}, {Benedetto}, {Beniwal}, {Bennett}, {Bentley}, {BenYaala},
  {Bergamin}, {Berger}, {Bernuzzi}, {Berry}, {Bersanetti}, {Bertolini},
  {Betzwieser}, {Beveridge}, {Bhandare}, {Bhardwaj}, {Bhattacharjee},
  {Bhaumik}, {Bilenko}, {Billingsley}, {Bini}, {Birney}, {Birnholtz},
  {Biscans}, {Bischi}, {Biscoveanu}, {Bisht}, {Biswas}, {Bitossi}, {Bizouard},
  {Blackburn}, {Blair}, {Blair}, {Blair}, {Bobba}, {Bode}, {Boer}, {Bogaert},
  {Boldrini}, {Bonavena}, {Bondu}, {Bonilla}, {Bonnand}, {Booker}, {Boom},
  {Bork}, {Boschi}, {Bose}, {Bose}, {Bossilkov}, {Boudart}, {Bouffanais},
  {Bozzi}, {Bradaschia}, {Brady}, {Bramley}, {Branch}, {Branchesi}, {Brau},
  {Breschi}, {Briant}, {Briggs}, {Brillet}, {Brinkmann}, {Brockill}, {Brooks},
  {Brooks}, {Brown}, {Brunett}, {Bruno}, {Bruntz}, {Bryant}, {Bulik}, {Bulten},
  {Buonanno}, {Buscicchio}, {Buskulic}, {Buy}, {Byer}, {Cadonati}, {Cagnoli},
  {Cahillane}, {Calder{\'o}n Bustillo}, {Callaghan}, {Callister}, {Calloni},
  {Cameron}, {Camp}, {Canepa}, {Canevarolo}, {Cannavacciuolo}, {Cannon}, {Cao},
  {Cao}, {Capocasa}, {Capote}, {Carapella}, {Carbognani}, {Carlin}, {Carney},
  {Carpinelli}, {Carrillo}, {Carullo}, {Carver}, {Casanueva Diaz}, {Casentini},
  {Castaldi}, {Caudill}, {Cavagli{\`a}}, {Cavalier}, {Cavalieri}, {Ceasar},
  {Cella}, {Cerd{\'a}-Dur{\'a}n}, {Cesarini}, {Chaibi}, {Chakravarti},
  {Chalathadka Subrahmanya}, {Champion}, {Chan}, {Chan}, {Chan}, {Chan},
  {Chan}, {Chandra}, {Chanial}, {Chao}, {Charlton}, {Chase},
  {Chassande-Mottin}, {Chatterjee}, {Chatterjee}, {Chatterjee}, {Chaturvedi},
  {Chaty}, {Chatziioannou}, {Chen}, {Chen}, {Chen}, {Chen}, {Chen}, {Chen},
  {Chen}, {Chen}, {Cheng}, {Cheong}, {Cheung}, {Chia}, {Chiadini}, {Chiang},
  {Chiarini}, {Chierici}, {Chincarini}, {Chiofalo}, {Chiummo}, {Cho}, {Cho},
  {Choudhary}, {Choudhary}, {Christensen}, {Chu}, {Chu}, {Chu}, {Chua},
  {Chung}, {Ciani}, {Ciecielag}, {Cie{\'s}lar}, {Cifaldi}, {Ciobanu}, {Ciolfi},
  {Cipriano}, {Cirone}, {Clara}, {Clark}, {Clark}, {Clarke}, {Clearwater},
  {Clesse}, {Cleva}, {Coccia}, {Codazzo}, {Cohadon}, {Cohen}, {Cohen},
  {Colleoni}, {Collette}, {Colombo}, {Colpi}, {Compton}, {Constancio}, {Conti},
  {Cooper}, {Corban}, {Corbitt}, {Cordero-Carri{\'o}n}, {Corezzi}, {Corley},
  {Cornish}, {Corre}, {Corsi}, {Cortese}, {Costa}, {Cotesta}, {Coughlin},
  {Coulon}, {Countryman}, {Cousins}, {Couvares}, {Coward}, {Cowart}, {Coyne},
  {Coyne}, {Creighton}, {Creighton}, {Criswell}, {Croquette}, {Crowder},
  {Cudell}, {Cullen}, {Cumming}, {Cummings}, {Cunningham}, {Cuoco},
  {Cury{\l}o}, {Dabadie}, {Dal Canton}, {Dall'Osso}, {D{\'a}lya}, {Dana},
  {DaneshgaranBajastani}, {D'Angelo}, {Danilishin}, {D'Antonio}, {Danzmann},
  {Darsow-Fromm}, {Dasgupta}, {Datrier}, {Datta}, {Dattilo}, {Dave}, {Davier},
  {Davies}, {Davis}, {Davis}, {Daw}, {Dean}, {DeBra}, {Deenadayalan},
  {Degallaix}, {De Laurentis}, {Del{\'e}glise}, {Del Favero}, {De Lillo}, {De
  Lillo}, {Del Pozzo}, {DeMarchi}, {De Matteis}, {D'Emilio}, {Demos}, {Dent},
  {Depasse}, {De Pietri}, {De Rosa}, {De Rossi}, {DeSalvo}, {De Simone},
  {Dhurandhar}, {D{\'\i}az}, {Diaz-Ortiz}, {Didio}, {Dietrich}, {Di Fiore}, {Di
  Fronzo}, {Di Giorgio}, {Di Giovanni}, {Di Giovanni}, {Di Girolamo}, {Di
  Lieto}, {Ding}, {Di Pace}, {Di Palma}, {Di Renzo}, {Divakarla}, {Dmitriev},
  {Doctor}, {D'Onofrio}, {Donovan}, {Dooley}, {Doravari}, {Dorrington},
  {Drago}, {Driggers}, {Drori}, {Ducoin}, {Dupej}, {Durante}, {D'Urso},
  {Duverne}, {Dwyer}, {Eassa}, {Easter}, {Ebersold}, {Eckhardt}, {Eddolls},
  {Edelman}, {Edo}, {Edy}, {Effler}, {Eguchi}, {Eichholz}, {Eikenberry},
  {Eisenmann}, {Eisenstein}, {Ejlli}, {Engelby}, {Enomoto}, {Errico}, {Essick},
  {Estell{\'e}s}, {Estevez}, {Etienne}, {Etzel}, {Evans}, {Evans}, {Ewing},
  {Fafone}, {Fair}, {Fairhurst}, {Farah}, {Farinon}, {Farr}, {Farr}, {Farrow},
  {Fauchon-Jones}, {Favaro}, {Favata}, {Fays}, {Fazio}, {Feicht}, {Fejer},
  {Fenyvesi}, {Ferguson}, {Fernandez-Galiana}, {Ferrante}, {Ferreira},
  {Fidecaro}, {Figura}, {Fiori}, {Fishbach}, {Fisher}, {Fittipaldi}, {Fiumara},
  {Flaminio}, {Floden}, {Fong}, {Font}, {Fornal}, {Forsyth}, {Franke},
  {Frasca}, {Frasconi}, {Frederick}, {Freed}, {Frei}, {Freise}, {Frey},
  {Fritschel}, {Frolov}, {Fronz{\'e}}, {Fujii}, {Fujikawa}, {Fukunaga},
  {Fukushima}, {Fulda}, {Fyffe}, {Gabbard}, {Gadre}, {Gair}, {Gais},
  {Galaudage}, {Gamba}, {Ganapathy}, {Ganguly}, {Gao}, {Gaonkar}, {Garaventa},
  {Garc{\'\i}a-N{\'u}{\~n}ez}, {Garc{\'\i}a-Quir{\'o}s}, {Garufi}, {Gateley},
  {Gaudio}, {Gayathri}, {Ge}, {Gemme}, {Gennai}, {George}, {Gerberding},
  {Gergely}, {Gewecke}, {Ghonge}, {Ghosh}, {Ghosh}, {Ghosh}, {Ghosh},
  {Giacomazzo}, {Giacoppo}, {Giaime}, {Giardina}, {Gibson}, {Gier}, {Giesler},
  {Giri}, {Gissi}, {Glanzer}, {Gleckl}, {Godwin}, {Goetz}, {Goetz}, {Gohlke},
  {Golomb}, {Goncharov}, {Gonz{\'a}lez}, {Gopakumar}, {Gosselin}, {Gouaty},
  {Gould}, {Grace}, {Grado}, {Granata}, {Granata}, {Grant}, {Gras}, {Grassia},
  {Gray}, {Gray}, {Greco}, {Green}, {Green}, {Gretarsson}, {Gretarsson},
  {Griffith}, {Griffiths}, {Griggs}, {Grignani}, {Grimaldi}, {Grimm}, {Grote},
  {Grunewald}, {Gruning}, {Guerra}, {Guidi}, {Guimaraes}, {Guix{\'e}},
  {Gulati}, {Guo}, {Guo}, {Gupta}, {Gupta}, {Gupta}, {Gustafson}, {Gustafson},
  {Guzman}, {Ha}, {Haegel}, {Hagiwara}, {Haino}, {Halim}, {Hall}, {Hamilton},
  {Hammond}, {Han}, {Haney}, {Hanks}, {Hanna}, {Hannam}, {Hannuksela},
  {Hansen}, {Hansen}, {Hanson}, {Harder}, {Hardwick}, {Haris}, {Harms},
  {Harry}, {Harry}, {Hartwig}, {Hasegawa}, {Haskell}, {Hasskew}, {Haster},
  {Hattori}, {Haughian}, {Hayakawa}, {Hayama}, {Hayes}, {Healy}, {Heidmann},
  {Heidt}, {Heintze}, {Heinze}, {Heinzel}, {Heitmann}, {Hellman}, {Hello},
  {Helmling-Cornell}, {Hemming}, {Hendry}, {Heng}, {Hennes}, {Hennig},
  {Hennig}, {Hernandez}, {Hernandez Vivanco}, {Heurs}, {Hild}, {Hill},
  {Himemoto}, {Hines}, {Hiranuma}, {Hirata}, {Hirose}, {Hochheim}, {Hofman},
  {Hohmann}, {Holcomb}, {Holland}, {Hollows}, {Holmes}, {Holt}, {Holz}, {Hong},
  {Hopkins}, {Hough}, {Hourihane}, {Howell}, {Hoy}, {Hoyland}, {Hreibi},
  {Hsieh}, {Hsu}, {Huang}, {Huang}, {Huang}, {Huang}, {Huang}, {Huang},
  {H{\"u}bner}, {Huddart}, {Hughey}, {Hui}, {Hui}, {Husa}, {Huttner},
  {Huxford}, {Huynh-Dinh}, {Ide}, {Idzkowski}, {Iess}, {Ikenoue}, {Imam},
  {Inayoshi}, {Ingram}, {Inoue}, {Ioka}, {Isi}, {Isleif}, {Ito}, {Itoh},
  {Iyer}, {Izumi}, {JaberianHamedan}, {Jacqmin}, {Jadhav}, {Jadhav}, {James},
  {Jan}, {Jani}, {Janquart}, {Janssens}, {Janthalur}, {Jaranowski}, {Jariwala},
  {Jaume}, {Jenkins}, {Jenner}, {Jeon}, {Jeunon}, {Jia}, {Jin}, {Johns},
  {Jones}, {Jones}, {Jones}, {Jones}, {Jones}, {Jonker}, {Ju}, {Jung}, {Jung},
  {Junker}, {Juste}, {Kaihotsu}, {Kajita}, {Kakizaki}, {Kalaghatgi},
  {Kalogera}, {Kamai}, {Kamiizumi}, {Kanda}, {Kandhasamy}, {Kang}, {Kanner},
  {Kao}, {Kapadia}, {Kapasi}, {Karat}, {Karathanasis}, {Karki}, {Kashyap},
  {Kasprzack}, {Kastaun}, {Katsanevas}, {Katsavounidis}, {Katzman}, {Kaur},
  {Kawabe}, {Kawaguchi}, {Kawai}, {Kawasaki}, {K{\'e}f{\'e}lian}, {Keitel},
  {Key}, {Khadka}, {Khalili}, {Khan}, {Khazanov}, {Khetan}, {Khursheed},
  {Kijbunchoo}, {Kim}, {Kim}, {Kim}, {Kim}, {Kim}, {Kim}, {Kimball}, {Kimura},
  {Kinley-Hanlon}, {Kirchhoff}, {Kissel}, {Kita}, {Kitazawa}, {Kleybolte},
  {Klimenko}, {Knee}, {Knowles}, {Knyazev}, {Koch}, {Koekoek}, {Kojima},
  {Kokeyama}, {Koley}, {Kolitsidou}, {Kolstein}, {Komori}, {Kondrashov},
  {Kong}, {Kontos}, {Koper}, {Korobko}, {Kotake}, {Kovalam}, {Kozak},
  {Kozakai}, {Kozu}, {Kringel}, {Krishnendu}, {Kr{\'o}lak}, {Kuehn}, {Kuei},
  {Kuijer}, {Kumar}, {Kumar}, {Kumar}, {Kumar}, {Kume}, {Kuns}, {Kuo}, {Kuo},
  {Kuromiya}, {Kuroyanagi}, {Kusayanagi}, {Kuwahara}, {Kwak}, {Lagabbe},
  {Laghi}, {Lalande}, {Lam}, {Lamberts}, {Landry}, {Landry}, {Lane}, {Lang},
  {Lange}, {Lantz}, {La Rosa}, {Lartaux-Vollard}, {Lasky}, {Laxen},
  {Lazzarini}, {Lazzaro}, {Leaci}, {Leavey}, {Lecoeuche}, {Lee}, {Lee}, {Lee},
  {Lee}, {Lee}, {Lee}, {Lehmann}, {Lema{\^\i}tre}, {Leonardi}, {Leroy},
  {Letendre}, {Levesque}, {Levin}, {Leviton}, {Leyde}, {Li}, {Li}, {Li}, {Li},
  {Li}, {Li}, {Lin}, {Lin}, {Lin}, {Lin}, {Lin}, {Linde}, {Linker}, {Linley},
  {Littenberg}, {Liu}, {Liu}, {Liu}, {Liu}, {Llamas}, {Llorens-Monteagudo},
  {Lo}, {Lockwood}, {London}, {Longo}, {Lopez}, {Lopez Portilla}, {Lorenzini},
  {Loriette}, {Lormand}, {Losurdo}, {Lott}, {Lough}, {Lousto}, {Lovelace},
  {Lucaccioni}, {L{\"u}ck}, {Lumaca}, {Lundgren}, {Luo}, {Lynam}, {Macas},
  {MacInnis}, {Macleod}, {MacMillan}, {Macquet}, {Maga{\~n}a Hernandez},
  {Magazz{\`u}}, {Magee}, {Maggiore}, {Magnozzi}, {Mahesh}, {Majorana},
  {Makarem}, {Maksimovic}, {Maliakal}, {Malik}, {Man}, {Mandic}, {Mangano},
  {Mango}, {Mansell}, {Manske}, {Mantovani}, {Mapelli}, {Marchesoni},
  {Marchio}, {Marion}, {Mark}, {M{\'a}rka}, {M{\'a}rka}, {Markakis},
  {Markosyan}, {Markowitz}, {Maros}, {Marquina}, {Marsat}, {Martelli},
  {Martin}, {Martin}, {Martinez}, {Martinez}, {Martinez}, {Martinovic},
  {Martynov}, {Marx}, {Masalehdan}, {Mason}, {Massera}, {Masserot},
  {Massinger}, {Masso-Reid}, {Mastrogiovanni}, {Matas}, {Mateu-Lucena},
  {Matichard}, {Matiushechkina}, {Mavalvala}, {McCann}, {McCarthy},
  {McClelland}, {McClincy}, {McCormick}, {McCuller}, {McGhee}, {McGuire},
  {McIsaac}, {McIver}, {McRae}, {McWilliams}, {Meacher}, {Mehmet}, {Mehta},
  {Meijer}, {Melatos}, {Melchor}, {Mendell}, {Menendez-Vazquez}, {Menoni},
  {Mercer}, {Mereni}, {Merfeld}, {Merilh}, {Merritt}, {Merzougui}, {Meshkov},
  {Messenger}, {Messick}, {Meyers}, {Meylahn}, {Mhaske}, {Miani}, {Miao},
  {Michaloliakos}, {Michel}, {Michimura}, {Middleton}, {Milano}, {Miller},
  {Miller}, {Miller}, {Miller}, {Millhouse}, {Mills}, {Milotti}, {Minazzoli},
  {Minenkov}, {Mio}, {Mir}, {Miravet-Ten{\'e}s}, {Mishra}, {Mishra}, {Mistry},
  {Mitra}, {Mitrofanov}, {Mitselmakher}, {Mittleman}, {Miyakawa}, {Miyamoto},
  {Miyazaki}, {Miyo}, {Miyoki}, {Mo}, {Moguel}, {Mogushi}, {Mohapatra},
  {Mohite}, {Molina}, {Molina-Ruiz}, {Mondin}, {Montani}, {Moore}, {Moraru},
  {Morawski}, {More}, {Moreno}, {Moreno}, {Mori}, {Morisaki}, {Moriwaki},
  {Mours}, {Mow-Lowry}, {Mozzon}, {Muciaccia}, {Mukherjee}, {Mukherjee},
  {Mukherjee}, {Mukherjee}, {Mukherjee}, {Mukund}, {Mullavey}, {Munch},
  {Mu{\~n}iz}, {Murray}, {Musenich}, {Muusse}, {Nadji}, {Nagano}, {Nagano},
  {Nagar}, {Nakamura}, {Nakano}, {Nakano}, {Nakashima}, {Nakayama}, {Napolano},
  {Nardecchia}, {Narikawa}, {Naticchioni}, {Nayak}, {Nayak}, {Negishi}, {Neil},
  {Neilson}, {Nelemans}, {Nelson}, {Nery}, {Neubauer}, {Neunzert}, {Ng}, {Ng},
  {Nguyen}, {Nguyen}, {Nguyen}, {Nguyen Quynh}, {Ni}, {Nichols}, {Nishizawa},
  {Nissanke}, {Nitoglia}, {Nocera}, {Norman}, {North}, {Nozaki}, {Nuttall},
  {Oberling}, {O'Brien}, {Obuchi}, {O'Dell}, {Oelker}, {Ogaki}, {Oganesyan},
  {Oh}, {Oh}, {Oh}, {Ohashi}, {Ohishi}, {Ohkawa}, {Ohme}, {Ohta}, {Okada},
  {Okutani}, {Okutomi}, {Olivetto}, {Oohara}, {Ooi}, {Oram}, {O'Reilly},
  {Ormiston}, {Ormsby}, {Ortega}, {O'Shaughnessy}, {O'Shea}, {Oshino},
  {Ossokine}, {Osthelder}, {Otabe}, {Ottaway}, {Overmier}, {Pace}, {Pagano},
  {Page}, {Pagliaroli}, {Pai}, {Pai}, {Palamos}, {Palashov}, {Palomba}, {Pan},
  {Pan}, {Panda}, {Pang}, {Pang}, {Pankow}, {Pannarale}, {Pant}, {Panther},
  {Paoletti}, {Paoli}, {Paolone}, {Parisi}, {Park}, {Park}, {Parker},
  {Pascucci}, {Pasqualetti}, {Passaquieti}, {Passuello}, {Patel}, {Pathak},
  {Patricelli}, {Patron}, {Paul}, {Payne}, {Pedraza}, {Pegoraro}, {Pele},
  {Pe{\~n}a Arellano}, {Penn}, {Perego}, {Pereira}, {Pereira}, {Perez},
  {P{\'e}rigois}, {Perkins}, {Perreca}, {Perri{\`e}s}, {Petermann},
  {Petterson}, {Pfeiffer}, {Pham}, {Phukon}, {Piccinni}, {Pichot},
  {Piendibene}, {Piergiovanni}, {Pierini}, {Pierro}, {Pillant}, {Pillas},
  {Pilo}, {Pinard}, {Pinto}, {Pinto}, {Piotrzkowski}, {Pirello}, {Pitkin},
  {Placidi}, {Planas}, {Plastino}, {Pluchar}, {Poggiani}, {Polini}, {Pong},
  {Ponrathnam}, {Popolizio}, {Porter}, {Poulton}, {Powell}, {Pracchia},
  {Pradier}, {Prajapati}, {Prasai}, {Prasanna}, {Pratten}, {Principe}, {Prodi},
  {Prokhorov}, {Prosposito}, {Prudenzi}, {Puecher}, {Punturo}, {Puosi},
  {Puppo}, {P{\"u}rrer}, {Qi}, {Quetschke}, {Quitzow-James}, {Raab},
  {Raaijmakers}, {Radkins}, {Radulesco}, {Raffai}, {Rail}, {Raja}, {Rajan},
  {Ramirez}, {Ramirez}, {Ramos-Buades}, {Rana}, {Rapagnani}, {Rapol}, {Ray},
  {Raymond}, {Raza}, {Razzano}, {Read}, {Rees}, {Regimbau}, {Rei}, {Reid},
  {Reid}, {Reitze}, {Relton}, {Renzini}, {Rettegno}, {Rezac}, {Ricci},
  {Richards}, {Richardson}, {Richardson}, {Riemenschneider}, {Riles},
  {Rinaldi}, {Rink}, {Rizzo}, {Robertson}, {Robie}, {Robinet}, {Rocchi},
  {Rodriguez}, {Rolland}, {Rollins}, {Romanelli}, {Romano}, {Romel},
  {Romero-Rodr{\'\i}guez}, {Romero-Shaw}, {Romie}, {Ronchini}, {Rosa}, {Rose},
  {Rosi{\'n}ska}, {Ross}, {Rowan}, {Rowlinson}, {Roy}, {Roy}, {Roy}, {Rozza},
  {Ruggi}, {Ryan}, {Sachdev}, {Sadecki}, {Sadiq}, {Sago}, {Saito}, {Saito},
  {Sakai}, {Sakai}, {Sakellariadou}, {Sakuno}, {Salafia}, {Salconi}, {Saleem},
  {Salemi}, {Samajdar}, {Sanchez}, {Sanchez}, {Sanchez}, {Sanchis-Gual},
  {Sanders}, {Sanuy}, {Saravanan}, {Sarin}, {Sassolas}, {Satari},
  {Sathyaprakash}, {Sato}, {Sato}, {Sauter}, {Savage}, {Sawada}, {Sawant},
  {Sawant}, {Sayah}, {Schaetzl}, {Scheel}, {Scheuer}, {Schiworski}, {Schmidt},
  {Schmidt}, {Schnabel}, {Schneewind}, {Schofield}, {Sch{\"o}nbeck}, {Schulte},
  {Schutz}, {Schwartz}, {Scott}, {Scott}, {Seglar-Arroyo}, {Sekiguchi},
  {Sekiguchi}, {Sellers}, {Sengupta}, {Sentenac}, {Seo}, {Sequino}, {Sergeev},
  {Setyawati}, {Shaffer}, {Shahriar}, {Shams}, {Shao}, {Sharma}, {Sharma},
  {Shawhan}, {Shcheblanov}, {Shibagaki}, {Shikauchi}, {Shimizu}, {Shimoda},
  {Shimode}, {Shinkai}, {Shishido}, {Shoda}, {Shoemaker}, {Shoemaker},
  {ShyamSundar}, {Sieniawska}, {Sigg}, {Singer}, {Singh}, {Singh}, {Singha},
  {Sintes}, {Sipala}, {Skliris}, {Slagmolen}, {Slaven-Blair}, {Smetana},
  {Smith}, {Smith}, {Soldateschi}, {Somala}, {Somiya}, {Son}, {Soni}, {Soni},
  {Sordini}, {Sorrentino}, {Sorrentino}, {Sotani}, {Soulard}, {Souradeep},
  {Sowell}, {Spagnuolo}, {Spencer}, {Spera}, {Srinivasan}, {Srivastava},
  {Srivastava}, {Staats}, {Stachie}, {Steer}, {Steinlechner}, {Steinlechner},
  {Stops}, {Stover}, {Strain}, {Strang}, {Stratta}, {Strunk}, {Sturani},
  {Stuver}, {Sudhagar}, {Sudhir}, {Sugimoto}, {Suh}, {Summerscales}, {Sun},
  {Sun}, {Sunil}, {Sur}, {Suresh}, {Sutton}, {Suzuki}, {Suzuki}, {Swinkels},
  {Szczepa{\'n}czyk}, {Szewczyk}, {Tacca}, {Tagoshi}, {Tait}, {Takahashi},
  {Takahashi}, {Takamori}, {Takano}, {Takeda}, {Takeda}, {Talbot}, {Talbot},
  {Tanaka}, {Tanaka}, {Tanaka}, {Tanaka}, {Tanaka}, {Tanasijczuk}, {Tanioka},
  {Tanner}, {Tao}, {Tao}, {Tapia San Mart{\'\i}n}, {Taranto}, {Tasson},
  {Telada}, {Tenorio}, {Terhune}, {Terkowski}, {Thirugnanasambandam}, {Thomas},
  {Thomas}, {Thompson}, {Thondapu}, {Thorne}, {Thrane}, {Tiwari}, {Tiwari},
  {Tiwari}, {Toivonen}, {Toland}, {Tolley}, {Tomaru}, {Tomigami}, {Tomura},
  {Tonelli}, {Torres-Forn{\'e}}, {Torrie}, {Tosta e Melo}, {T{\"o}yr{\"a}},
  {Trapananti}, {Travasso}, {Traylor}, {Trevor}, {Tringali}, {Tripathee},
  {Troiano}, {Trovato}, {Trozzo}, {Trudeau}, {Tsai}, {Tsai}, {Tsang}, {Tsang},
  {Tsao}, {Tse}, {Tso}, {Tsubono}, {Tsuchida}, {Tsukada}, {Tsuna}, {Tsutsui},
  {Tsuzuki}, {Turbang}, {Turconi}, {Tuyenbayev}, {Ubhi}, {Uchikata},
  {Uchiyama}, {Udall}, {Ueda}, {Uehara}, {Ueno}, {Ueshima}, {Unnikrishnan},
  {Uraguchi}, {Urban}, {Ushiba}, {Utina}, {Vahlbruch}, {Vajente}, {Vajpeyi},
  {Valdes}, {Valentini}, {Valsan}, {van Bakel}, {van Beuzekom}, {van den
  Brand}, {Van Den Broeck}, {Vander-Hyde}, {van der Schaaf}, {van Heijningen},
  {Vanosky}, {van Putten}, {van Remortel}, {Vardaro}, {Vargas}, {Varma},
  {Vas{\'u}th}, {Vecchio}, {Vedovato}, {Veitch}, {Veitch}, {Venneberg},
  {Venugopalan}, {Verkindt}, {Verma}, {Verma}, {Veske}, {Vetrano},
  {Vicer{\'e}}, {Vidyant}, {Viets}, {Vijaykumar}, {Villa-Ortega}, {Vinet},
  {Virtuoso}, {Vitale}, {Vo}, {Vocca}, {von Reis}, {von Wrangel}, {Vorvick},
  {Vyatchanin}, {Wade}, {Wade}, {Wagner}, {Walet}, {Walker}, {Wallace},
  {Wallace}, {Walsh}, {Wang}, {Wang}, {Wang}, {Ward}, {Warner}, {Was},
  {Washimi}, {Washington}, {Watchi}, {Weaver}, {Webster}, {Weinert},
  {Weinstein}, {Weiss}, {Weller}, {Wellmann}, {Wen}, {We{\ss}els}, {Wette},
  {Whelan}, {White}, {Whiting}, {Whittle}, {Wilken}, {Williams}, {Williams},
  {Williamson}, {Willis}, {Willke}, {Wilson}, {Winkler}, {Wipf}, {Wlodarczyk},
  {Woan}, {Woehler}, {Wofford}, {Wong}, {Wu}, {Wu}, {Wu}, {Wu}, {Wysocki},
  {Xiao}, {Xu}, {Yamada}, {Yamamoto}, {Yamamoto}, {Yamamoto}, {Yamamoto},
  {Yamashita}, {Yamazaki}, {Yang}, {Yang}, {Yang}, {Yang}, {Yang}, {Yap},
  {Yeeles}, {Yelikar}, {Ying}, {Yokogawa}, {Yokoyama}, {Yokozawa}, {Yoo},
  {Yoshioka}, {Yu}, {Yu}, {Yuzurihara}, {Zadro{\.z}ny}, {Zanolin}, {Zeidler},
  {Zelenova}, {Zendri}, {Zevin}, {Zhan}, {Zhang}, {Zhang}, {Zhang}, {Zhang},
  {Zhang}, {Zhao}, {Zhao}, {Zhao}, {Zhao}, {Zhou}, {Zhou}, {Zhu}, {Zhu},
  {Zimmerman}, {Zlochower}, {Zucker}, \& {Zweizig}}]{LVK21}
{The LIGO Scientific Collaboration}, {the Virgo Collaboration}, {the KAGRA
  Collaboration}, {et~al.} 2021, arXiv e-prints, arXiv:2111.03634.
\newblock \doarXiv{2111.03634}

\bibitem[{{Umbreit} {et~al.}(2012){Umbreit}, {Fregeau}, {Chatterjee}, \&
  {Rasio}}]{Umbreit2012U}
{Umbreit}, S., {Fregeau}, J.~M., {Chatterjee}, S., \& {Rasio}, F.~A. 2012,
  \apj, 750, 31, \dodoi{10.1088/0004-637X/750/1/31}

\bibitem[{{Weatherford} {et~al.}(2021){Weatherford}, {Fragione}, {Kremer},
  {Chatterjee}, {Ye}, {Rodriguez}, \& {Rasio}}]{Weatherford21}
{Weatherford}, N.~C., {Fragione}, G., {Kremer}, K., {et~al.} 2021, \apjl, 907,
  L25, \dodoi{10.3847/2041-8213/abd79c}

\bibitem[{{Woosley}(2017)}]{Woosley2017}
{Woosley}, S.~E. 2017, \apj, 836, 244, \dodoi{10.3847/1538-4357/836/2/244}

\bibitem[{{Woosley}(2019)}]{Woosley2019}
---. 2019, \apj, 878, 49, \dodoi{10.3847/1538-4357/ab1b41}

\bibitem[{{Zevin} \& {Holz}(2022)}]{Zevin22}
{Zevin}, M., \& {Holz}, D.~E. 2022, arXiv e-prints, arXiv:2205.08549.
\newblock \doarXiv{2205.08549}

\bibitem[{Zwart \& McMillan(2002)}]{Portegies_Zwart_2002}
Zwart, S. F.~P., \& McMillan, S. L.~W. 2002, The Astrophysical Journal, 576,
  899, \dodoi{10.1086/341798}

\end{thebibliography}

\end{document}